\documentclass[12pt]{article}
\usepackage{amsmath}
\usepackage{graphicx,psfrag,epsf}
\usepackage{enumerate}
\usepackage{natbib}
\usepackage{amsfonts}
\usepackage{color}
\usepackage{hyperref}
\usepackage{float}
\usepackage{bbm}
\usepackage{array}
\usepackage{booktabs}
\usepackage{subcaption}
\usepackage{cleveref}
\usepackage{caption,setspace,verbatim}
\usepackage{algorithm2e}
\usepackage{siunitx}
\usepackage{pdflscape} 
\usepackage{bm}
\usepackage{listings}
\usepackage{makecell}
\usepackage{setspace}
\usepackage{indentfirst}
\usepackage{caption}
\usepackage{multirow}

\newcommand{\blind}{0}

\begin{document}

\def\spacingset#1{\renewcommand{\baselinestretch}%
{#1}\small\normalsize} \spacingset{1}


\if0\blind
{
  \title{\bf Time-Varying Functional Cox Model}
  \author{Hongyu Du\thanks{The authors gratefully acknowledge}\hspace{0.2cm}and Andrew Leroux\\
University of Colorado Anschutz Medical Campus}
  \maketitle
} \fi

\if1\blind
{
  \bigskip
  \bigskip
  \bigskip
  \begin{center}
    {\LARGE\bf Title}
\end{center}
  \medskip
} \fi

\bigskip
\begin{abstract}
In this work, we propose two novel approaches for estimating the time-varying effects of functional predictors under a linear functional Cox model framework. The time-varying functional Cox model allows for domain specific time-varying associations of a functional predictor completely observed at baseline. The approaches both use penalized regression splines for estimating a coefficient surface that is smooth in both the functional domain and event time process. Methods are implemented in high-quality, fast, stable software for penalized spline regression. The first approach, suitable for small-to-medium-sized datasets, utilizes the connection between the Cox partial likelihood and the Poisson likelihood, providing valid estimation and inference. The second approach employs a landmark approach for approximating the time-varying effect of interest, which can substantially decrease the computational burden for larger datasets and/or high-dimensional functional predictors. The two models, the Time-Varying Functional Linear Cox Model (TV-FLCM) and the corresponding landmark model, the Time-Varying Landmark Functional Linear Cox Model (TV-FLCM-L), both address the issue of violations of the proportional hazards assumption for the association between a functional predictor measured at baseline and a time-to-event outcome subject to right censoring. The model is linear in the functional predictor but allows the association to vary non-linearly in both the functional domain and the follow-up time as a bivariate smooth coefficient. The development of the methods was motivated by the study of the association between all-cause mortality and diurnal patterns of motor activity estimated in 1-minute epochs (where motor activity is the functional predictor and time of day is the functional domain) in the 2011-2014 National Health and Nutrition Survey (NHANES). In this large dataset (N=4445, dimension of the functional predictor=1440), the first method is computationally infeasible on standard laptops, necessitating the development of a complementary landmark approach. Combining the computational efficiency of the landmark approach to estimating time-varying effects with powerful, high-quality, open-source software for performing non-parametric regression using penalized splines with automated smoothing parameter selection via the {\ttfamily mgcv} package allows the approach to scale extremely well computationally. We assess the performance of our methods in a comprehensive simulation study, showing that both methods obtain high accuracy in estimating the shape of the functional coefficient, while the landmark approach attains orders of magnitude faster computation times with only a small loss of accuracy in estimation. In our simulation study, the non-landmark approach attains nominal coverage probability, indicating good inference properties of the method for the scenarios considered. Inference for the landmark approach was not assessed due to the bias induced by the nature of the approach. Sensitivity to landmark modeling choices (landmark times and window lengths) is assessed. We apply our method to estimate a time-varying functional effect of objectively measured diurnal patterns of physical activity on mortality in the NHANES 2011-2014, showing an attenuation of diurnal effects over an 8-year follow-up period.
\end{abstract}

\noindent%

\spacingset{1.45}
\section{Introduction}
\label{sec:intro}


With advances in technology, high-dimensional datasets, particularly functional data, are increasingly common. Unlike traditional datasets where observations are points, functional data treats each observation as a function, typically modeled as stochastic Gaussian processes \citep{wang2020intuitive}. Functional data analysis has emerged as a vital area of statistics, offering tools to study complex relationships within such data \citep{JO2005,stadtmuller2014introduction,wang2016functional}. Functional regression, a key tool in this domain, includes models such as scalar-on-function, function-on-scalar, and function-on-function regression \citep{chiou2004functional,malfait2003historical,greven2011longitudinal,cai2006prediction,greven2017general}. Time-to-event analysis with functional predictors is a relatively new area of research \citep{kong2018flcrm, yang2021weighted, qu2016optimal, gellar2015cox, lee2015bflcrm, leroux2020statistical}. Here, we extend the linear function Cox regression model by allowing for functional effects that vary over time (e.g., a violation of the proportional hazards assumption in Cox regression). This Time-Varying Functional Linear Cox Model (TV-FLCM) addresses violations of the proportional hazards assumption by allowing functional effects to vary smoothly over the functional domain and time \citep{tian2005cox, zhang2018time, winnett2001miscellanea, perperoglou2014cox, tan2012time, andrinopoulou2018improved}. The development of our method is motivated by the accelerometry data NHANES 2011-2104 \citep{leroux2019organizing}, where the daily patterns of physical activity continuously measured at high frequency (80hz) and summarized in 1-minute epochs are the functional predictor of interest \citep{leroux2018rnhanesdata}, and the time to all-cause mortality is the result. 

Given the computational challenges of high-dimensional data, we also propose a complementary Landmark Model (TV-FLCM-L), which approximates time-varying effects using partitioned follow-up periods. This approach significantly reduces computational burden while retaining accuracy, making it suitable for large datasets. In \cite{leroux2020statistical}'s paper, they proposed a similar model where the functional term is simultaneously controlled by both survival time $t$ and functional time $u$ as well. They explored two methods to tackle the model: the joint longitudinal and survival model \citep{tsiatis1995modeling, tsiatis2004joint, faucett1996simultaneously, crowther2013joint, wulfsohn1997joint, andersen1982cox} and the landmark approach \citep{rizopoulos2017dynamic, van2007dynamic}. The model can be categorized as a joint model, as the trajectory of $u$ is inherently regulated by time $t$ \citep{laird1982random, greven2011longitudinal, guo2002functional}. Other methods that can smoothly address time-varying effects and cumulative effects with leads and lags could also be applied to the model \citep{bender2019penalized}, and these will be explored in future work.

Motivated by its application to the NHANES dataset, our study explores the relationship between diurnal activity patterns and all-cause mortality over an eight-year follow-up. Through simulations and real-world data, we demonstrate the effectiveness and computational efficiency of our models. In this paper, Section~\ref{hflm:sec:methods} outlines the methodology of the TV-FLCM-L and its estimation through penalized partial likelihood. Section~\ref{hflm:sec:Landmark} delves into the precise steps for utilizing the `mgcv' package for computational purposes. In Section~\ref{hflm:sec:NHANES}, we apply our model to real-world NHANES datasets. Section~\ref{hflm:sec:simulation} conducts an extensive simulation study to evaluate and compare the relative performance of these novels. Section \ref{hflm:sec:discussion} gives the discussion of the paper. The supplementary material extends the theoretical foundation of Functional Principal Component Analysis (FPCA) for this model and covers extensions regarding dynamic predictions, smoothing parameters, the full likelihood approach, and additive models. 


\section{Methodology}
\label{hflm:sec:methods}

\noindent We begin by introducing the notation necessary to describe our method. Let $T_i$ denotes the time to event (survival time) for subject $i$ and $C_i$ denotes the right censoring time. We observe $Y_i =\text{min}(T_i,C_i)$ and let $\delta_i=I(T_i \le C_i)$ be the event indicator. For simplicity of presentation, consider a single scalar covariate whose information is available at baseline $x_i$, and a single square integrable functional predictor $\{Z_i(u): u \in \mathcal{U}\}$, whose information is similarly available at baseline. In practice, the functional predictor is observed on a discrete set of points in the functional domain $\mathcal{U}$. Note that the method we propose here readily extends to multiple scalar and functional predictors. As we are modeling the association between predictors and time-to-event using the Cox framework, we propose an additive model for the subject-specific log hazard as follows
\begin{align*}
\log\lambda_i(t|X_i,\{Z_i(u): u \in \mathcal{U}\}) = \log\lambda_0(t) + \eta_i(t)
\end{align*}
where $\lambda_0(t)$ is the baseline hazard function, left unspecified, and $\eta_i(t)$ contains the possibly time-specific information on the effects of the scalar and functional covariates on the subject-specific hazard function. In a standard Cox regression model with a scalar covariate, $\eta_i(t) = \eta_i = x_i \beta$, to build up to the Time-varying Functional Linear Cox Model, we first introduce the Cox model with time-varying effects in Section~\ref{hflm:subsec:TVCOX}, followed by the linear functional Cox model in Section~\ref{hflm:subsec:FLCM}. We then merge these two ideas to present the Time-varying Functional Linear Cox Model (TV-FLCM) in Section~\ref{hflm:subsec:mainmodel_full} and a computationally efficient approximation to the TV-FLCM using a landmark approach for the estimation \citep{van2007dynamic} in Section~\ref{hflm:subsec:mainmodel}.

\subsection{Time-Varying Effects in a Cox Model}
\label{hflm:subsec:TVCOX}

The proportional hazards assumption is a key assumption of the Cox model. When the association between the covariate and log hazard is not constant (i.e. the effect depends on $t$), an extension of the Cox model allows for time-varying effects to address the violation of the proportional hazards assumption. Specifically, the time-varying Cox model is of the form
\begin{align*}
    \log\lambda_i(t|X_i) &= \log\lambda_0(t) + X_i\beta(t)
\end{align*}
where $\eta_i$ from the standard Cox model now depends on $t$, $\eta_i(t) =  x_i\beta(t)$. The time-varying effect, $\beta(t)$, may be modeled parametrically (e.g., using polynomial terms), with the functional form suggested by the Shoenfeld residuals \citep{winnett2001miscellanea}, or estimated using penalized splines. Assuming a parametric form for $\beta$ (i.e., no penalization) and no tied event times, the Cox partial log-likelihood takes the form
\begin{align*}
\text{pl}(\beta) &= \sum_{i=1}^N d_i\left\{\eta_i(T_i)-ln\left(\sum_{\{j:T_j \ge T_i\} }e^{\eta_j(T_i)}\right)\right\}
\end{align*}
Evaluating this log-likelihood requires the value of the covariate effect for each event time for each participant in the risk set at that time. Using this insight, estimation of time-varying effects in a Cox model can be performed using standard Cox regression software by constructing a new dataset that contains covariate information for participants at each unique event time who are in the risk set, along with event indicators for whether each had an observed event at that specific time. \cite{JSSv061c01} provides a reference for how to estimate such models using standard {\ttfamily R} and {\ttfamily SAS} packages/functions. Alternatively, one may exploit the connection between the Poisson and Cox partial likelihoods to estimate a Cox model with time-varying effects, the formula for which is given in \cite{10.2307/2346901}. The Poisson/GLM approach has been used successfully to estimate Cox models with relatively complex association structures and non-linear effects estimated using penalized regression splines \citep{doi:10.1177/1471082X17748083,10.1093/biostatistics/kxy003}.

\subsection{Functional Linear Cox Model}
\label{hflm:subsec:FLCM}

The linear functional Cox model has been proposed in the literature to estimate the domain-specific association of a functional covariate, $\{Z_i(u): u \in \mathcal{U}\}$, with an individual's risk. The model takes the form
\begin{align}
    \log\lambda_i(t|X_i, \{Z_i(u): u \in \mathcal{U}\}) &= \log\lambda_0(t) + X_i\beta(t) + \int_{\cal{U}}Z_i(u) \gamma(u) du
\end{align}

Numerous studies have explored different versions of the functional linear Cox model \citep{kong2018flcrm, yang2021weighted, qu2016optimal, gellar2015cox, lee2015bflcrm, leroux2020statistical}, which include a linear functional term of the form $\int_{\cal{U}}Z_i(u) \gamma(u)du$. \cite{kong2018flcrm} and \cite{yang2021weighted}'s papers consider the simplest functional covariate $\int_\mathcal{U} Z_i(u)\gamma(u)du$ in the Cox model, employing parameter expansion and FPCA. They express $Z_i(u)$ as $\mu(u) + \sum_{k=1}^{\infty} \xi_{ik}\phi_k(u)$ and $\gamma(u)$ as $\sum_{k=1}^{\infty} \gamma_k\phi_k(u)$. This transformation allows the functional linear Cox model to be converted into a multivariate Cox model, facilitating the use of partial likelihood to estimate $\hat{\xi}_{ik}$ and $\hat{\gamma}_k$. The estimated Cox model is then used to derive the estimated hazard function or survival function. Additionally, both papers incorporate a testing procedure to assess the significance of the functional effect $\gamma(u)$ being zero.

In \cite{lee2015bflcrm}'s study, a Bayesian approach is introduced for estimating the functional linear Cox model. This involves assigning prior distributions to all parameters in the model and estimating these parameters through their posterior distributions. \cite{qu2016optimal}'s work, on a related note, provides a comprehensive exploration of the theory and inference associated with the functional Cox model. In \cite{gellar2015cox}'s study, the functional effect $\gamma(u)$ is expanded using penalized B-splines, and the model is estimated using penalized partial likelihood with an added penalty term.

\subsection{Time-varying Functional Linear Cox Model}
\label{hflm:subsec:mainmodel_full}

The Time-varying Functional Linear Cox Model (TV-FLCM) is a model used to investigate the general Cox model with a time-varying functional term. Its formula links estimation using results from ~\ref{hflm:subsec:TVCOX} combined with ~\ref{hflm:subsec:FLCM}
\begin{align}
\log\lambda_i(t|X_i, \{Z_i(u): u \in \mathcal{U}\}) &=\log\lambda_0(t)+X_i\beta+\int_\mathcal{U} Z_i(u)\gamma(u,t)du\label{eq:tv_flcm_full}
\end{align}

\begin{figure}[!ht]
  \centering
  \begin{subfigure}{0.32\textwidth}
    \includegraphics[width=\textwidth]{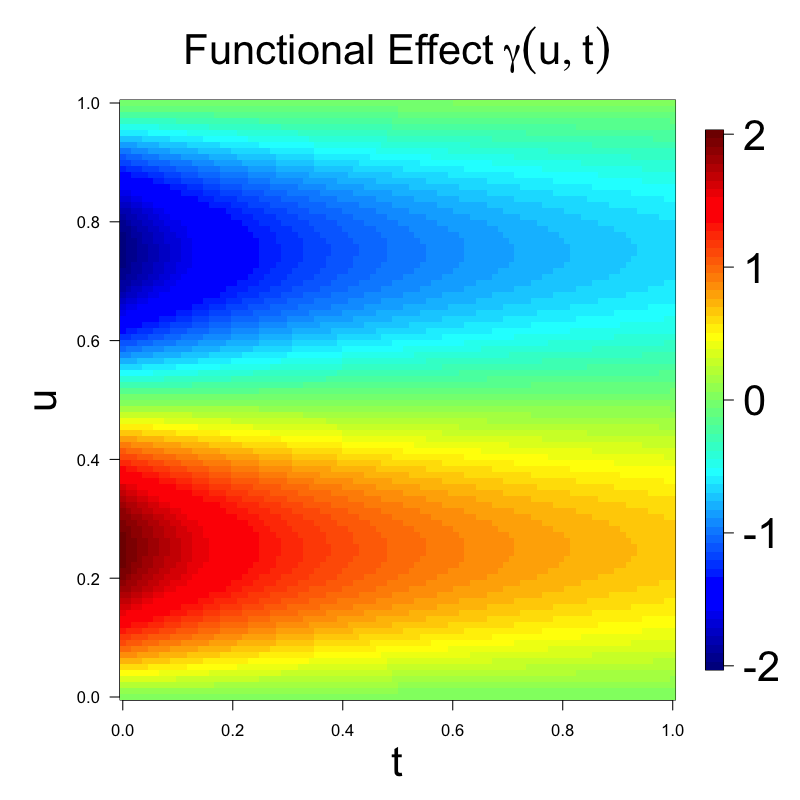}
    \caption{}
    \label{fig:subfig1_1}
  \end{subfigure}
  \begin{subfigure}{0.32\textwidth}
    \includegraphics[width=\textwidth]{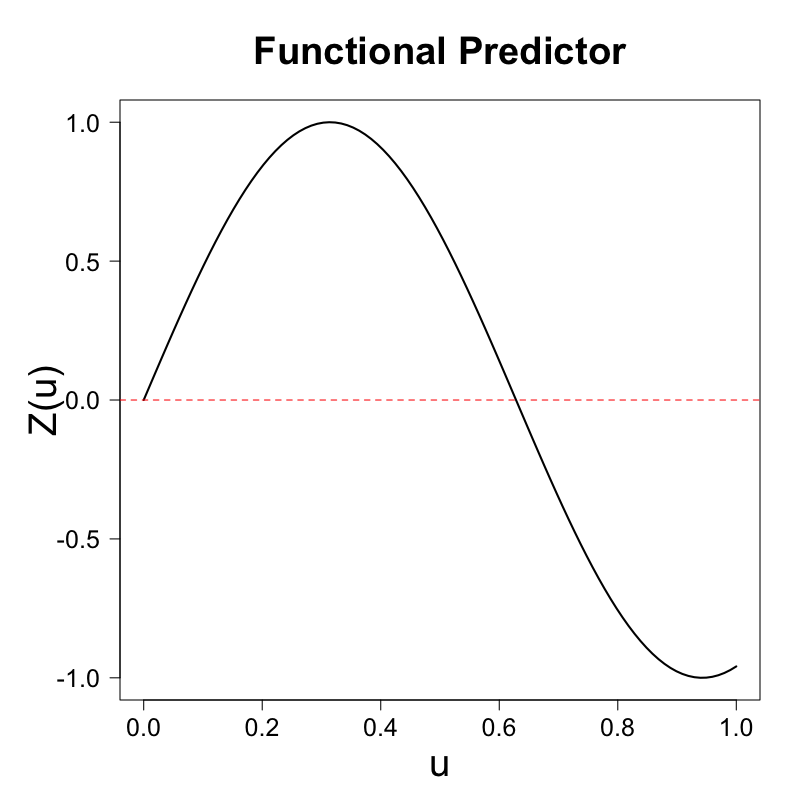}
    \caption{}
    \label{fig:subfig1_2}
  \end{subfigure}
  \begin{subfigure}{0.32\textwidth}
    \includegraphics[width=\textwidth]{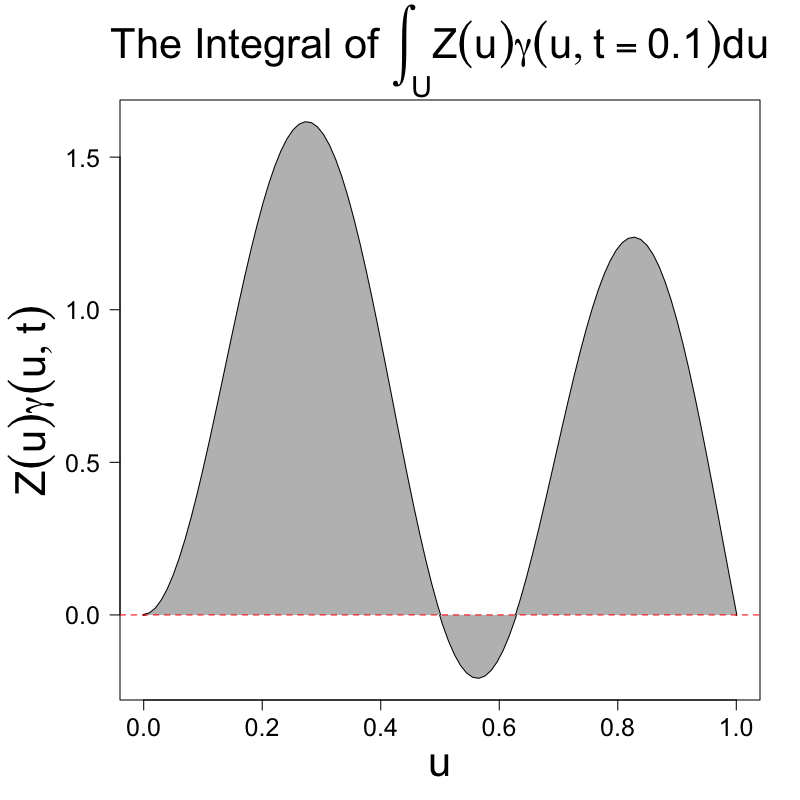}
    \caption{}
    \label{fig:subfig1_3}
  \end{subfigure}
  \caption{(a) Functional effect $\gamma(u,t)=sin(2*\pi*u)/(t+0.5)$. (b) Functional predictor $Z(u) = sin(5*u)$. (c)The integral of functional effect times functional predictor when $t=0.1$.}
  \label{fig:whole1}
\end{figure}

TV-FLCM allows the effect of the functional covariate on risk to vary both in the functional domain ($u$) and over time ($t$) for each point in the survival domain. Figure \ref{fig:whole1} presents an illustration of a hypothetical bivariate effect. In this example, we consider the simulated functional effect \(\gamma(u, t) = \frac{\sin(2 \pi u)}{t + 0.5}\) in panel \ref{fig:subfig1_1} and the functional predictor \(\sin(5 u)\) in panel \ref{fig:subfig1_2}. Since the functional effect \(\gamma(u, t)\) is controlled by both \(u\) and \(t\), the plot for the functional effect is presented as a 2D plot or heatmap, where different colors represent different values of \(\gamma(u, t)\). On the other hand, the functional predictor \(Z(u)=\sin(5 u)\) is only controlled by \(u\), which makes it a curve. The panel \ref{fig:subfig1_3} illustrates the integral when \(t = 0.1\), specifically \(\int_{\mathcal{U}} Z(u) \gamma(u, t=0.1) \, du\).

\subsection{Landmark Linear Functional Cox Model}
\label{hflm:subsec:mainmodel}

The primary challenge in using existing software to estimate TV-FLCM is the computation time. Assuming there are no ties, the datasets used to evaluate the correct log-likelihood contain $N^* = \sum_{i=1}^N d_i \sum_{\{j:Y_j \geq Y_i\}}$ rows. Penalized Cox regression using a tensor product basis expansion for $\gamma$ with $K_u$ and $K_t$ marginal basis functions for the functional domain and time directions results in a design matrix with $K_u*K_t$ columns. With two smoothing parameters to be estimated, the computational complexity of these models is $O(n^2)$. In our data application, this amounts to a design matrix with more than 1 million rows and 1440 columns.  When estimating the TV-FLCM, we must consider all subjects who have survived up to each event time. By extending the functional term using tensor product splines, the entire Cox model is transformed into a Poisson regression model. In general, even with a dataset of only 5000 subjects with various event times, the total number of rows for the design matrix in the Poisson regression model can easily reach 1 million. Each row of the design matrix represents the survival information for an individual subject at a specific event time. This data scale is generally infeasible for existing software to handle on standard computers.  On a MacBook Pro 2019 with the 2.6 GHz 6-Core Intel Core i7 processor and memory of 16 GB, the estimation of the model~\eqref{eq:tv_flcm_full} was very hard and time-consuming.   

Alternatively, a landmark approach may be used to estimate an approximation of the model~\eqref{eq:tv_flcm_full} while substantially reducing computational burden. The landmark method \citep{van2007dynamic, rizopoulos2017dynamic} is an effective technique for approximating time-varying effects that involve partitioning the follow-up time into smaller intervals (bins) and estimating a separate (i.e. time-varying) effect in each of these bins. Smoothness across bins can be imposed and smoothness can be controlled using penalized regression splines, as was done for model~\eqref{eq:tv_flcm_full}. Thus, we propose the Time-varying Functional Linear Cox Model using a landmark approach (TV-FLCM-L). 

In the landmark approach, we choose a set $L$ landmark times $\boldsymbol{s}=\left\{s_1,\ldots,s_L\right\}$ and $L$ corresponding prediction window lengths $\boldsymbol{w}=\left\{w_1,\ldots,w_L\right\}$. Often, landmark times are chosen to be evenly spaced, with constant prediction window lengths ($w_1 = ,\ldots, w_L = w$), forming a partition of $(0, T_{\text{max}}]$. Then, the TV-FLCM-L takes a form similar to the TV-FLCM, specifically
\begin{align}
\log\lambda_i(t|X_i,Z_i(u),s_l)=\log\lambda_0(t|s_l)+X_i\beta(s_l)+\int_\mathcal{U}Z_i(u)\gamma(u,s_l)du \label{eq:LLFCM} 
\end{align}
for $s_l < t \le s_l+w_l$. Note some key differences between TV-FLCM and TV-FLCM-L presented as model~\eqref{eq:LLFCM} above. First, the TV-FLCM-L conditions on survival up to the landmark time $s_l$. Second, TV-FLCM-L allows for a time-varying effect through the landmark times (note that $\gamma$ is now a function of $s_l$ and not $t$). Third, TV-FLCM-L assumes the proportional hazards assumption in the prediction window since $\gamma$ does not change for $t \in (s_l, s_l + w]$. Fourth, the model assumes a landmark-specific baseline hazard function which we leave unspecified. 

\subsection{Estimation of $\gamma(u,t)$}
\label{hflm:subsec:estimation}
Penalized spline smoothing, \citep{ruppert2003semiparametric, wood2017generalized}, and its connection with mixed effects modeling provides a powerful inferential platform for nonparametric regression modeling. Thus, pairing penalized spline smoothing and functional modeling \citep{ivanescu2015penalized, goldsmith2011penalized} provides a modern, easy-to-implement, extendable framework for functional data analysis. In TV-FLCM-L, we focus on estimating the coefficient $\beta(s_l)$ and the functional effect $\gamma(u,s_l)$ at each landmark time. At each landmark time, the coefficient $\beta(s_l)$ is a scalar value. Thus, our primary attention is directed towards the functional effect $\gamma(u,s_l)$. In more general cases, we consider the functional effect $\gamma(u,t)$. In \cite{leroux2020statistical}'s paper, they have the constraint $u \le t$, resulting in the functional effect surface resembling a triangle. For this scenario, it is appropriate to use thin plate regression splines to expand $\gamma(u,s_l)$ \citep{wood2003thin}. However, in our model, $u$ is entirely independent of $t$, which results in a rectangular surface. Therefore, we should consider tensor product splines to expand the functional effect \citep{wood2006low}.

For the choice of basis, we prefer to add penalty terms, often referred to as P-splines \citep{perperoglou2019review, eilers1996flexible}. In particular, we contemplate employing penalized tensor product splines in our model. Next, let's delve into the approximation of $\gamma(u,s_l)$ using two spline bases separately for $u$ and $s_l$, which can be expressed as $\bm{B(s_l)}=\left\{B_1(s_l),B_2(s_l),...,B_{K_s}(s_l)\right\}$ and $\bm{B(u)}=\left\{B_1(u),B_2(u),...,B_{K_u}(u)\right\}$
so that 
\begin{align*}
\gamma(u,s_l)=\sum_{j=1}^{K_u}\sum_{k=1}^{K_s}\xi_{j,k}B_{j}(u)B_{k}(s_l)
\end{align*}
where $B_j(\cdot)$ and $B_k(\cdot)$ represent two univariate splines defined on the domains of $u$ and $s_l$ respectively. The parameters $\left\{\xi_{j,k}:j=1,2,...,K_u;k=1,2,...,K_s\right\}$ correspond to the spline coefficients. More theory about the penalized tensor product splines can be seen in \cite{wood2006low}'s paper, and we provide more details about our situation in the supplementary material. For the functional domain $\mathcal{U}$, we utilize cyclic cubic regression splines, while for the survival time $T$, we employ cubic regression splines. It is essential to carefully choose the number of knots $K_s$ and $K_u$, an insufficient number may result in a curve that is not smooth enough, while an excessive number can lead to overfitting. The functional term now can be represented as,
\begin{align*}
\int_{\mathcal{U}}Z_i(u)\gamma(u,s_l)du=\sum_{j=1}^{K_u}\sum_{k=1}^{K_s}\xi_{j,k}\int_\mathcal{U}Z_i(u)B_j(u)B_k(s_l)du
\end{align*}
Thus, without penalization, this is a standard Cox model with predictors $\int_{\mathcal{U}}Z_i(u)B_j(u)B_k(s_l)du$ and associated parameters $\xi_{j,k}$.  However, in practice, $Z_i(\cdot)$ is not continuously observed and may be measured with noise. Because $Z_i$ is densely measured, we could numerically approximate the integral. If we divide range $\mathcal{U}$ into $v$ small intervals, then,
\begin{align*}
\sum_v\delta_vZ_i(v)\sum_{j=1}^{K_u}\sum_{k=1}^{K_s}\xi_{j,k}B_j(v)B_k(s_l)\approx \int_\mathcal{U}Z_i(u)\gamma(u,s_l)du
\end{align*}

On the other hand, the nonfunctional term $X_i\beta(s_l)$ is still controlled by $s_l$. Therefore,
\begin{align*}
    X_i\sum_{k_1=1}^{K_1}\xi_{k_1}\phi_{k_1}(s_l)\approx X_i\beta(s_l)
\end{align*}
The choices of tuning parameters $K_s$, $K_u$, and $K_1$ are crucial, as discussed earlier. Typically, the tuning parameters $K_s$ and $K_u$ for the functional term are larger than the tuning parameter $K_1$ for the nonfunctional term. Moreover, all three tuning parameters should not exceed the number of landmark times $L$.

Now, let us consider more deeply the choice of the tuning parameters of the base functions $B(s_l)$ and $B(u)$. On the one hand, we need to choose $K_s$ and $K_u$ large enough for a high degree of flexibility and good fitness. On the other hand, we need to estimate the smooth functional datum. Hence, we need to make a trade-off between the lack of data fit and smoothness \citep{perperoglou2019review}. The method to solve this problem is the penalized log-likelihood. The general form is
$pl = l_{\gamma}(u_1,y_1,...,u_n,y_n)-\lambda \int_\mathcal{U}\left\{D^2Z(u)\right\}^2du$, where the penalty term $\int_\mathcal{U}\left\{D^2Z(u)\right\}^2du$ is a roughness penalty that becomes small if the spline function is smooth. A rough spline estimate that has a high value of $l_{\gamma}(u_1,y_1,...,u_n,y_n)$ and is close to the data values results in a high value of $\int_\mathcal{U}\left\{D^2Z(u)\right\}^2du$. Therefore, we should maximize the penalized log-likelihood to use the smoothing parameter $\lambda$ controlling the trade-off between model fit, measured by the first term, and variability of the function (smoothness), measured by the second term.

The simplest method to estimate likelihood is the `separate landmark model', where a separate regression model is fitted to the landmark dataset, defined by the landmark time $s_l$ and window $w_l$. This approach results in $L$ separate regression models.

If we let $\bm{\theta}=[\beta, \xi]$, where $\bm{\xi}=\left\{\xi_1, \xi_2, ..., \xi_{j,k}\right\}$ and
\begin{align*}
\eta_i(s_l)=X_i\beta(s_l)+\sum_{j=1}^{K_u}\sum_{k=1}^{K_s}\left[\xi_{j,k}\int_\mathcal{U}Z_i(u)B_j(v)B_k(s_l)du\right]
\end{align*}
then the separate landmark model is fit by maximizing penalized partial log-likelihood separately for each landmark time $s_l$,
\begin{align*}
     \mbox{ppl}(\beta, \gamma, s_l)=\sum_{i:s_l<T_i\le s_l+w_l}d_i\left\{\eta_i-ln\left(\sum_{T_j \ge T_i}e^{\eta_i}\right)\right\}-P_{s_l}(\gamma)
\end{align*}
The above equation is the standard formulation of the log partial likelihood \citep{efron1977efficiency, cox1975partial} minus a penalty term $P_{s_l}(\gamma)$. The coefficients can be estimated by maximizing the penalized partial log-likelihood using a Newton-Raphson procedure.

However, a limitation of the separate landmark models is that some of the models may not be estimable due to a lack of information, especially when landmark times are close to the baseline or far into the future. Therefore, we should consider the super landmark model, and this results in the pseudo-penalized partial log-likelihood,
\begin{align*}
     \mbox{psppl}(\beta, \gamma, s)=\sum_{l=1}^L\sum_{i:s_l<T_i\le s_l+w_l}d_i\left\{\eta_i-ln\left(\sum_{T_j \ge T_i}e^{\eta_i}\right)\right\}-\left\{P(\gamma)+P(\beta)\right\}
\end{align*}
Here, $\beta(s_l)$ can be smoothed by adding the penalty $P(\beta)$. The function $\gamma(\cdot,\cdot)$ is considered a bivariate smooth function, and the smoothing penalty $P(\gamma)$ is a bivariate penalty. The penalty term can be expressed as $P(\gamma)=\lambda_1 \bm{\theta}^{'}\bm{D_1}\bm{\theta}+\lambda_2 \bm{\theta}^{'}\bm{D_2}\bm{\theta}$, where $\bm{D_1}$ and $\bm{D_2}$ are symmetric non-negative definite penalty matrices. The penalty term has two smoothing parameters because we use tensor product splines here and each parameter is responsible for one direction. The model fitting is performed by stratifying on landmark times, and the next section shows how to fit these models using the $\mathbf{mgcv::gam}$ function in R.

\subsection{Identifiability}
\label{hflm:subsec:identifiability}
The TV-FLCM model (both landmark and non-landmark models) suffers from two sources of non-identifiability. The first is the confounding of the baseline hazard with the coefficient function. Consider the TV-FLCM-L model, though the result applies to the non-landmark model as well. Note that we re-write our model as follows
\begin{align*}
    \log(\lambda_i(t|X_i,{Z_i(u): u \in \mathcal{U}}, s_l)) &= \log \lambda_0(t|s_l) + X_i \beta(s_l) + \int_\mathcal{U} Z_i(u) \gamma(u,s_l)du \\
    & =\left\{\log \lambda_0(t|s_l)+g(s_l)\right\}+X_i\beta(s_l) \\
    &+ \int_\mathcal{U} \left\{Z_i(u)-\mu_z(s_l,u)\right\} \gamma(u,s_l)du
\end{align*}
and $g(s_l)=\int_\mathcal{U}\mu_z(s_l,u)\gamma(s_l,u)du$ is confounded with the baseline hazard. We address this issue by mean centering the functional predictor for each landmark (or unique event time in the non-landmark model) as $\Tilde{Z_i}(u,s_l)=Z_i(u)-\left\{\sum_{i: T_i > s_l}Z_i(u)\right\}/\left\{i: T_i>s_l\right\}$.

Additionally, the non-identifiability problem over the estimable domain could be solved by imposing constraints
\begin{align*}
    \sum_{i=1}^N Z_i(u)\gamma(u, s_l) =0
\end{align*}
for each $u \in \mathcal{U}$ \citep{cui2021additive}. These constraints restrict the functional term to a unique form within the range $Z_i(u)$, thus ensuring identifiability in the area of interest. The restriction can be implemented directly by the \textbf{ti} function with \textbf{mc = T} of the \textbf{gam} function in R package \textbf{mgcv}.

\subsection{Smoothing Parameter Selection}
\label{hflm:subsec:Smoothing}
Smoothing parameter selection is a critical component of nonparametric functional regression with penalized splines and is an open problem in the context of landmark models. There have been many methods to perform smoothing parameter selection, including \textbf{GCV}, \textbf{AIC} \citep{hurvich1998smoothing}, \textbf{EPIC}, \textbf{REML} \citep{ruppert2003semiparametric}, and \textbf{CVL} \citep{gellar2014variable}. Here we choose \textbf{REML} for the estimation of the smoothing parameter \citep{wood2016smoothing}. This approach is suitable for landmark models, as it implicitly assumes the independence of the observations across strata.


\section{Landmark Dataset Computation}
\label{hflm:sec:Landmark}

\subsection{Dataset Decomposition}
\label{hflm:subsec:LMdata}

Before introducing how to use $\mathbf{mgcv::gam}$ to do analysis, we need to understand how the landmark dataset is created and structured. For each landmark time $s_l$, the dataset contains the information for all individuals still alive and not censored at $s_l$. These datasets are then stacked to create a stacked landmark dataset. Similarly as the considerations in \cite{leroux2020statistical}'s paper, we still need four matrices containing: 1) the time-varying predictors; 2)the times when the time-varying predictors were observed; 3) the landmark time; and 4) the numeric integration multipliers. We provide an example to show how this works. In this example, we consider two subjects. The landmark times for the first subject are $\bm{s}=\left\{0,1,2,3,4\right\}$, and the corresponding windows are $\bm{w}=\left\{1,1,1,1,1\right\}$. The landmark times for the second subject are $\bm{s}=\left\{0,1,2,3\right\}$, and the corresponding windows are $\bm{w}=\left\{1,1,1,1\right\}$. The functional term will be estimated using the Riemann integration. The whole dataset for this example can be seen in Table \ref{tab:table1}.

Based on Table \ref{tab:table1}, the second column $\mathbf{T}$ is associated with event times. The example subjects died at 4.5 years and 3.5 years. The third column $\mathbf{d}$ is the event indicator. The fourth column $\mathbf{X}$ stands for the fixed value covariate, which means $X_1=7$ and $X_2=4$. The fifth column, labeled $\mathbf{svec}$, is the landmark times. The sixth and seventh columns $\mathbf{umat}$ and $\mathbf{zmat}$ represent the matrix version of the time it took to observe the functional predictor and the specific values of $Z_i(u)$. The matrix $\mathbf{smat}$ corresponds to the matrix versions of the landmark times. In Leroux's model, the matrix $\mathbf{zmat}$ will be $0$ entries if $\mathbf{umat} > \mathbf{smat}$. The next matrix, $\mathbf{lmat}$, contains the numeric integration factors. The final matrix $\mathbf{zlmat}$ contains the element-wise product of the time-varying predictor $\mathbf{zmat}$ and the numeric integration factor $\mathbf{lmat}$.

\subsection{Fitting TV-FLCM-L in R}
\label{hflm:subsec:fittingTLFCM}
Based on the stacked dataset above, TV-FLCM-L can be fit through the $\mathbf{gam}$ function in R. For the estimation of $\beta(s_l)$ and $\gamma(s_l,\cdot)$, we use penalized splines. Now, we will go into detail about the important connections and key software components to fit models from the simplest to the most complex model. The dataset in Table \ref{tab:table1} will be stored in the data frame \textbf{data\_lm}. The \textbf{gam} function in R for the separated landmark model is shown below:

{
\begin{center}
\begin{verbatim}
gam(event_time_yrs ~ ti(umat, by=Zlmat, bs=c("cc", "cr"),
k=c(5,5), mc=c(T,F)), family=cox.ph, weights=d, data=data_lm)
\end{verbatim}
\end{center}
}

In this function, \textbf{data\_lm} is the row-bind dataset of all subjects at one landmark time. The \textbf{family=cox.ph} maximizes the Cox partial likelihood corresponding to the observed outcome. The \textbf{weights=d} uses the current dataset. The \textbf{bs=c("cc", "cr")} and \textbf{k=c(5,5)} stands for the number of 5 knots and cyclic cubic regression and cubic regression splines for the $s_l$ and $u$, respectively. Returning to the construction of our linear predictor, the term \textbf{ti(umat, by=zlmat, k=c(5,5), bs=c("cc", "cr"))} specifies a numeric approximation of the functional term, $\int_\mathcal{U} Z_i(u)\gamma(s_l,u)du$. Actually, the function \textbf{ti()} is very friendly: (1) adds a tensor product spline basis to the linear predictor and (2) includes a corresponding penalty term to the log-likelihood to be maximized. In our model, you can try different weighted sums of a spline basis expansion of the \textbf{umat} matrix using different numbers of the k basis function in the \textbf{ti()} function. These weights are determined by the \textbf{zlmat} matrix. The \textbf{zlmat} is a matrix that contains $Z_i(u)$ multiplied element-wise by a matrix of numeric integration factors. Therefore, the \textbf{ti()} function adds the term $\sum_v\delta_vZ_i(v)\sum_{j=1}^{5}\sum_{k=1}^{5}\xi_{j,k}B_j(v)B_k(s_l)\approx \int_\mathcal{U}Z_i(u)\gamma(u,s_l)du$ to the linear predictor, where $\delta_vZ_i(v)$ is the $v^{th}$ column of \textbf{zlmat} for subject $i$ and $\delta_v$ is the corresponding numeric integration factor.

Now, let us focus on the super landmark model. The model fitting by \textbf{gam} function in R is:
{
\singlespacing
\begin{verbatim}
gam(cbind(event_time_yrs, s_vec) ~ s(svec, k=5, by=X) +
     ti(smat, umat, by=Zlmat, bs=c("cr", "cc"), k=c(10,10), mc=c(T,F)),
     family=cox.ph, weights=d, data=data_lm, method = "REML")
\end{verbatim}
}
The code given extends the ideas discussed regarding fitting individual models by employing a slightly modified syntax. Unlike the iterative approach of fitting models at various landmark times, fitting the comprehensive super landmark model entails utilizing the entire landmark dataset without requiring such iterations. In the context of the \textbf{gam} function in R, the output changes from $\textbf{event\_time\_yrs}$ to $\textbf{cbind(event\_time\_yrs, s\_vec)}$. This adjustment fits a Cox model stratified based on landmark times, enabling each landmark time to have its unique baseline hazard.

The linear predictor has been modified to allow smooth variations of both $\beta(s_l)$ and $\gamma(s_l,u)$ between landmark times. When considering the time-invariant predictor $\textbf{s(svec, k=5, by=X)}$, it generates a thin plate spline basis expansion of the landmark time multiplied by $\textbf{X}$. This results in the addition of $X_i\sum_{k_1=1}^5\epsilon_{k_1}\phi_{k_1}(s_l)=X_i\beta(s_l)$ to the linear predictor. The second penalty on $\beta(s_l)$ is automatically incorporated into the log-likelihood. Now, let us pay attention to the functional term, created using the code \textbf{ti(smat, umat, by=Zlmat, bs=c("cc", "cr"), k=c(10,10),mc=c(T, F))}. Compared to the functional term in a separate model, this model contains the landmark matrix $\textbf{smat}$ in addition to $\textbf{umat}$. This code adds the term $\sum_v\delta_vZ_i(v)\sum_{j=1}^{10}\sum_{k=1}^{10}\xi_{j,k}B_j(v)B_k(s_l) \approx \int_\mathcal{U}Z_i(u)\gamma(u,s_l)du$ to the linear predictor.


\section{Application to NHANES Dataset}
\label{hflm:sec:NHANES}
\subsection{Background and Introduction}
\label{hflm:subsec:backgroudand introduction}

In this section, we will apply the model TV-FLCM-L to the \textbf{NHANES} dataset to explore the correlation between mortality rate and the functional predictor denoted by \textbf{MIMS}, which means Monitor Independent Movement Summary. NHANES, a comprehensive survey conducted by the CDC, collects health and nutrition data from a diverse sample of the US population.

The NHANES dataset can be broadly categorized into three parts: covariate data, mortality data, and accelerometry data. Covariate data comprises demographic and personal information collected through questionnaires, encompassing details such as age, education, gender, smoking habits, history of heart attack, cancer, alcohol consumption, diabetes, hypertension, and more. Mortality data includes crucial variables such as event times and event indicators. The functional predictor \textbf{MIMS} is derived from accelerometry data and obtained using hip-worn or wrist-worn accelerometers. Each participant wore the accelerometer continuously for 7 days, with data recorded at minute intervals each day, resulting in $1440 (24\times60)$ records for each individual per day. Therefore, if 1000 participants are considered for this study, the dimensions of the \textbf{MIMS} matrix would be $(7000 \times 1440)\cdot N$, where $N$ represents the number of other variables, yielding a very extensive dataset. In \cite{leroux2019organizing}'s paper, this NHANES dataset is presented in a long format, making analysis challenging. Fortunately, they reformatted this long-format dataset, where each row in the covariate matrix represents a participant per day. Consequently, the previously mentioned dataset can be transformed into dimensions of $7000*(1400+N)$.

Although the long-format dataset has been transformed into a shorter format, the 1440 observations per subject per day still pose challenges for visualization and analysis. To mitigate complexity and simplify the dataset, various strategies can be employed, including but not limited to (1) Mean or Total Activity Count: Aggregating the activity count to obtain the mean or total count per day; (2) Mean or Total Log-Transformed Activity Count: Summarizing the log-transformed activity count by calculating the mean or total per day; (3) Total Sedentary Time: Calculating the overall sedentary time for each subject by adding the sedentary minutes; (4) Total minutes of moderate / vigorous physical activity: adding the minutes spent in moderate or vigorous physical activity to obtain an overall measure of the intensity of physical activity. Although these strategies are effective in simplifying the dataset, it is important to note that they can result in information loss due to a significant reduction in dimensionality. Another popular method to further reduce complexity with minimal information loss is Functional Principal Component Analysis (FPCA). In FPCA, the high degree of skewness in the activity count data is first reduced by applying a transformation to the data at each minute using $x=log(1+a)$, where $a$ denotes the activity count. This transformation has the added benefit of transformating $0$ into $0$. Now, let $J_i$ be the number of days of accelerometry data for subject $i=1,..., N$ and let $J=\sum_{i=1}^NJ_i$ be the total number of days of data. The new covariate data matrix, $\bm{X}$, is $J\times1440$ dimensional. The whole process has been performed before performing a specific analysis. In the preprocessed dataset, one subject has only one line.

\subsection{Specific Dataset and Model}
\label{hflm:subsec:specificdatasetandmodel}
\begin{figure}[!ht]
  \centering
  \begin{subfigure}{0.32\textwidth}
    \includegraphics[width=\textwidth]{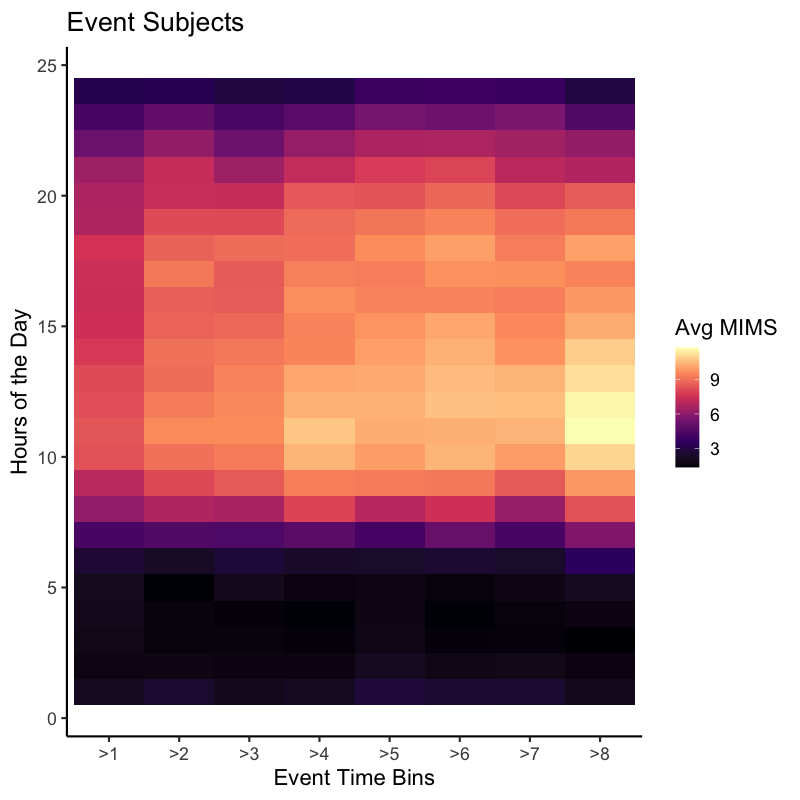}
    \caption{}
    \label{fig:subfig3_1}
  \end{subfigure}
  \begin{subfigure}{0.32\textwidth}
    \includegraphics[width=\textwidth]{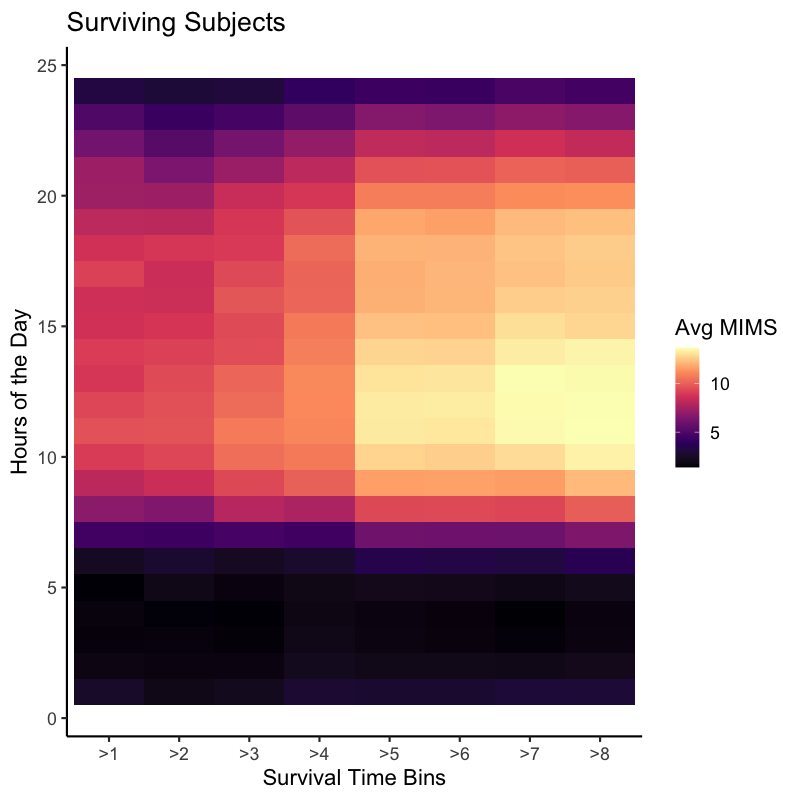}
    \caption{}
    \label{fig:subfig3_2}
  \end{subfigure}
  \begin{subfigure}{0.32\textwidth}
    \includegraphics[width=\textwidth]{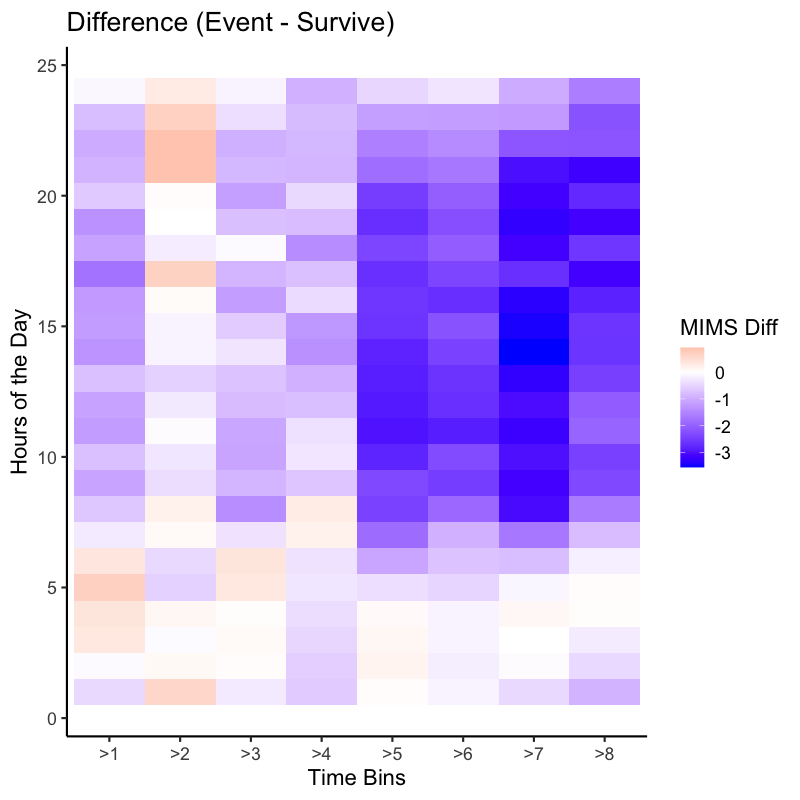}
    \caption{}
    \label{fig:subfig3_3}
  \end{subfigure}
  \caption{(a) Heatmap showing the average MIMS over 24 hours for subjects experiencing events within each time interval. (b) Heatmap showing the average MIMS over 24 hours for subjects who survived beyond each time point. (c) Heatmap displaying the difference in values between panel (a) and panel (b).}
  \label{fig:whole3}
\end{figure}

In our application, the dataset utilized comprises 42 variables, considering the functional covariate `MIMS' as one variable. This dataset includes information from 4445 subjects, all between 50 and 80 years of age. In particular, there are no missing values in the variables: age, event times, event indicators, and functional covariates for all minutes. Figure \ref{fig:whole3} provides a comparative analysis of the average MIMS over 24 hours for subjects with events and those who survived beyond each time point. The panel \ref{fig:subfig3_1} reveals that subjects experiencing events exhibit higher MIMS values during daytime hours, particularly between hours 10 and 20, suggesting increased activity levels during this period. Furthermore, subjects with longer event times tend to maintain higher average MIMS values throughout the day, indicating a possible correlation between prolonged event times and sustained activity. The panel \ref{fig:subfig3_2} shows a similar pattern of elevated MIMS values during the daytime; however, surviving subjects consistently display higher MIMS values, especially in the later time bins, which may reflect their overall greater physical activity and resilience over time. The panel \ref{fig:subfig3_3} highlights the differences between the two groups, with negative values predominating, particularly in the later time bins and during peak activity hours. This indicates that surviving subjects generally exhibit higher MIMS values compared to those experiencing events, at most hours of the day and at time intervals, suggesting a notable disparity in activity levels between the two groups.

In our model, the functional term is derived from the `MIMS' matrix, encompassing data recorded entirely before the survival experiment. The scalar terms used in the model are age and gender. Thus, the final model can be represented as
\begin{align*}
    \log\lambda_i(t|\mbox{age},\mbox{gender}, Z_i(u), s_l)  & =\log\lambda_0(t|s_l)+\mbox{age}_i\beta_1(s_l)  \\ & +\mbox{gender}_i\beta_2(s_l)+\int_{\mathcal{U}}Z_i(u)\gamma(u,s_l)du
\end{align*}
where $\bm{Z(u)}$ is the `MIMS' matrix, $1 \le i \le 4445$, and $s_l$ is the specific landmark time. 

To estimate $\gamma(u,t)$, we could make use of various windows, landmarks, tuning parameters, and overlaps. Here, we examine and present three instances. First, we set the window at $w=0.8$ and the tuning parameters at $k_1=k_2=10$. We set windows with varied values in Example 2. The window has values $(0.5, 0.25, 1.2, 0.8, 1.25, 1, 0.5, 0.75, 1, 0.65, 0.1, 0.5) + 0.3$, meaning that a distinct window length is allowed for each landmark time, and the windows may overlap (+0.3). In this case, the tuning parameters that we employ are $k_1=k_2=5$. In the third case, we similarly set the tuning parameters $k_1=k_2=5$ and set all window lengths equal to $\infty$.

\subsection{Outcomes}
\label{hflm:subsec:outcomes}
\begin{figure}[!ht]
  \centering
  \begin{subfigure}{0.32\textwidth}
    \includegraphics[width=\textwidth]{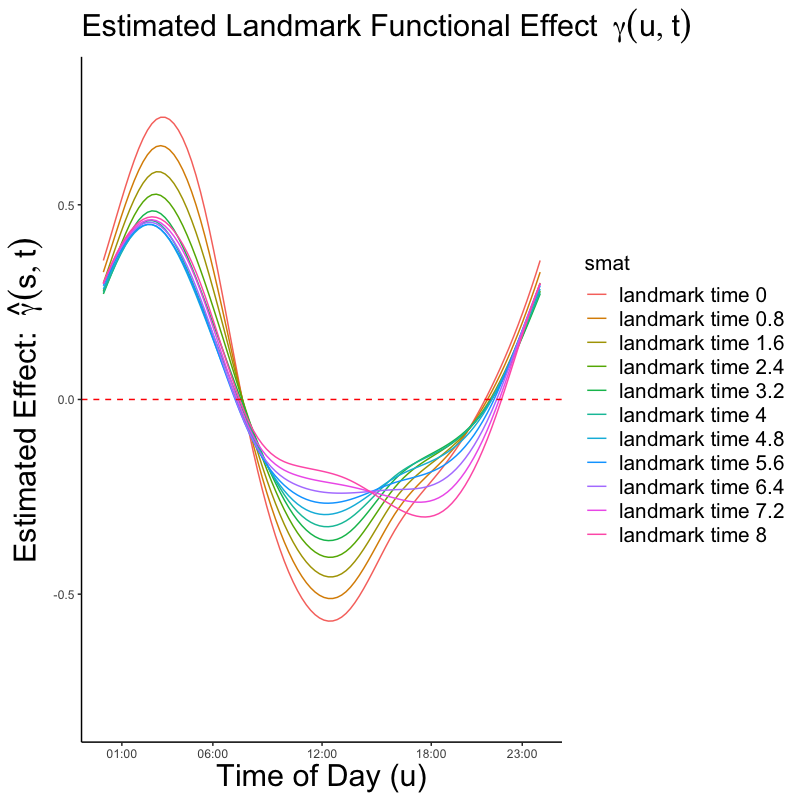}
    \caption{}
    \label{fig:subfig2_1}
  \end{subfigure}
  \begin{subfigure}{0.32\textwidth}
    \includegraphics[width=\textwidth]{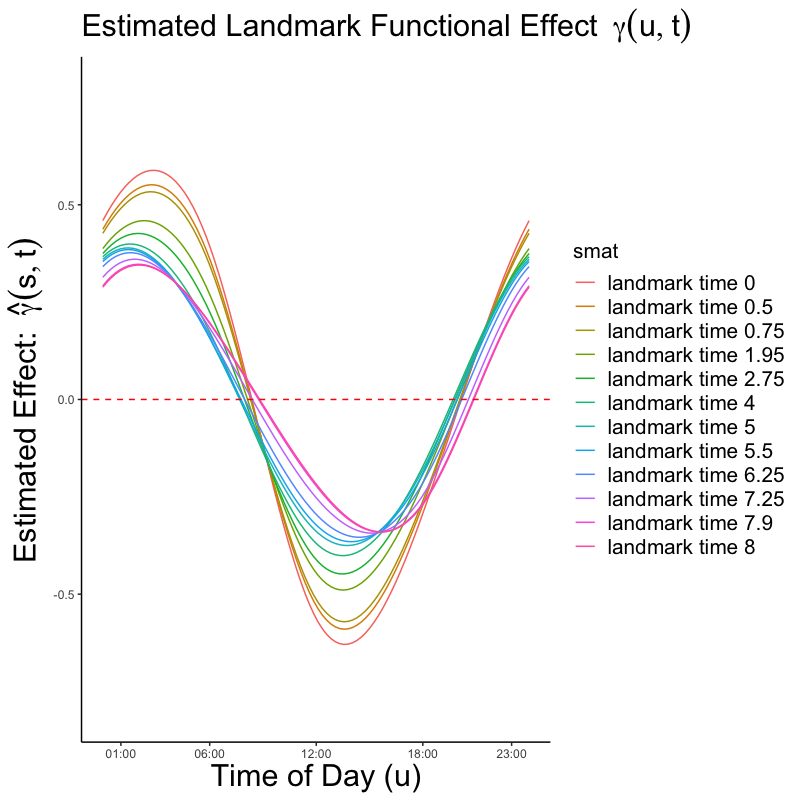}
    \caption{}
    \label{fig:subfig2_2}
  \end{subfigure}
  \begin{subfigure}{0.32\textwidth}
    \includegraphics[width=\textwidth]{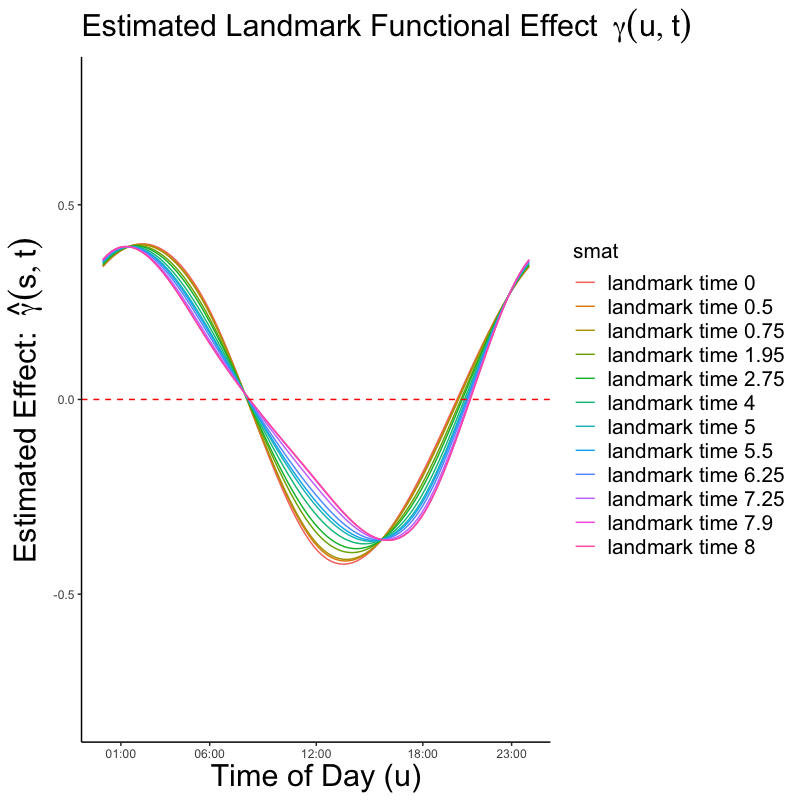}
    \caption{}
    \label{fig:subfig2_3}
  \end{subfigure}
  \caption{Estimated functional effects in NHANES dataset application. In panel (a), the windows are $w=0.8$ and tunning parameters are $k_1=k_2=10$. In panel (b), the windows are $(0.5, 0.25, 1.2, 0.8, 1.25, 1, 0.5, 0.75, 1, 0.65, 0.1, 0.5)$ + 0.3 and tunning parameters are $k_1=k_2=5$. In panel (c), the windows are $w=\infty$ and tunning parameters are $k_1=k_2=5$.}
  \label{fig:whole2}
\end{figure}

We create visual representations illustrating the estimated time-varying functional effects at different landmark times by curves for all three scenarios. The curve plots for all three scenarios are available from panel \ref{fig:subfig2_1} to panel \ref{fig:subfig2_3} in Figure \ref{fig:whole2}. Our primary focus is on examining the curves associated with each landmark time. As shown in Figure \ref{fig:whole2}, all the curves exhibit remarkable smoothness and their overall shapes demonstrate notable stability. For each panel, the negative values appear approximately between 8:00 am and 9:00 pm. If the value of $\gamma(u,s_l)$ is negative, then the term $\int_{\mathcal{U}}Z_i(u)\gamma(u,s_l)du$ will be negative since $Z_i(u)$ is always greater than 0. Then, between 8:00 am and 9:00 pm, the subjects have a lower hazard rate and a higher survival rate. This also corresponds to the fact that subjects have higher survival rates daily and exercising during the day, especially in the afternoon, is best for the body. Within each panel, we could observe that the scale of the pink curve is consistently smaller than that of the red curve. The reason is that the pink curve includes information on the latest landmark times, involving subjects far from the beginning of the study. Consequently, this results in a lower functional effect. In contrast, the red curve always incorporates information from the earliest landmark times containing the most intensive and comprehensive data. The panel \ref{fig:subfig2_3} shows the smallest scale for all curves compared to the other two. This is because it includes windows extending to infinity, resulting in each landmark time incorporating weaker information from subjects who are later in the study. Based on these outcomes, it is concluded that the TV-FLCM-L model performs well for the NHANES dataset application.


\section{Simulation Study}
\label{hflm:sec:simulation}

\subsection{Data Generating Mechamism}
\label{hflm:subsec:Datagene}
In this section, we perform simulations to evaluate the performance of estimating the functional effect in the TV-FLCM-L. We simplify the model by considering only one functional covariate and no scalar covariate, resulting in the TV-FLCM-L expressed as
\begin{align*}
    \log\lambda_i(t|X_i,Z_i(u), s_l)=\log\lambda_0(t|s_l)+\int_\mathcal{U}Z_i(u)\gamma(u,s_l)du
\end{align*}
We simulate both survival time and functional time on the interval $(0,1]$. For the survival time, we divide it into 100 small intervals. For the functional domain, we divide it based on frequencies $J = 100$. The number of subjects observed is chosen as $\textbf{N}=(2000, 3000, 4000)$. The three shapes of the simulated functional effect are $\gamma(u,t)=sin(2\pi u)/(t+0.5)$, $\gamma(u,t)=sin(2\pi u)/(t/2+1)$, and $\gamma(u,t)=10cos\left\{4\pi(t-u)\right\}$. We choose two windows for the landmark times, either $w=0.04$ or $w=\infty$. For both cases, the set of landmark times used is $\bm{S}=\left\{0, 0.04, 0.08,...,0.96\right\}$. Therefore, we have only one frequency for the functional domain, three subject observations, and two windows. We have $1\times3\times2=6$ cases in each simulation. Additionally, we conduct more simulations based on different functional effects and also create more plots for the survival distribution, the censored distribution, and survival curves of the subjects. These cases will be detailed in the supplementary materials. For all scenarios, we simulate 500 times and use their mean values for estimation. The functional predictor is simulated as $Z_i(u)=\sum_{k=1}^{10}b_{ki}\phi_k(u)$, where $\phi_k(u)$ are cubic B-spline basis functions and $b_{ki} \sim N(0,\Sigma_b)$ where $\Sigma_b$ is defined by $var(b_{ki})=4$ for $k=1,...,10$ and $cor(b_{li},b_{mi})=0.3$ for $l \ne m$. We also need to consider that the functional effect is measured with bias, $Z_{i, real}(u)=Z_i(u)+\epsilon_i(u)$ where $\epsilon_i \sim N(0,0.25^2)$. The functional predictor prior is not smoothed before conducting model fitting for the landmark models.

Now, since we have simulated some basic parameters, the next step of the simulation is to simulate survival times. Firstly, we obtain subject-specific survival curves $S_i(t|Z_i(u))$. The $S(\cdot)$ is approximated by numeric integration on a fine grid $t_{sim}$ on the interval $(0,1]$. Here, we use $\text{sim}=100$. Thus, survival times are obtained as $S^{-1}(\mathcal{U}) = inf\left\{t:t\in t_{sim},S_i(t|Z_i(u))\le \mathcal{U}\right\}$ where $\mathcal{U}$ is a uniform random variable on $[0,1]$. 
This simulation approach is inspired by the papers by \cite{bender2005generating} and \cite{baek2021survival}. If the scenario involves a time-varying covariate, the simulation of the survival function can be conducted using \cite{austin2012generating}'s method, but this situation will not be researched in this paper. Lastly, to introduce censoring, survival times are censored by simulating censoring times $C_i$ as the minimum of $1$ and an exponential random variable with a mean of $1$.

The primary metric for assessing the model's performance is its ability to estimate the true functional effect. We employ the criterion that compares the original 2D plot with the estimated 2D plot. If the two plots exhibit similar shapes and scales, then the estimated functional effect provides a reliable approximation of the actual functional effect.

\subsection{Evaluation Criteria}
\label{hflm:subsec:sim_criteria}

We evaluate the performance of our models in terms of accuracy of estimation for $\gamma$, inference (TV-FLCM only), and computation time. Specifically, for $B$ simulated datasets, for each simulation scenario, we assess estimation accuracy as average mean integrated squared error (AMSE) as 
\begin{align*}
\text{AMSE} &= B^{-1}\sum_{b=1}^B \int_\mathcal{U} \int_T (\hat{\gamma}^b(u, t) -\gamma(u, t) )^2 du dt\;.
\end{align*}
Inference is evaluated using pointwise empirical coverage rates for pointwise Wald 95\% confidence intervals
\begin{align*}
\text{CI}^b(u,t) &= B^{-1}\sum_{b=1}^B 1(\gamma(u,t) \in \hat{\gamma}(u,t) \pm \text{SE}(\hat{\gamma}(u,t))\;,
\end{align*}
and the average coverage across $u,t$
\begin{align*}
\text{CI}^b &= B^{-1}\sum_{b=1}^B \int_\mathcal{U} \int_T \text{CI}^b(u,t) du dt\;,
\end{align*}
where integration is approximated by averaging across $u \in \boldsymbol{u}^{\text{pred}}$, $t \in \boldsymbol{t}^{\text{pred}}$ ($\boldsymbol{u}^{\text{pred}}$ and $\boldsymbol{t}^{\text{pred}}$ are grids within the range [0,1], divided into 100 intervals) for both $\text{AMSE}$ and $\text{CI}^b$.  Finally, the computation time is reported as the average computation time between the simulated datasets.

\subsection{Results}
\label{hflm:subsec:result}

The results of the simulations of three distinct functions are shown in Tables~\ref{tab:table2}, \ref{tab:table3}, and \ref{tab:table4}, illustrating the estimated functional effects. Each of these tables presents the following results: The average $\hat{\gamma}$ across the simulated datasets for TV-FLCM-L with $w=0.04$ (top row of heatmaps), TV-FLCM-L with $w=\infty$ (second row of heatmaps), and TV-FLCM (third row of heatmaps). The empirical coverage probabilities in the pointwise direction ($\text{CI}^b(u,t)$) are presented in the fourth row of heatmaps. Below the heatmaps, AMSE for TV-FLCM-L with $w=0.04$, TV-FLCM-L with $w=\infty$, and TV-FLCM are shown along with the average coverage rate $\text{CI}^b$. The results are discussed in the following.

\paragraph{Estimation Accuracy} Comparing AMSE for the functional coefficient (Tables~\ref{tab:table2}, \ref{tab:table3}, and \ref{tab:table4}, bottom), we see that, as expected, the TV-FLCM (Poisson approach) consistently provides the highest accuracy when compared to the two TV-FLCM-L models (landmark approach). However, the AMSE associated with the landmark approach with a small prediction window ($w=0.04$) is not substantially higher than the TV-FLCM, with an increased AMSE ranging from 5.6\% to 91.3\% across sample sizes and assumed functional form for $\gamma(\cdot,\cdot)$.

\paragraph{Inference} Looking at the fourth row of heatmaps in Tables~\ref{tab:table2}-~\ref{tab:table4}, we find that, for the sample sizes and functional forms for $\gamma(\cdot,\cdot)$ considered in our simulation study, TV-FLCM achieves approximately nominal coverage when averaged over $u, t$. However, the patterns of over/under coverage vary with the shape of $\gamma$, with under-coverage tending to occur where the functions are largest in absolute magnitude. Furthermore, for $\gamma(u,t) = 10\text{cos}(4\pi (t-u))$ (Table~\ref{tab:table4}), we see that there is a substantial undercoverage for $t$ close to $0$. This can possibly be explained by the comparatively large proportion of individuals who have an observed event. On average, 15\% of simulated individuals have an event before $t=0.05$, reflected in the steep decline in individual survival curves implied by the data-generating mechanism (Table~\ref{tab:table6} in the supplementary material, column 3).

\paragraph{Computation time} We observed that Poisson regression models require significantly longer computation times compared to landmark models under identical simulation conditions. To maintain feasibility, we limited our testing to sample sizes of 4000 for each case. Across all sample sizes and the assumed functional form for $\gamma(\cdot,\cdot)$, the Poisson regression models require 10 to 42 times longer computation times than landmark models (Tables~\ref{tab:table2}-~\ref{tab:table5}, Computation Time).


\section{Discussion}
\label{hflm:sec:discussion}

This study introduces the Time-Varying Functional Linear Cox Model (TV-FLCM) and the corresponding landmark model (TV-FLCM-L), a novel integration of functional regression techniques and the Cox regression with time-varying effects. A key feature of this approach is the temporal independence of the functional domain preceding the survival time. Compared to traditional Poisson methods, the landmark approach offers computational advantages by leveraging the existing statistical software infrastructure and providing greater modeling flexibility.

The intrinsic structure of TV-FLCM-L is elucidated through an expansion of the functional term using tensor product splines. Model estimation is performed via a penalized partial likelihood approach using the \texttt{mgcv} package in R, making the method readily accessible using the same standard statistical software that supports the \texttt{refund} package \citep{crainiceanu2024functional, refund2023}. Extensive real-world data analyses and simulations demonstrate the efficacy and practical utility of the TV-FLCM-L in estimating functional effects. Moreover, even with large sample sizes, computational results reveal the efficiency, underscoring the landmark approach's advantages over conventional Cox modeling techniques.

However, several avenues for future research remain: (1) developing more precise confidence interval estimations, including bootstrap-based methods; (2) extending the model to multivariate or multilevel contexts to account for multiple domains or hierarchical data structures; and (3) developing rigorous asymptotic theory and expanded testing procedures. In summary, the TV-FLCM-L represents a significant methodological advance, providing a methodological framework for addressing a common issue in time-to-event analysis, that of time-varying effects, in the context of the linear functional Cox model.

{
\spacingset{1.0}
\bibliographystyle{agsm}
\small
\bibliography{ref}
}

\newpage

\begin{table}[!ht]
\begin{center}
\caption{Landmark Dataset}
\label{tab:table1}
\setlength{\tabcolsep}{1.5mm}
\begin{tabular}{c|c|c|c|c|c|c|c|c|c}
\toprule 
ID& T& d& X& svec& umat& zmat& smat& lmat& zlmat\\
\midrule 
1& 1& 0& 7& 0& 0 \hspace{0.05cm} 2 \hspace{0.05cm} 4 \hspace{0.05cm} 6& 1 \hspace{0.05cm} 0.3 \hspace{0.05cm} 0.7 \hspace{0.05cm} 1.1 &0 \hspace{0.05cm} 0 \hspace{0.05cm} 0 \hspace{0.05cm} 0&2 \hspace{0.05cm} 2 \hspace{0.05cm} 2 \hspace{0.05cm} 2&2 \hspace{0.05cm} 0.6 \hspace{0.05cm} 1.4 \hspace{0.05cm} 2.2\\

1& 2& 0& 7& 1& 0 \hspace{0.05cm} 2 \hspace{0.05cm} 4 \hspace{0.05cm} 6& 1 \hspace{0.05cm} 0.3 \hspace{0.05cm} 0.7 \hspace{0.05cm} 1.1 &1 \hspace{0.05cm} 1 \hspace{0.05cm} 1 \hspace{0.05cm} 1&2 \hspace{0.05cm} 2 \hspace{0.05cm} 2 \hspace{0.05cm} 2&2 \hspace{0.05cm} 0.6 \hspace{0.05cm} 1.4 \hspace{0.05cm} 2.2\\

1& 3& 0& 7& 2& 0 \hspace{0.05cm} 2 \hspace{0.05cm} 4 \hspace{0.05cm} 6& 1 \hspace{0.05cm} 0.3 \hspace{0.05cm} 0.7 \hspace{0.05cm} 1.1 &2 \hspace{0.05cm} 2 \hspace{0.05cm} 2 \hspace{0.05cm} 2&2 \hspace{0.05cm} 2 \hspace{0.05cm} 2 \hspace{0.05cm} 2&2 \hspace{0.05cm} 0.6 \hspace{0.05cm} 1.4 \hspace{0.05cm} 2.2\\

1& 4& 0& 7& 3& 0 \hspace{0.05cm} 2 \hspace{0.05cm} 4 \hspace{0.05cm} 6& 1 \hspace{0.05cm} 0.3 \hspace{0.05cm} 0.7 \hspace{0.05cm} 1.1 &3 \hspace{0.05cm} 3 \hspace{0.05cm} 3 \hspace{0.05cm} 3&2 \hspace{0.05cm} 2 \hspace{0.05cm} 2 \hspace{0.05cm} 2&2 \hspace{0.05cm} 0.6 \hspace{0.05cm} 1.4 \hspace{0.05cm} 2.2\\

1& 4.5& 1& 7& 4&0 \hspace{0.05cm} 2 \hspace{0.05cm} 4 \hspace{0.05cm} 6& 1 \hspace{0.05cm} 0.3 \hspace{0.05cm} 0.7 \hspace{0.05cm} 1.1&4 \hspace{0.05cm} 4 \hspace{0.05cm} 4 \hspace{0.05cm} 4&2 \hspace{0.05cm} 2 \hspace{0.05cm} 2 \hspace{0.05cm} 2&2 \hspace{0.05cm} 0.6 \hspace{0.05cm} 1.4 \hspace{0.05cm} 2.2\\

\midrule  
2& 1& 0& 4& 0& 0 \hspace{0.05cm} 2 \hspace{0.05cm} 4 \hspace{0.05cm} 6& 1.2 \hspace{0.05cm} 0.2 \hspace{0.05cm} 0.6 \hspace{0.05cm} 1.5 &0 \hspace{0.05cm} 0 \hspace{0.05cm} 0 \hspace{0.05cm} 0&2 \hspace{0.05cm} 2 \hspace{0.05cm} 2 \hspace{0.05cm} 2&2.4 \hspace{0.05cm} 0.4 \hspace{0.05cm} 1.2 \hspace{0.05cm} 3\\

2& 2& 0& 4& 1& 0 \hspace{0.05cm} 2 \hspace{0.05cm} 4 \hspace{0.05cm} 6& 1.2 \hspace{0.05cm} 0.2 \hspace{0.05cm} 0.6 \hspace{0.05cm} 1.5 &1 \hspace{0.05cm} 1 \hspace{0.05cm} 1 \hspace{0.05cm} 1&2 \hspace{0.05cm} 2 \hspace{0.05cm} 2 \hspace{0.05cm} 2&2.4 \hspace{0.05cm} 0.4 \hspace{0.05cm} 1.2 \hspace{0.05cm} 3\\

2& 3& 0& 4& 2& 0 \hspace{0.05cm} 2 \hspace{0.05cm} 4 \hspace{0.05cm} 6& 1.2 \hspace{0.05cm} 0.2 \hspace{0.05cm} 0.6 \hspace{0.05cm} 1.5 &2 \hspace{0.05cm} 2 \hspace{0.05cm} 2 \hspace{0.05cm} 2&2 \hspace{0.05cm} 2 \hspace{0.05cm} 2 \hspace{0.05cm} 2&2.4 \hspace{0.05cm} 0.4 \hspace{0.05cm} 1.2 \hspace{0.05cm} 3\\

2& 3.5& 0& 4& 3& 0 \hspace{0.05cm} 2 \hspace{0.05cm} 4 \hspace{0.05cm} 6& 1.2 \hspace{0.05cm} 0.2 \hspace{0.05cm} 0.6 \hspace{0.05cm} 1.5 &3 \hspace{0.05cm} 3 \hspace{0.05cm} 3 \hspace{0.05cm} 3&2 \hspace{0.05cm} 2 \hspace{0.05cm} 2 \hspace{0.05cm} 2&2.4 \hspace{0.05cm} 0.4 \hspace{0.05cm} 1.2 \hspace{0.05cm} 3\\

\bottomrule 
\end{tabular}
\end{center}
\end{table}

\newpage

\begin{table}[!ht]
\caption{The results of the estimated functional effects based on two landmark approaches and Poisson regression model, heatmaps for confidence intervals, average mean squared errors, coverage rates, and computation times for the simulations of $sin(2\pi u)/(t+0.5)$. The simulations are performed on different sample sizes (N=2000, 3000 or 4000)}
\label{tab:table2}
\label{tab:sim_sin}
\scalebox{0.95}{
\begin{tabular}{*{5}{c}}
\toprule
True Effect & Methods & \multicolumn{3}{c}{Sample Sizes} \\
\cmidrule(lr){3-5}\cmidrule(lr){1-1}\cmidrule(lr){2-2}
& & N=2000 & N=3000 & N=4000 \\
\cmidrule(lr){3-3}\cmidrule(lr){4-4}\cmidrule(lr){5-5}
\includegraphics[width=0.2\textwidth]{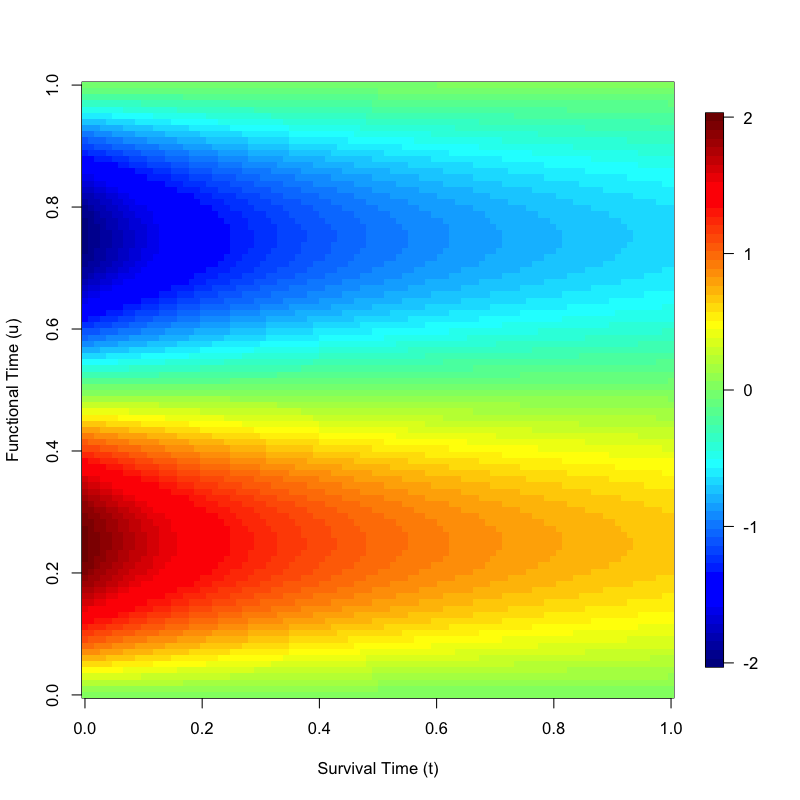} & w=0.04 & \includegraphics[width=0.2\textwidth]{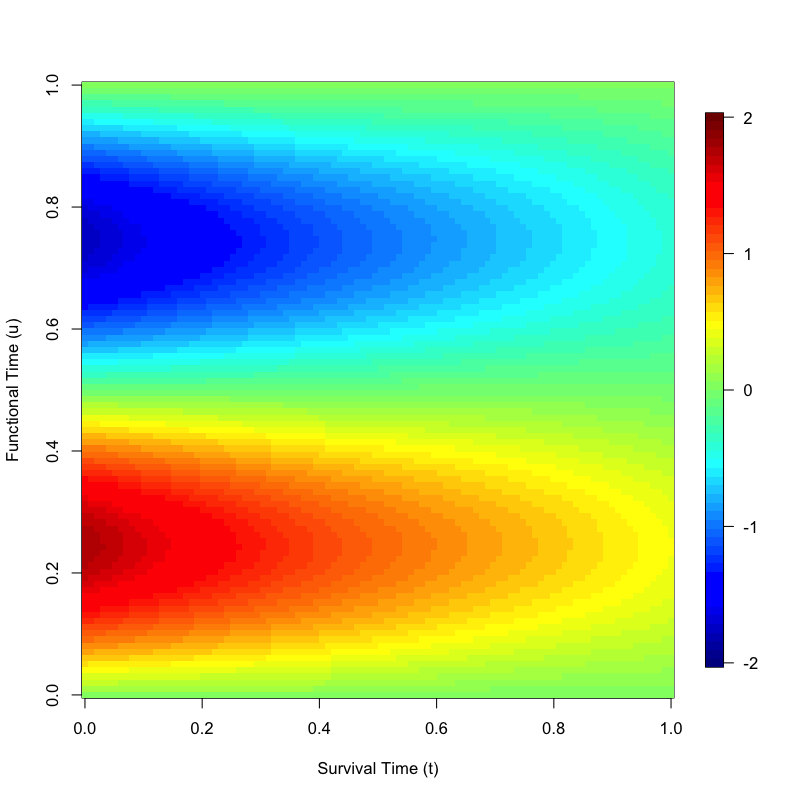} & \includegraphics[width=0.2\textwidth]{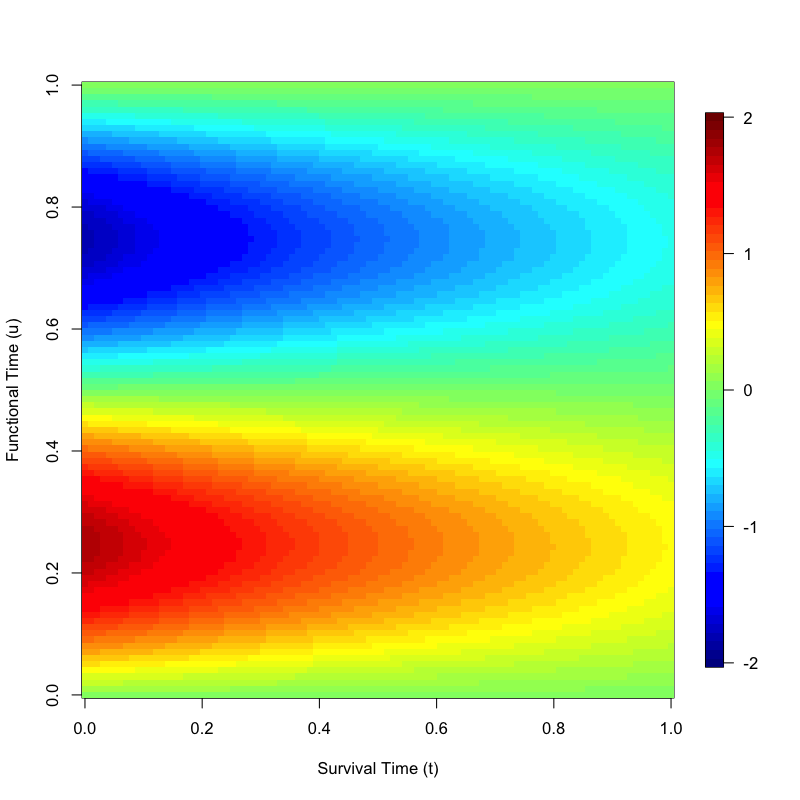} & \includegraphics[width=0.2\textwidth]{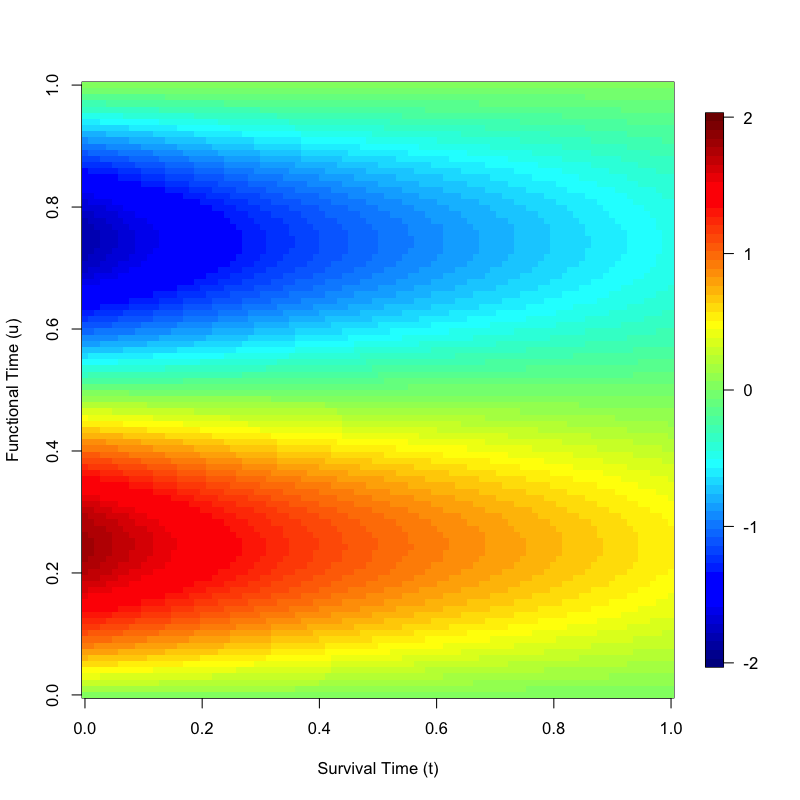} \\
 
 & w=$\infty$ & \includegraphics[width=0.2\textwidth]{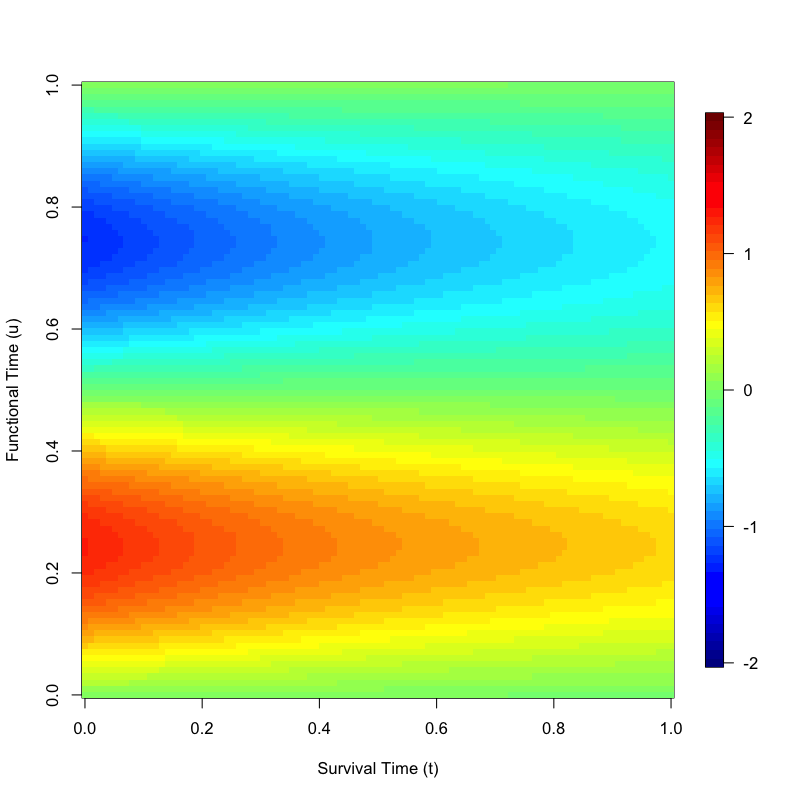} & \includegraphics[width=0.2\textwidth]{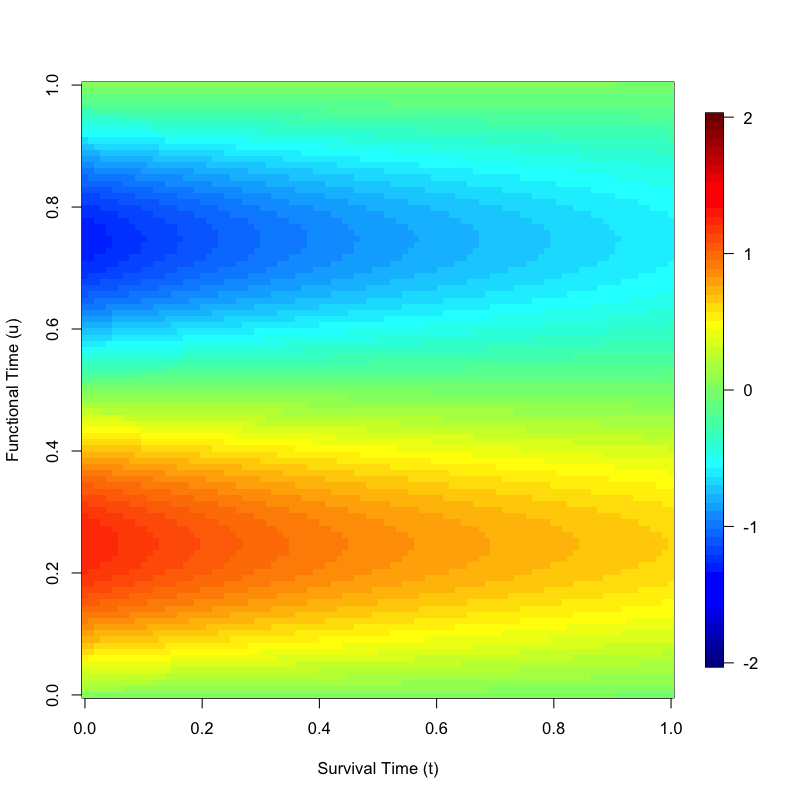} & \includegraphics[width=0.2\textwidth]{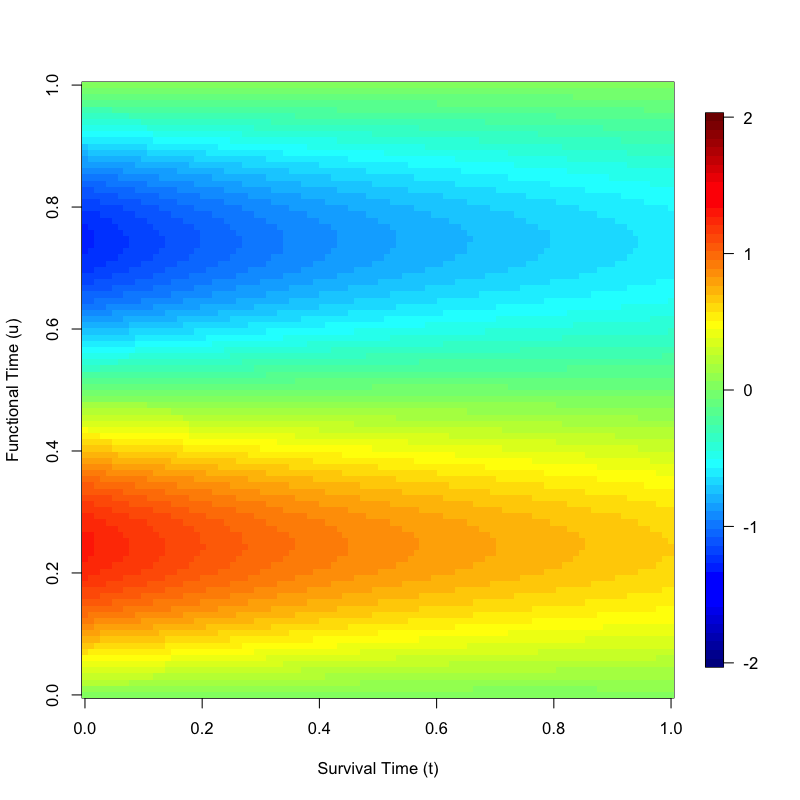} \\

 & Poisson & \includegraphics[width=0.2\textwidth]{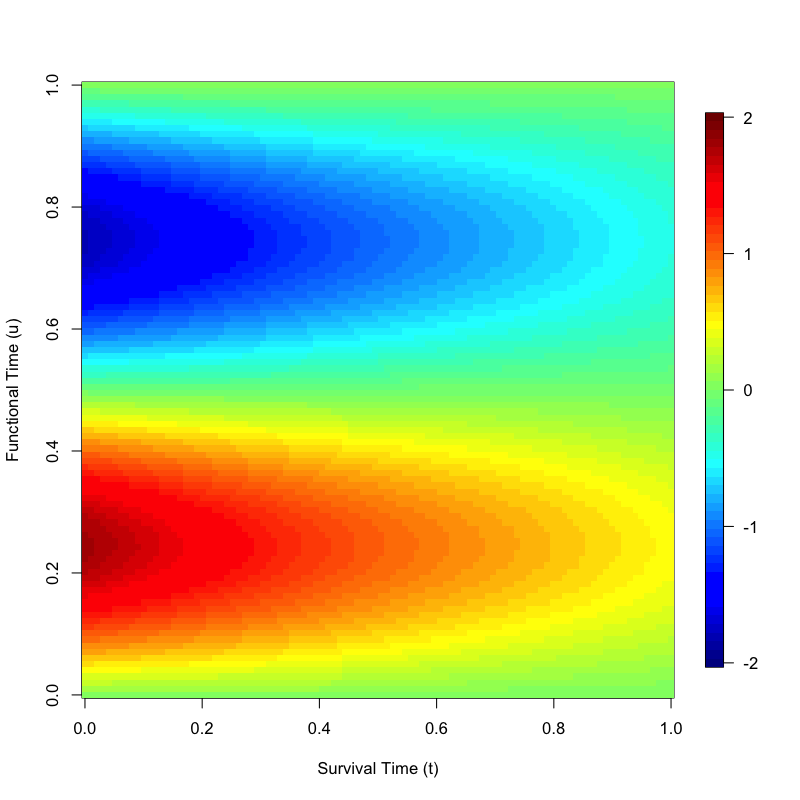} & \includegraphics[width=0.2\textwidth]{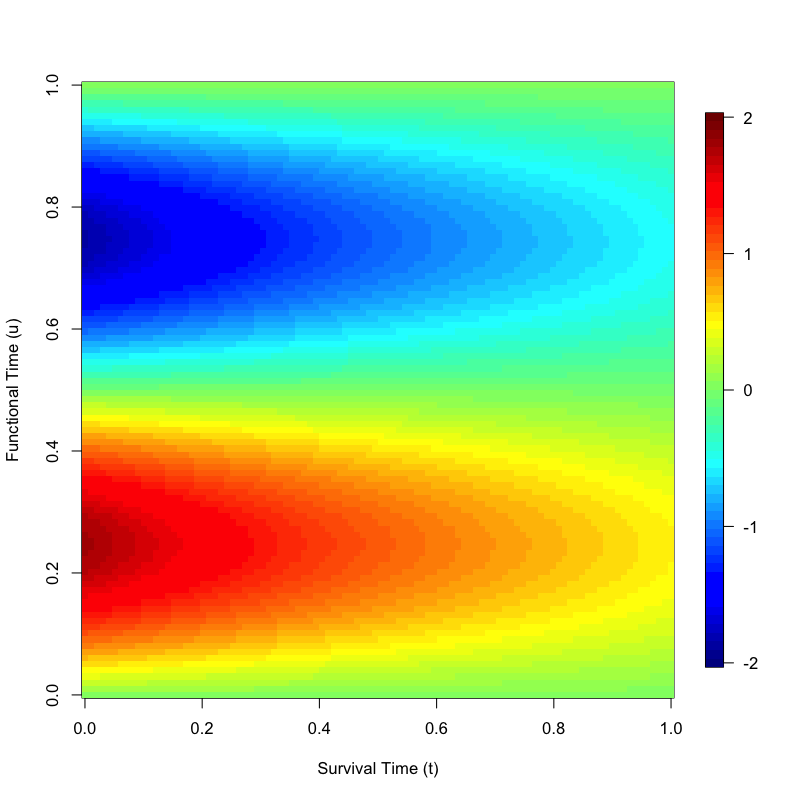} & \includegraphics[width=0.2\textwidth]{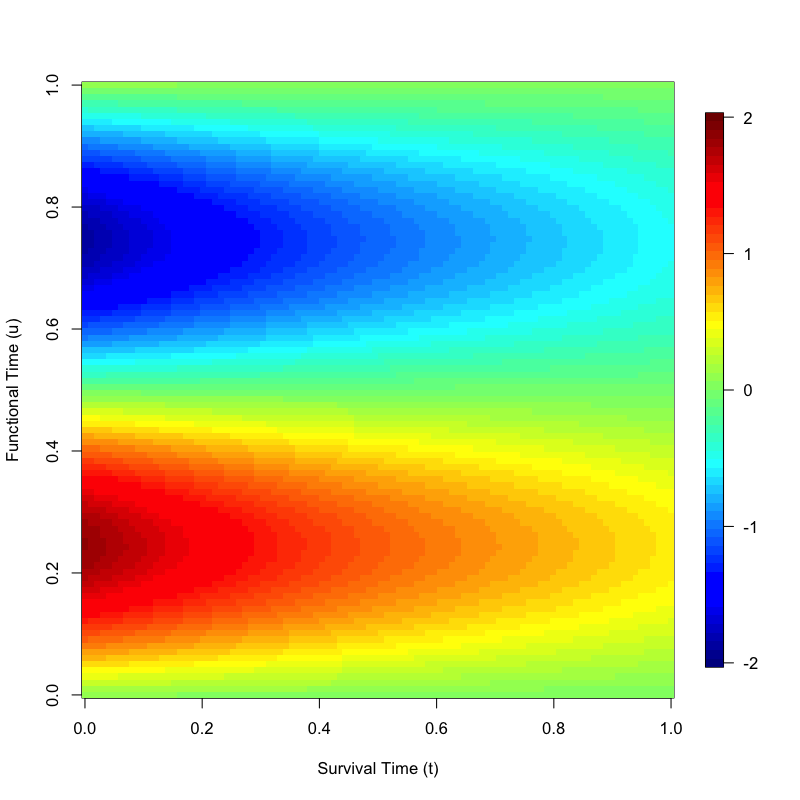} \\
 
 & CI & \includegraphics[width=0.2\textwidth]{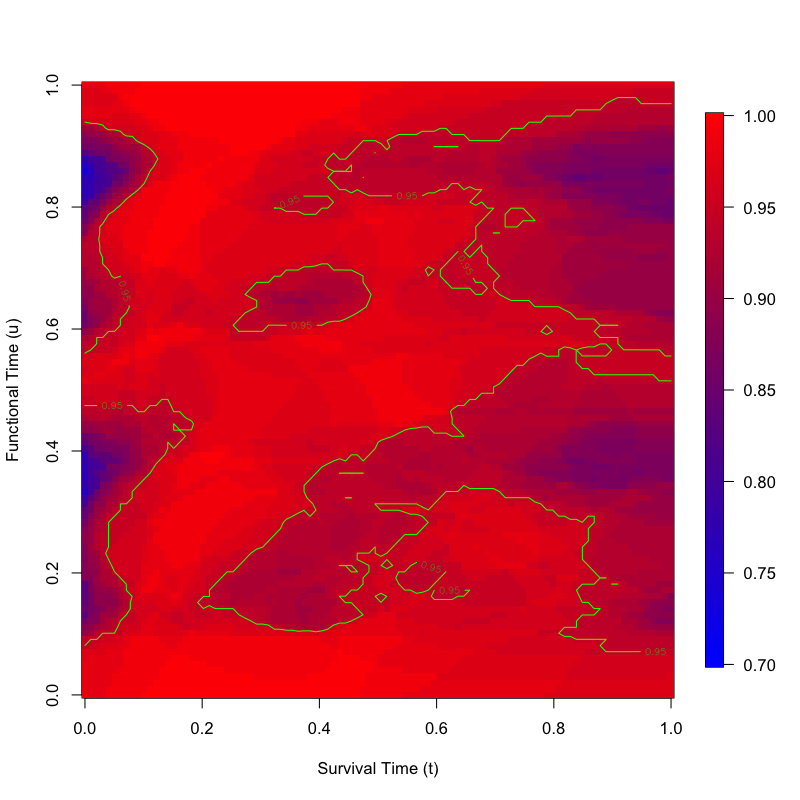} & \includegraphics[width=0.2\textwidth]{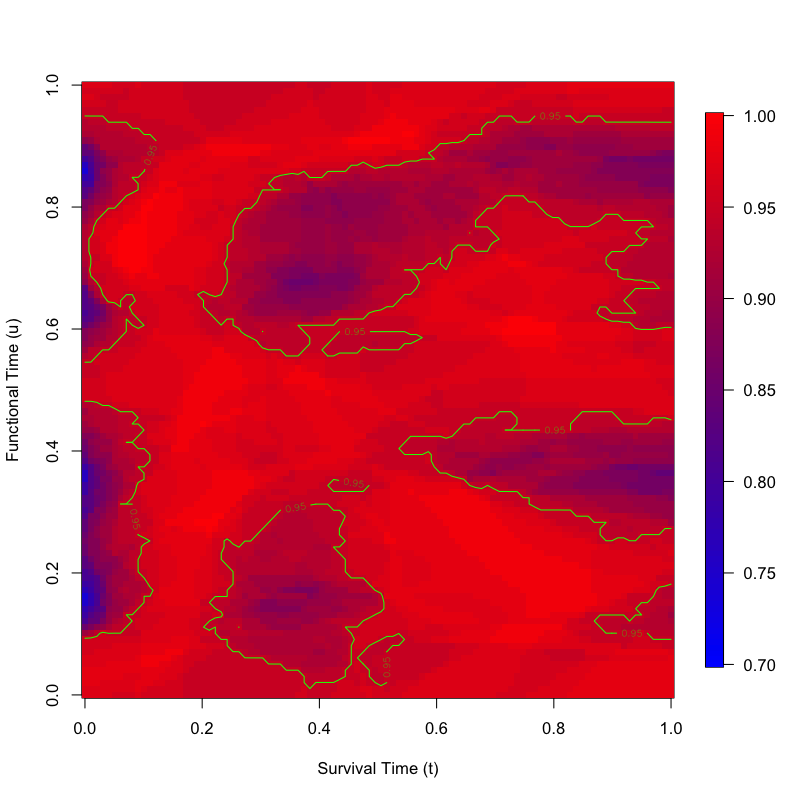} & \includegraphics[width=0.2\textwidth]{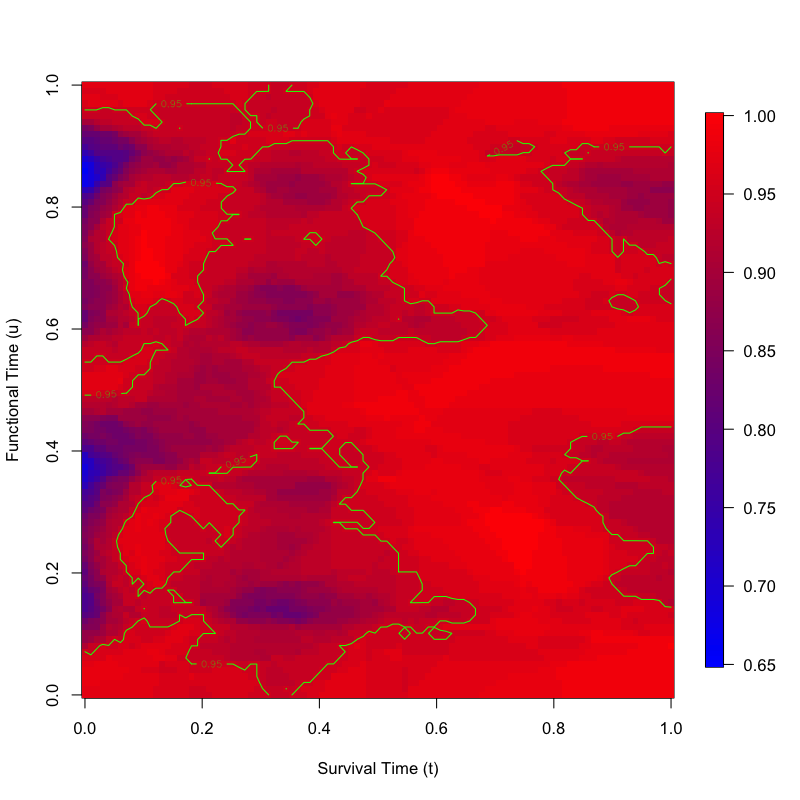} \\

 \midrule 

 Tests & Methods & N=2000 & N=3000 & N=4000 \\
 \cmidrule(lr){1-1}\cmidrule(lr){2-2}\cmidrule(lr){3-3}\cmidrule(lr){4-4}\cmidrule(lr){5-5}

 AMSE & w=0.04 & 0.050 & 0.033 & 0.029 \\

 & w=$\infty$ & 0.139 & 0.101 & 0.089 \\

 & Poisson & 0.045 & 0.031 & 0.026 \\
 
 Coverage Rate & CI & 95.1\% & 95.0\% & 95.0\% \\
 \midrule
 Computation Time & w=0.04 & 10 seconds & 12 seconds & 19 seconds \\

& Poisson & 172 seconds & 336 seconds & 280 seconds \\
 \bottomrule
\end{tabular}
}
\end{table}

\newpage

\begin{table}[!ht]
\caption{The results of the estimated functional effects based on two landmark approaches and Poisson regression model, heatmaps for confidence intervals, average mean squared errors, coverage rates, and computation times for the simulations of $sin(2\pi u)/(t/2+1)$. The simulations are performed on different sample sizes (N=2000, 3000 or 4000)}
\label{tab:table3}
\scalebox{0.95}{
\begin{tabular}{*{5}{c}}
\toprule
True Effect & Methods & \multicolumn{3}{c}{Sample Sizes} \\
\cmidrule(lr){3-5}\cmidrule(lr){1-1}\cmidrule(lr){2-2}
& & N=2000 & N=3000 & N=4000 \\
\cmidrule(lr){3-3}\cmidrule(lr){4-4}\cmidrule(lr){5-5}
\includegraphics[width=0.2\textwidth]{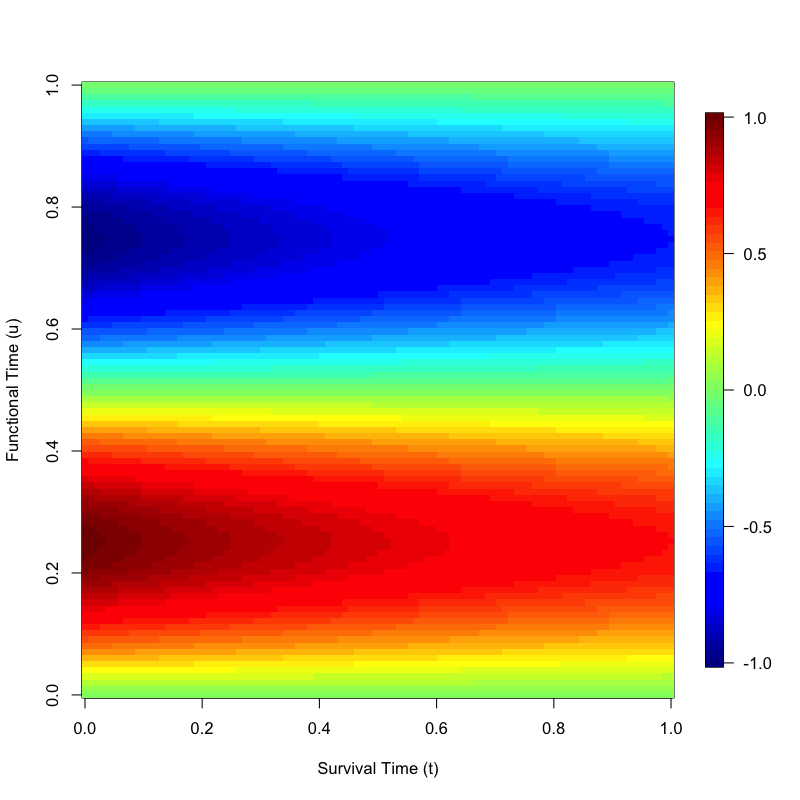} & w=0.04 & \includegraphics[width=0.2\textwidth]{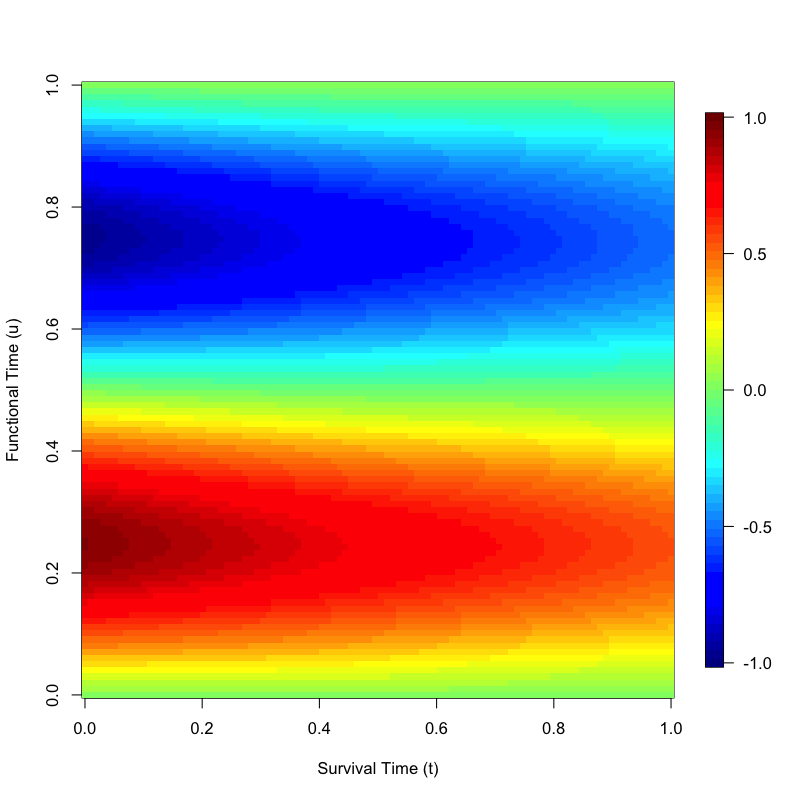} & \includegraphics[width=0.2\textwidth]{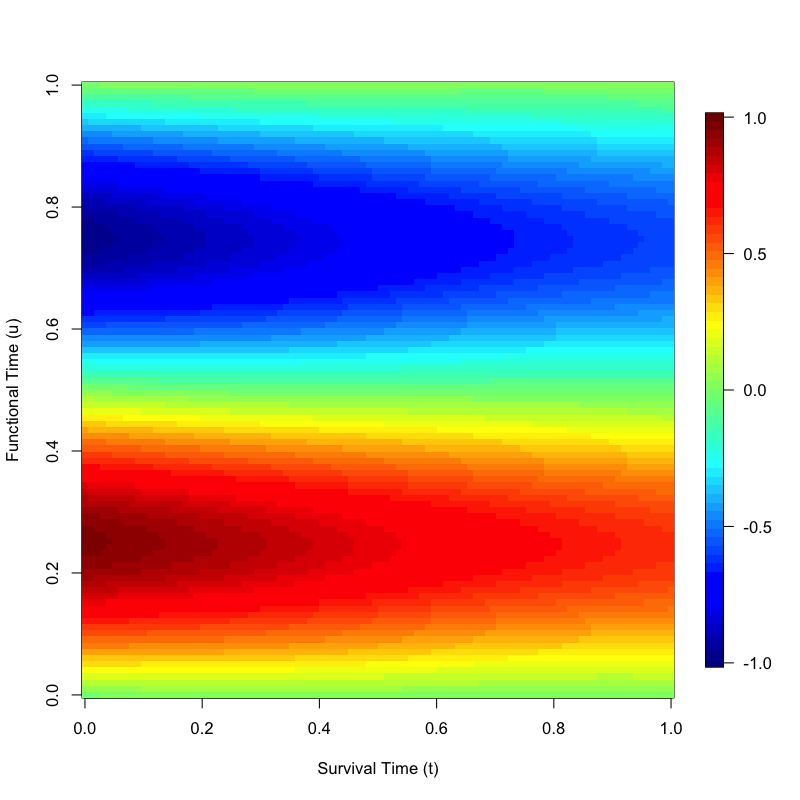} & \includegraphics[width=0.2\textwidth]{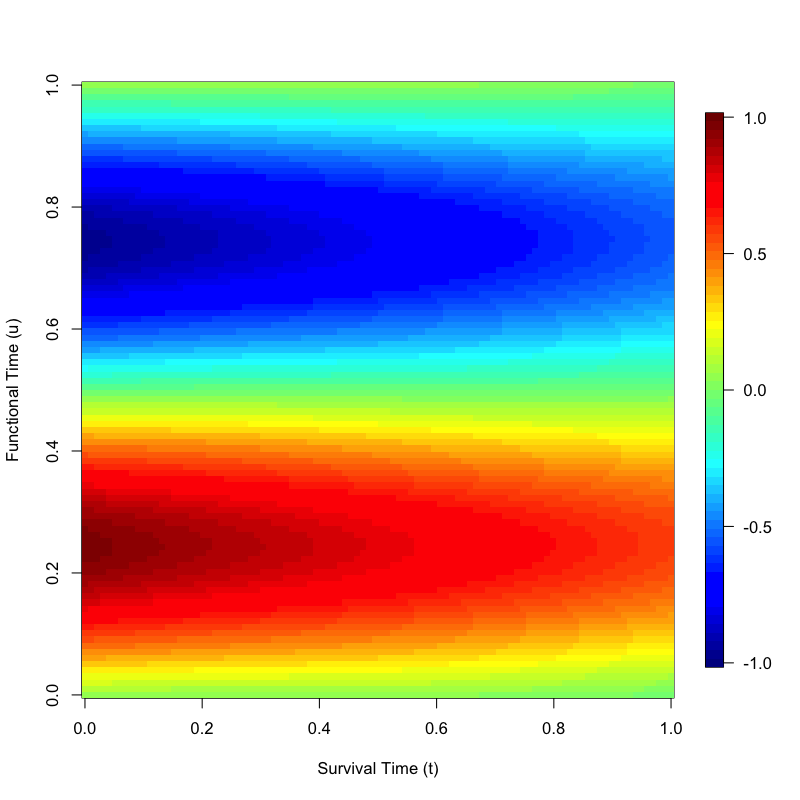} \\
 
 & w=$\infty$ & \includegraphics[width=0.2\textwidth]{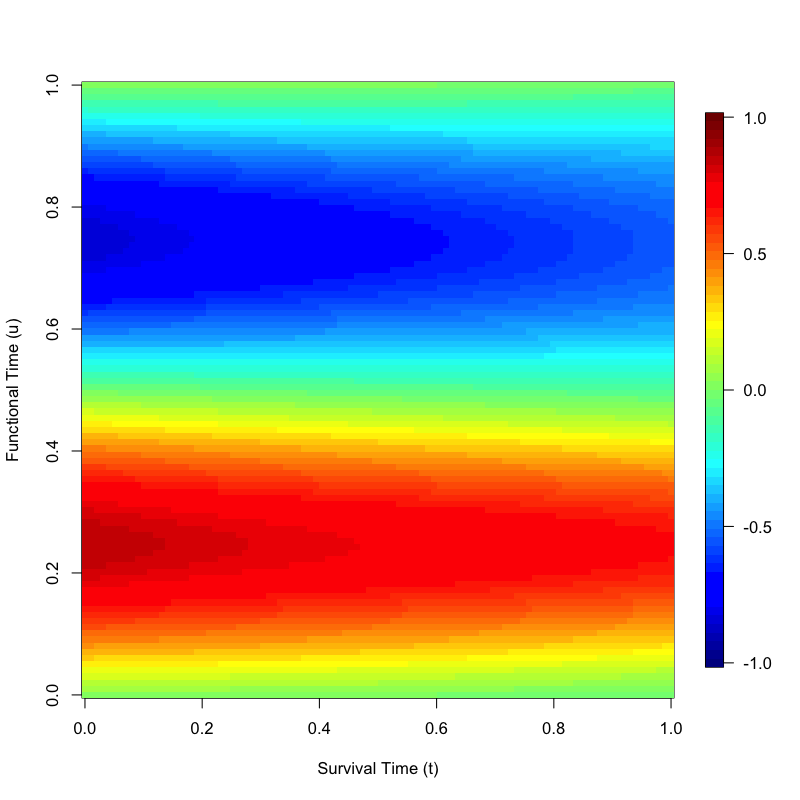} & \includegraphics[width=0.2\textwidth]{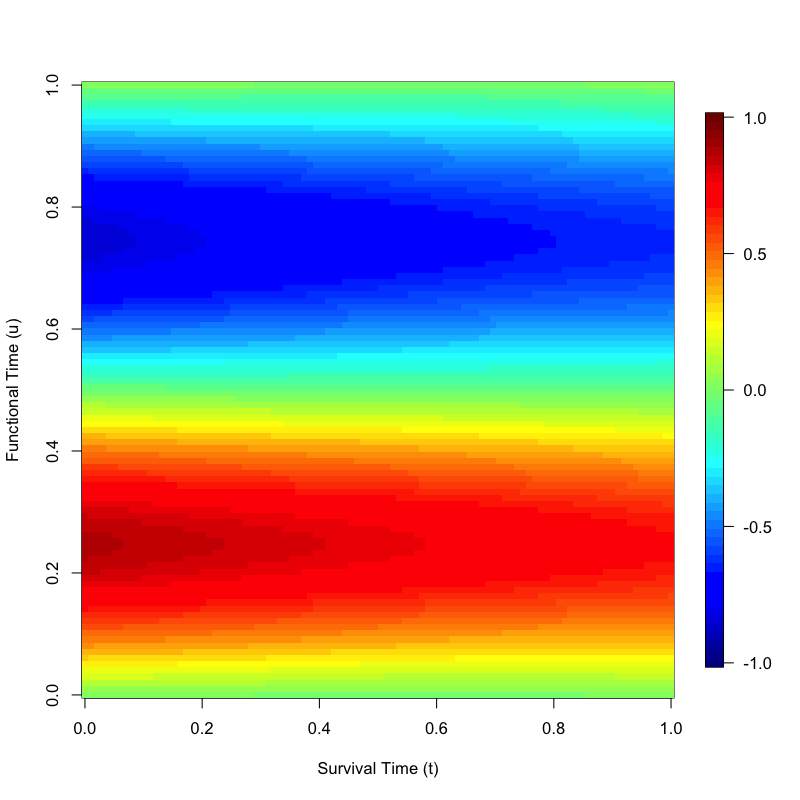} & \includegraphics[width=0.2\textwidth]{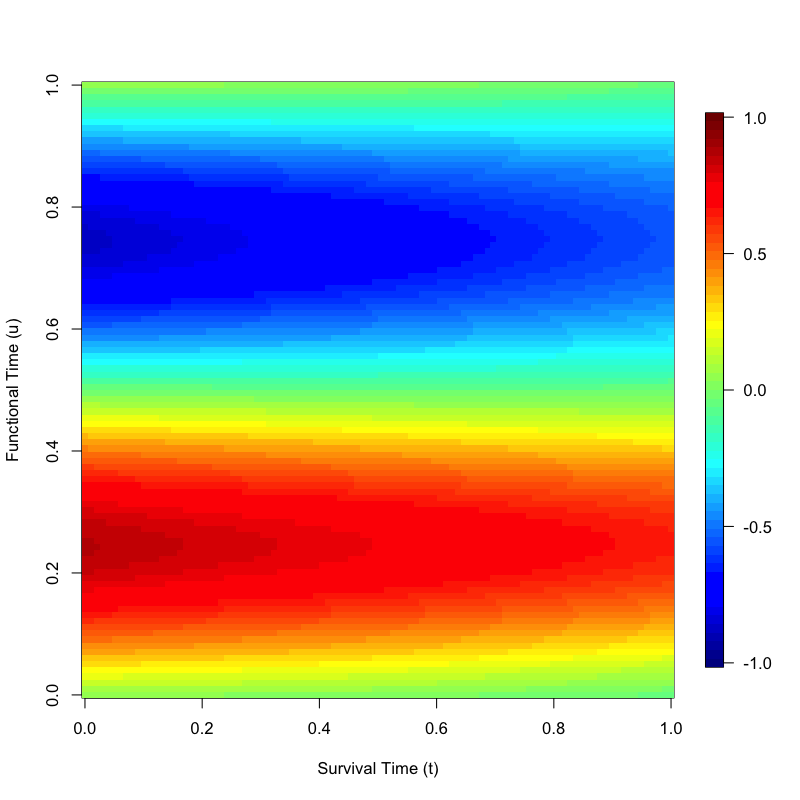} \\

 & Poisson & \includegraphics[width=0.2\textwidth]{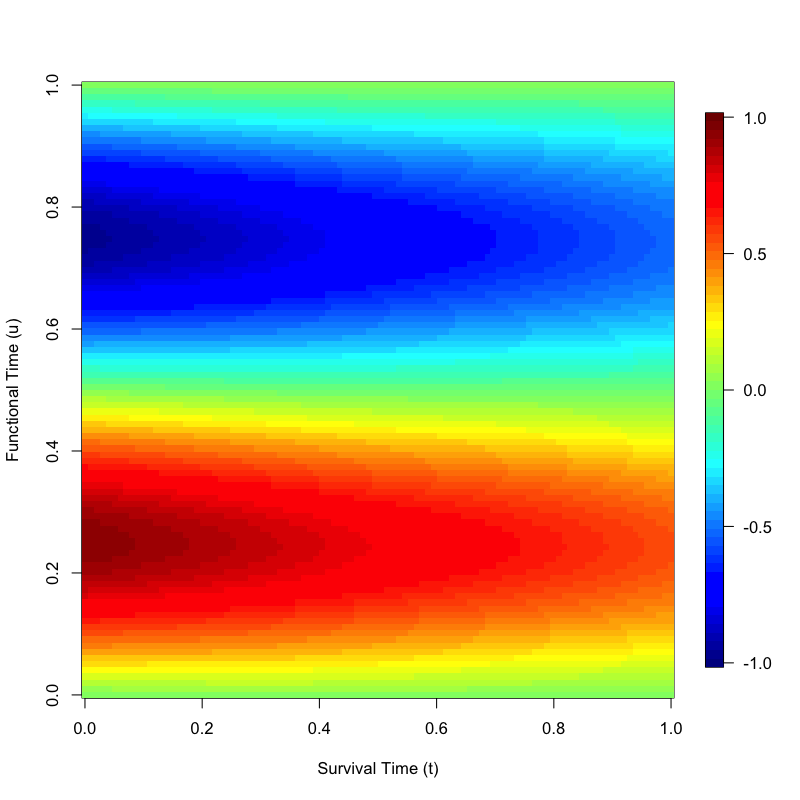} & \includegraphics[width=0.2\textwidth]{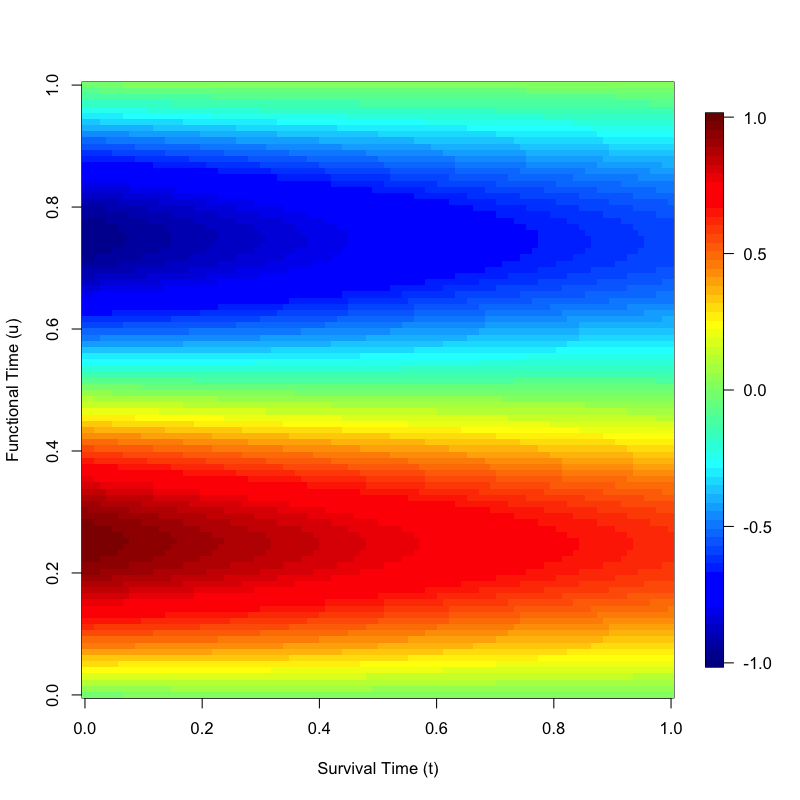} & \includegraphics[width=0.2\textwidth]{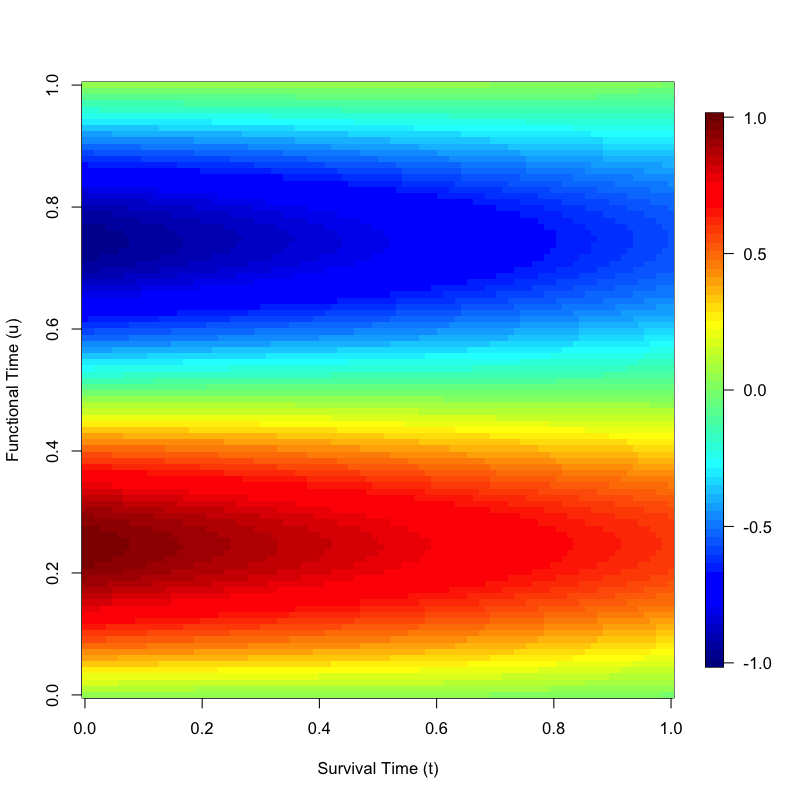} \\
 
 & CI & \includegraphics[width=0.2\textwidth]{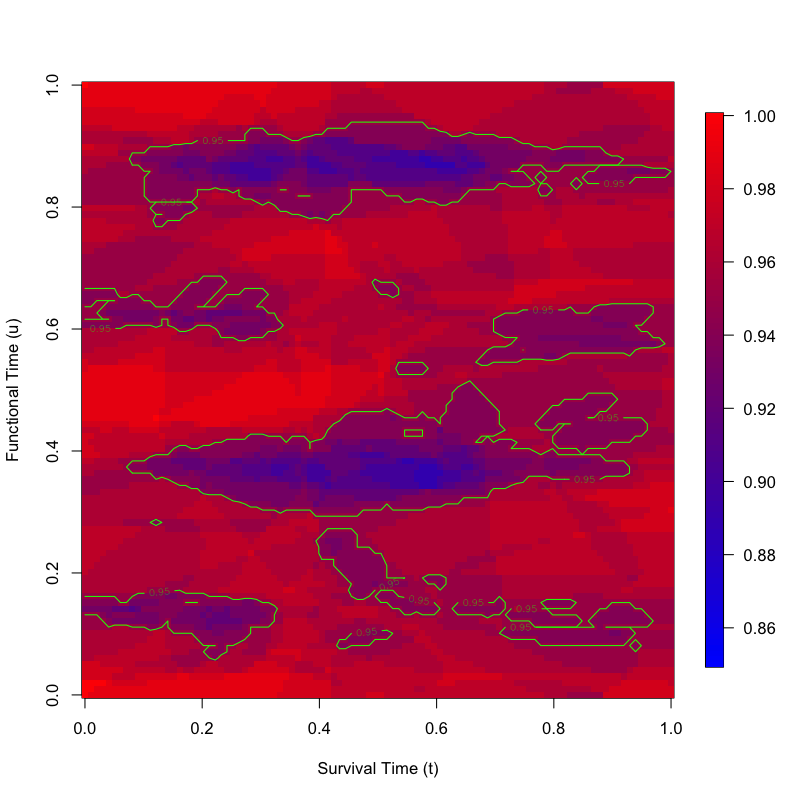} & \includegraphics[width=0.2\textwidth]{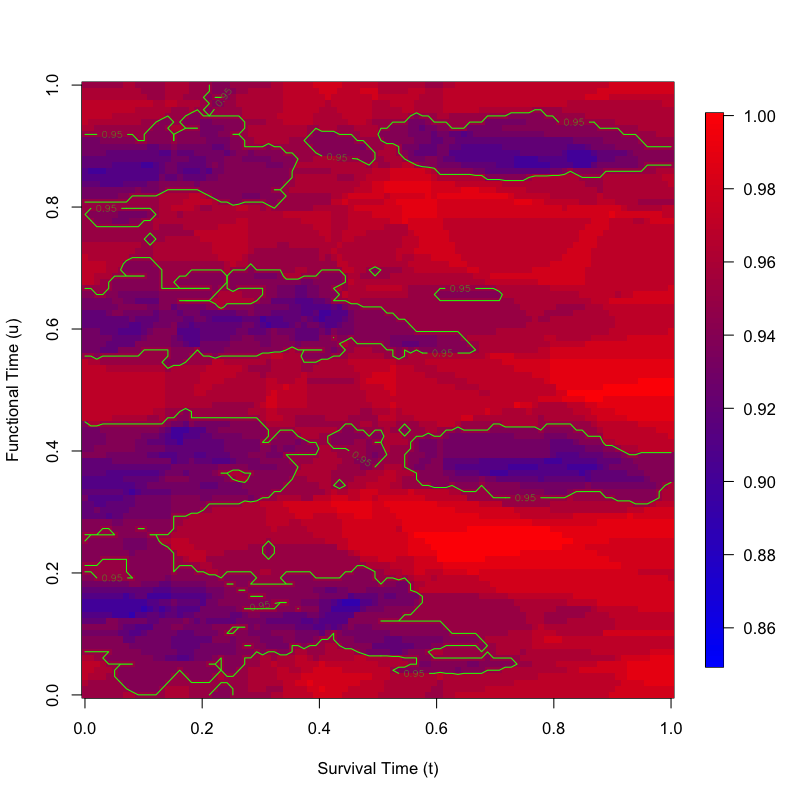} & \includegraphics[width=0.2\textwidth]{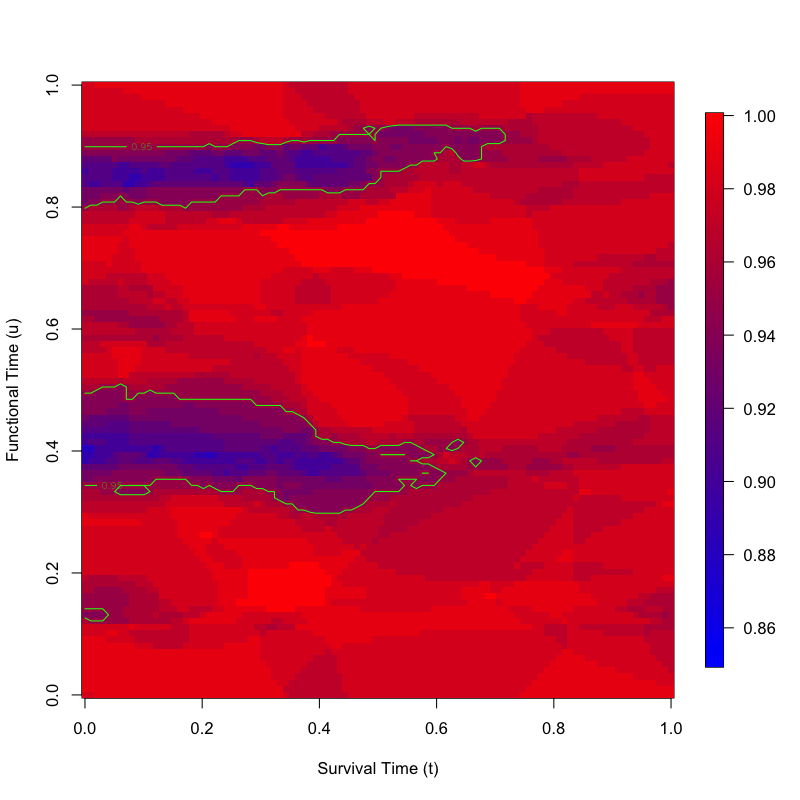} \\

 \midrule 

 Tests & Methods & N=2000 & N=3000 & N=4000 \\
 \cmidrule(lr){1-1}\cmidrule(lr){2-2}\cmidrule(lr){3-3}\cmidrule(lr){4-4}\cmidrule(lr){5-5}

 AMSE & w=0.04 & 0.037 & 0.026 & 0.019 \\

 & w=$\infty$ & 0.087 & 0.055 & 0.038 \\

 & Poisson & 0.034 & 0.024 & 0.018 \\
 
 Coverage Rate & CI & 95.6\% & 95.4\% & 95.3\% \\
 \midrule
 Computation Time & w=0.04 & 9 seconds & 16 seconds & 25 seconds \\

& Poisson & 374 seconds & 378 seconds & 329 seconds \\
 \bottomrule
\end{tabular}
}
\end{table}

\newpage

\begin{table}[!ht]
\caption{The results of the estimated functional effects based on two landmark approaches and Poisson regression model, heatmaps for confidence intervals, average mean squared errors, coverage rates, and computation times for the simulations of $10cos\left\{4\pi(t-u)\right\}$. The simulations are performed on different sample sizes (N=2000, 3000 or 4000)}
\label{tab:table4}
\scalebox{0.95}{
\begin{tabular}{*{5}{c}}
\toprule
True Effect & Methods & \multicolumn{3}{c}{Sample Sizes} \\
\cmidrule(lr){3-5}\cmidrule(lr){1-1}\cmidrule(lr){2-2}
& & N=2000 & N=3000 & N=4000 \\
\cmidrule(lr){3-3}\cmidrule(lr){4-4}\cmidrule(lr){5-5}
\includegraphics[width=0.2\textwidth]{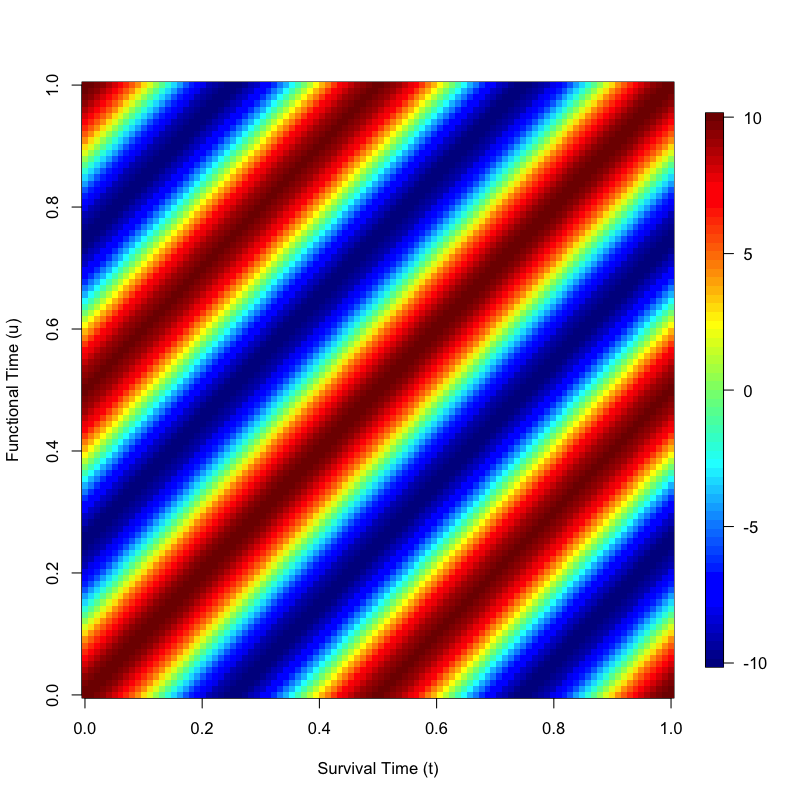} & w=0.04 & \includegraphics[width=0.2\textwidth]{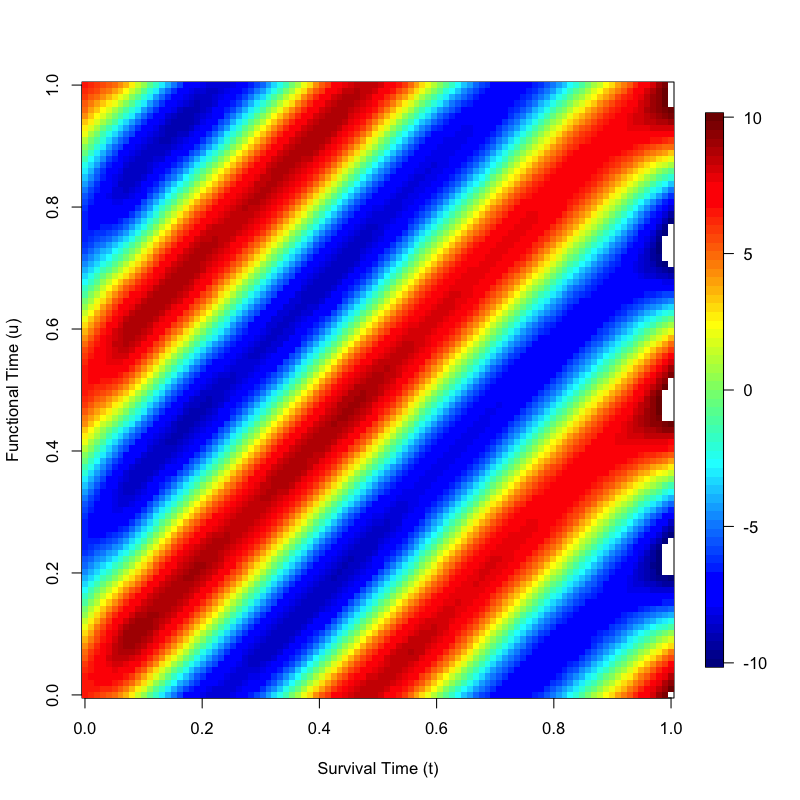} & \includegraphics[width=0.2\textwidth]{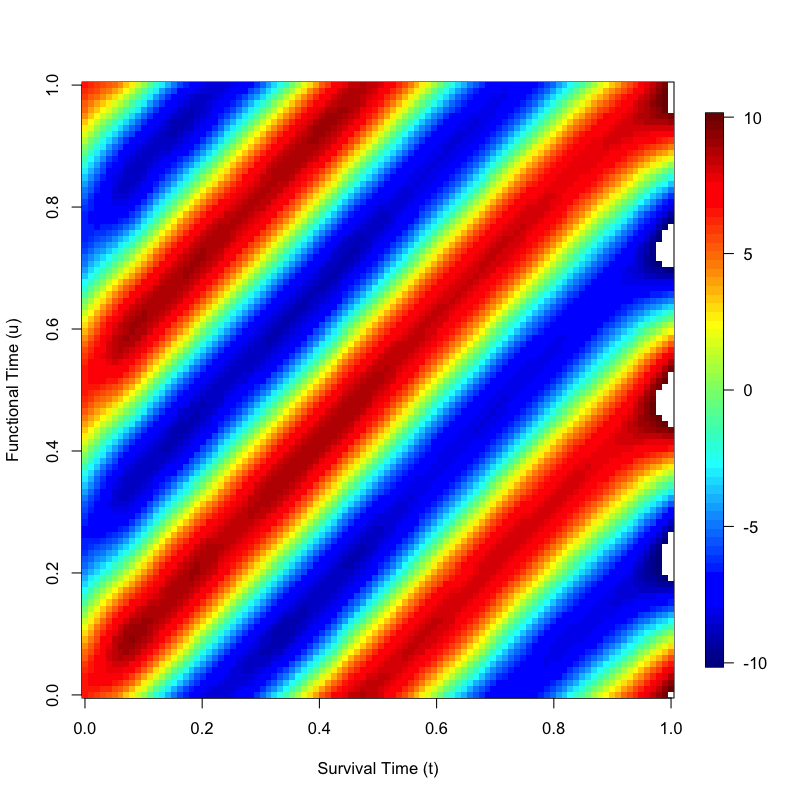} & \includegraphics[width=0.2\textwidth]{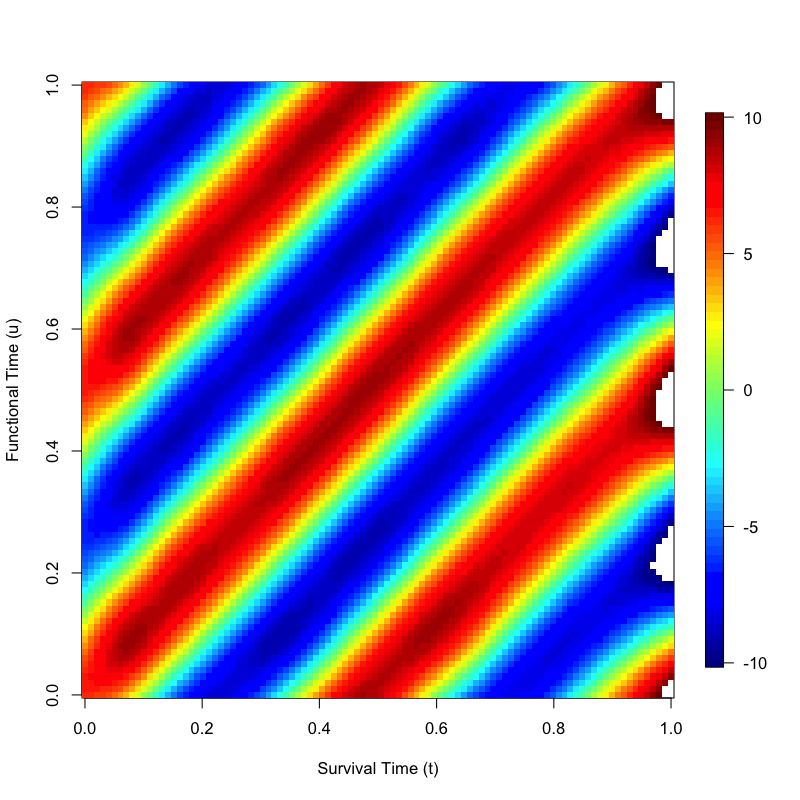} \\
 
 & w=$\infty$ & \includegraphics[width=0.2\textwidth]{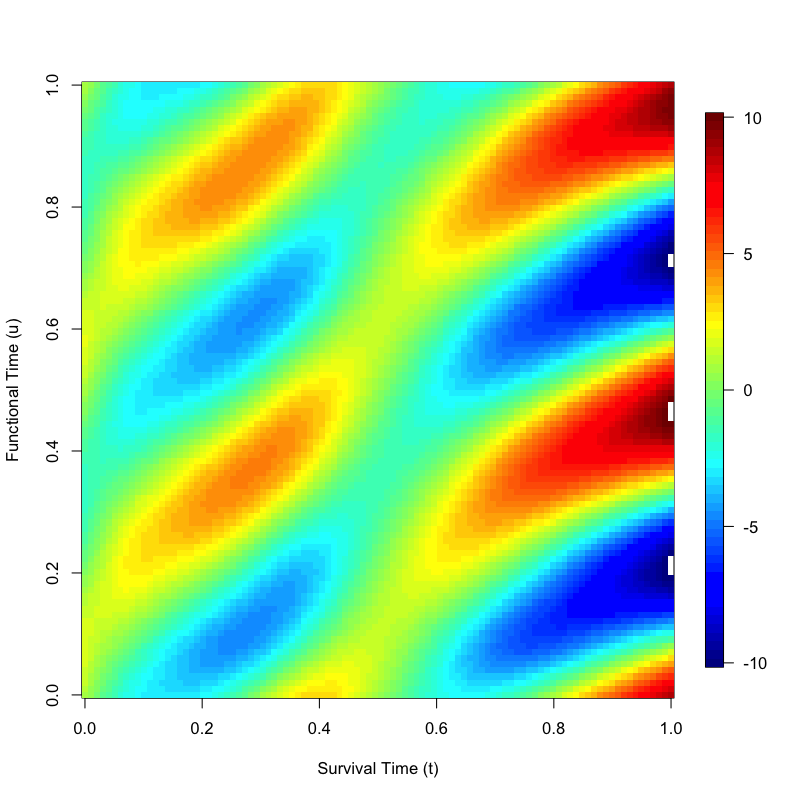} & \includegraphics[width=0.2\textwidth]{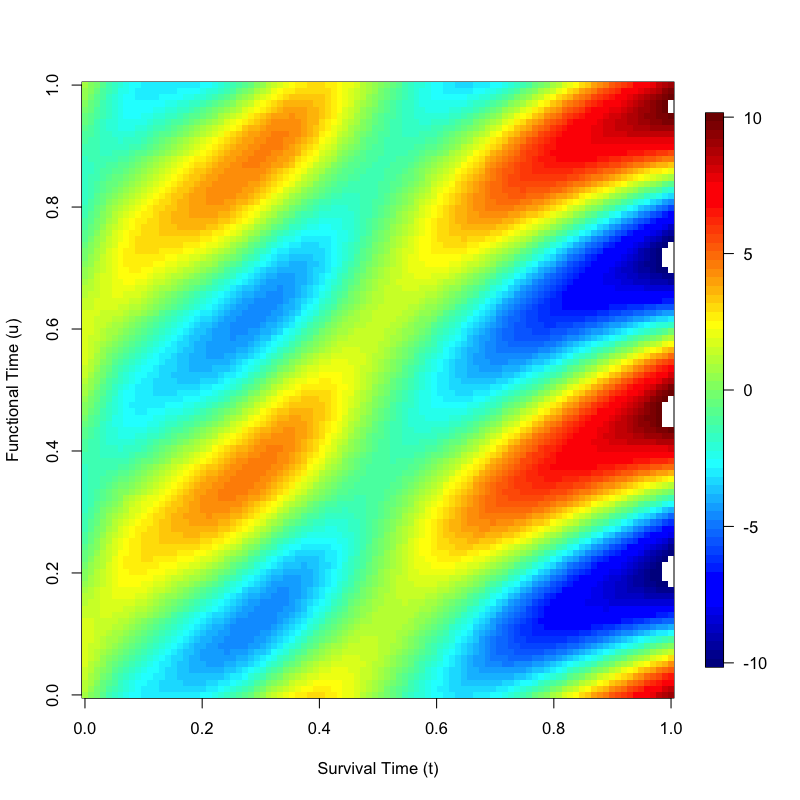} & \includegraphics[width=0.2\textwidth]{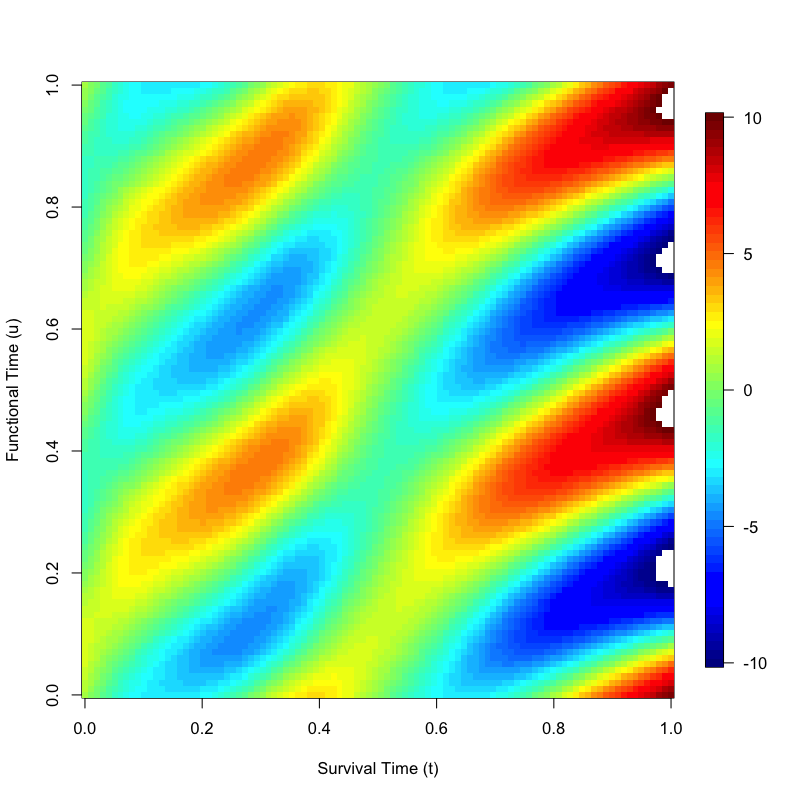} \\

 & Poisson & \includegraphics[width=0.2\textwidth]{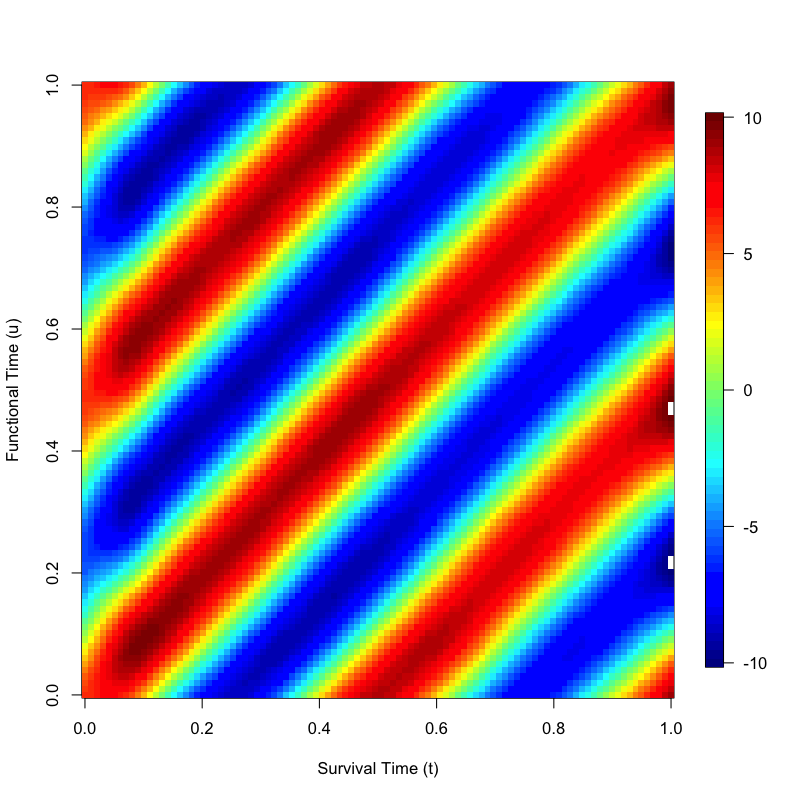} & \includegraphics[width=0.2\textwidth]{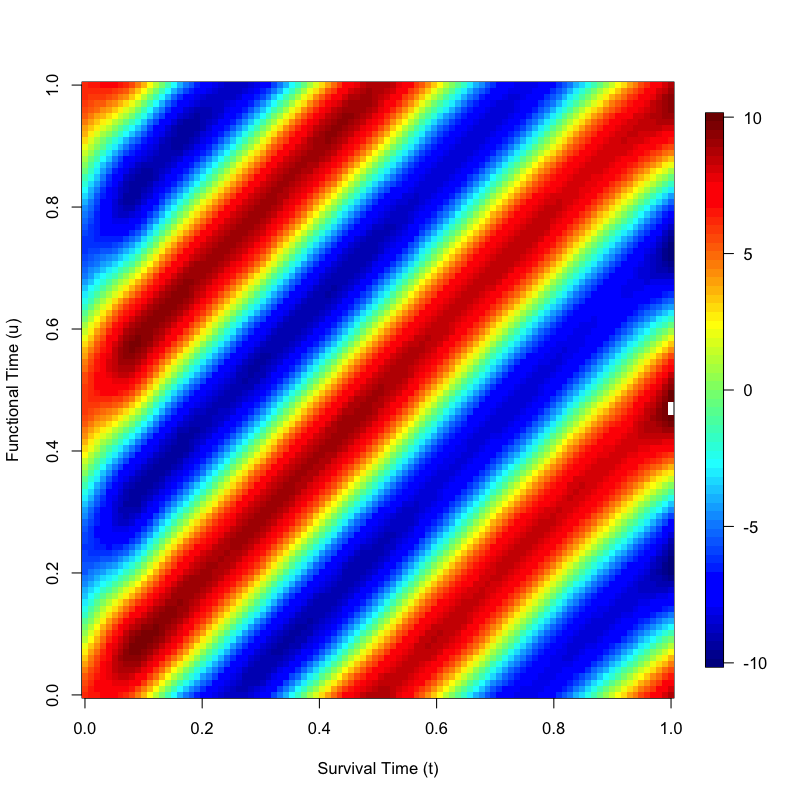} & \includegraphics[width=0.2\textwidth]{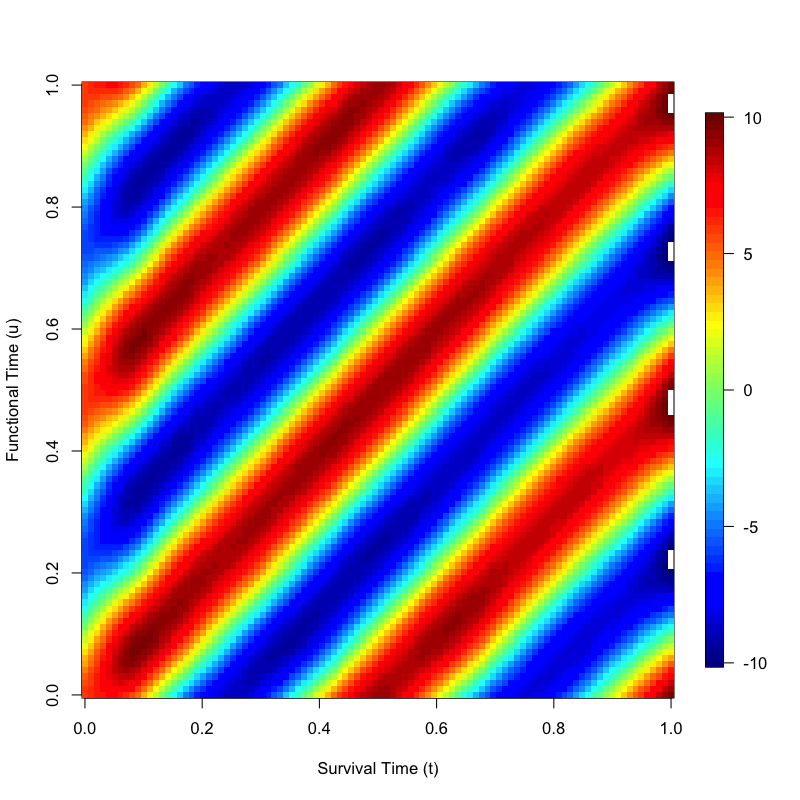} \\
 
 & CI & \includegraphics[width=0.2\textwidth]{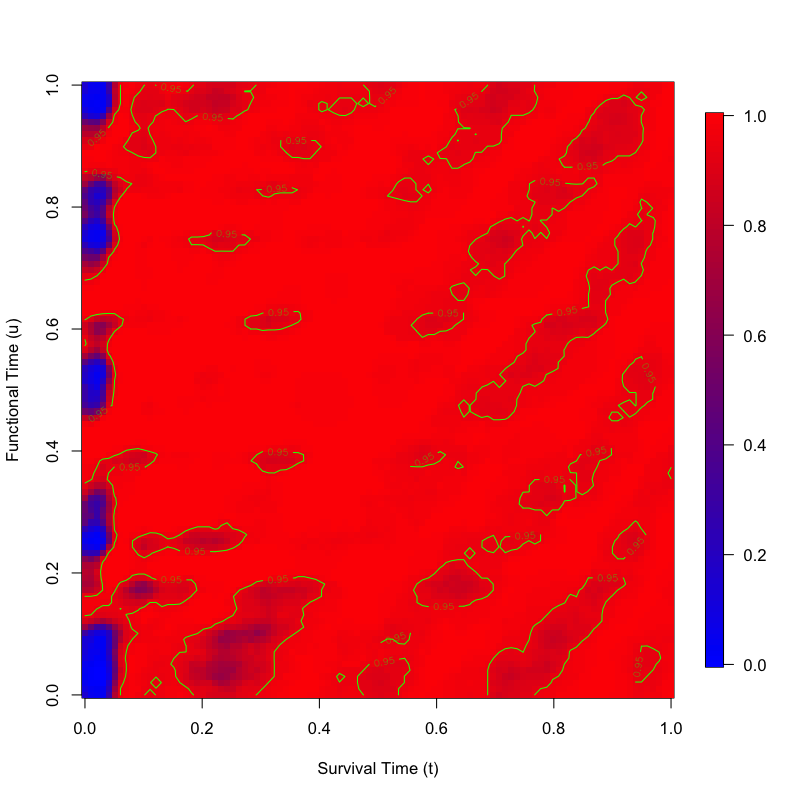} & \includegraphics[width=0.2\textwidth]{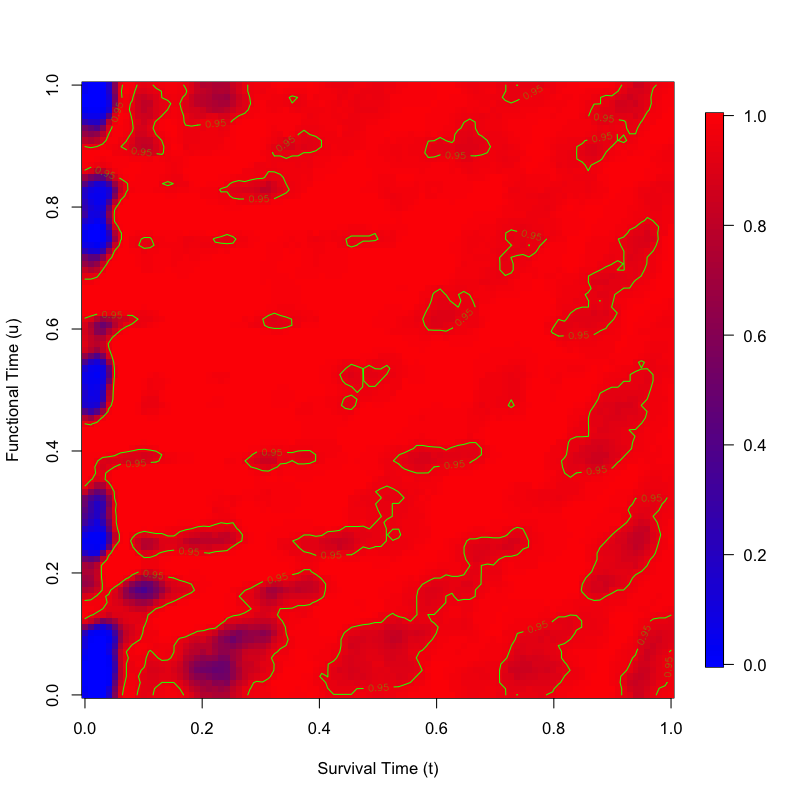} & \includegraphics[width=0.2\textwidth]{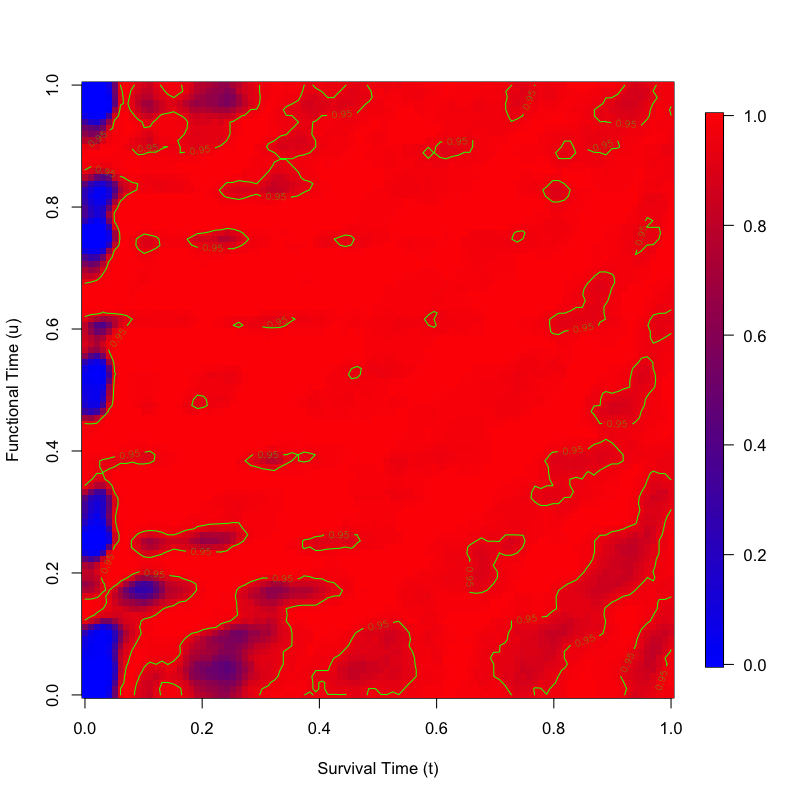} \\

 \midrule 

 Tests & Methods & N=2000 & N=3000 & N=4000 \\
 \cmidrule(lr){1-1}\cmidrule(lr){2-2}\cmidrule(lr){3-3}\cmidrule(lr){4-4}\cmidrule(lr){5-5}

 AMSE & w=0.04 & 5.563 & 5.051 & 4.782 \\

 & w=$\infty$ & 44.28 & 44.17 & 44.11 \\

 & Poisson & 3.533 & 2.900 & 2.510 \\
 
 Coverage Rate & CI & 95.1\% & 94.3\% & 94.0\% \\
 \midrule
 Computation Time & w=0.04 & 17 seconds & 25 seconds & 31 seconds \\

& Poisson & 168 seconds & 290 seconds & 379 seconds \\
 \bottomrule
\end{tabular}
}
\end{table}



\clearpage
\section{Supplementary material}
\label{hflm:sec:Supplementary}
\subsection{Additional Simulation Results}
In Table \ref{tab:table5}, we explore a more intricate scenario, $\gamma(u,t) = \cos(2\pi(t^3-2/(u^2+1)))$. In this complex case, TV-FLCM-L demonstrates its ability to capture the underlying structure and critical features of the estimated functional effects. However, it is important to acknowledge that some finer details, particularly in regions with extreme values, may be overlooked. To address this, we evaluate three different sample sizes: 2000, 3000, and 4000. Increasing the sample size in such intricate functions improves the accuracy of the overall shape and reduces the averaged mean squared error (AMSE) for both the landmark and Poisson methods. Furthermore, the Poisson regression method achieves accurate coverage rates, further enhancing its robustness. In summary, regardless of the functional effect's complexity, the TV-FLCM-L consistently delivers meaningful insights and reliable inferences, underscoring its versatility in addressing diverse and challenging functional relationships.

In Table \ref{tab:table6}, we present survival curves, histograms of surviving subjects, and histograms of censored subjects for the four simulated functional effects in two different sample sizes, 2000 and 3000. All four functions demonstrate well-defined survival curves. The survival curves for the first, second, and fourth functions are remarkably similar. In contrast, the third survival curve reveals a notable divergence, indicating that while some subjects survived until the study's conclusion, a significant proportion succumbed during the analysis period. This discrepancy arises because, in the third function, a scaling factor of 10 was introduced at the onset of the functional effect, thereby amplifying the function's magnitude and increasing the likelihood of subject events occurring earlier in the survival phase. Across the histograms of survival distributions: most subjects survive at the beginning of the study, and survival rates gradually decrease towards the end. In particular, the third function shows a distinct reduction in the number of surviving subjects during the interval from 0.8 to 1.0. Regarding the censored distributions, all four plots show a substantial concentration of censored data at the final time point (1.0). Additionally, the third function displays a notable clustering of censored data at the study's outset. These observations provide valuable insight into the dynamics of survival and censoring within each functional effect.

\newpage

\begin{table}[!ht]
\caption{The results of the estimated functional effects based on two landmark approaches and Poisson regression model, heatmaps for confidence intervals, average mean squared errors, coverage rates, and computation times for the simulations of $cos(2\pi(t^3-2/(u^2+1)))$. The simulations are performed on different sample sizes (N=2000, 3000 or 4000)}
\label{tab:table5}
\scalebox{0.95}{
\begin{tabular}{*{5}{c}}
\toprule
True Effect & Methods & \multicolumn{3}{c}{Sample Sizes} \\
\cmidrule(lr){3-5}\cmidrule(lr){1-1}\cmidrule(lr){2-2}
& & N=2000 & N=3000 & N=4000 \\
\cmidrule(lr){3-3}\cmidrule(lr){4-4}\cmidrule(lr){5-5}
\includegraphics[width=0.2\textwidth]{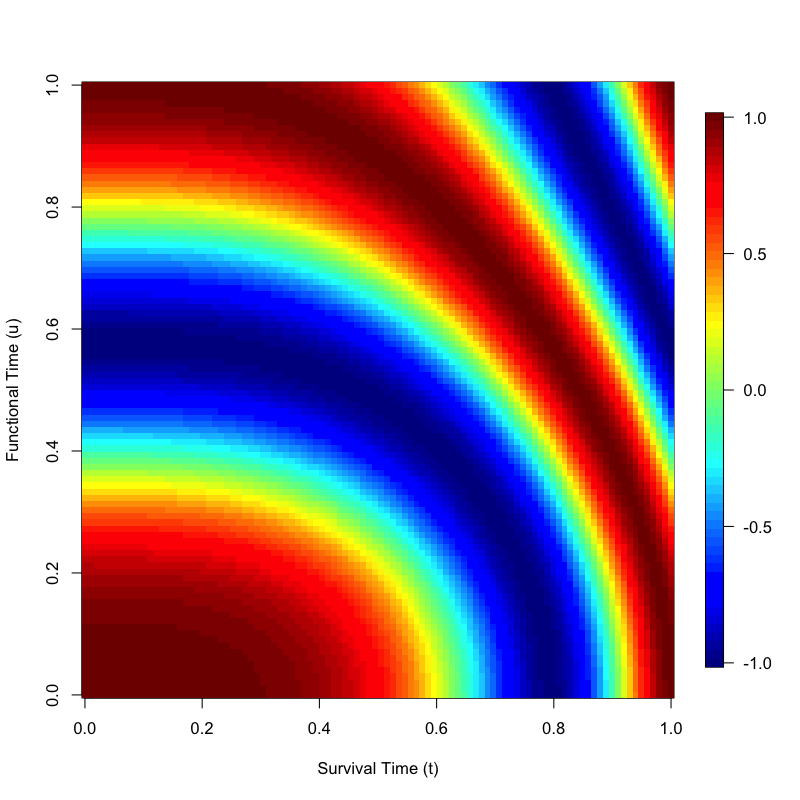} & w=0.04 & \includegraphics[width=0.2\textwidth]{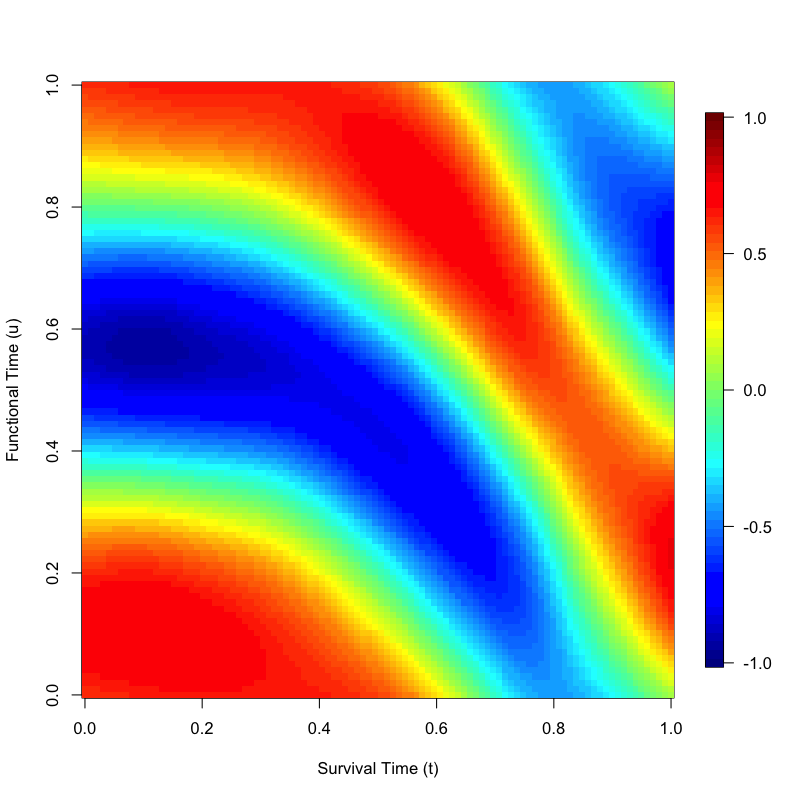} & \includegraphics[width=0.2\textwidth]{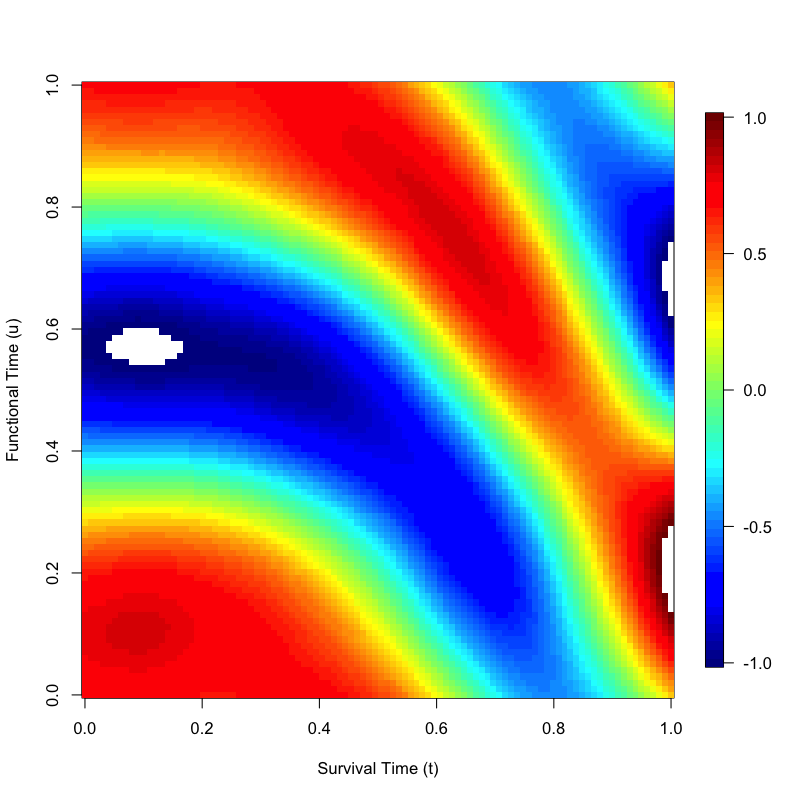} & \includegraphics[width=0.2\textwidth]{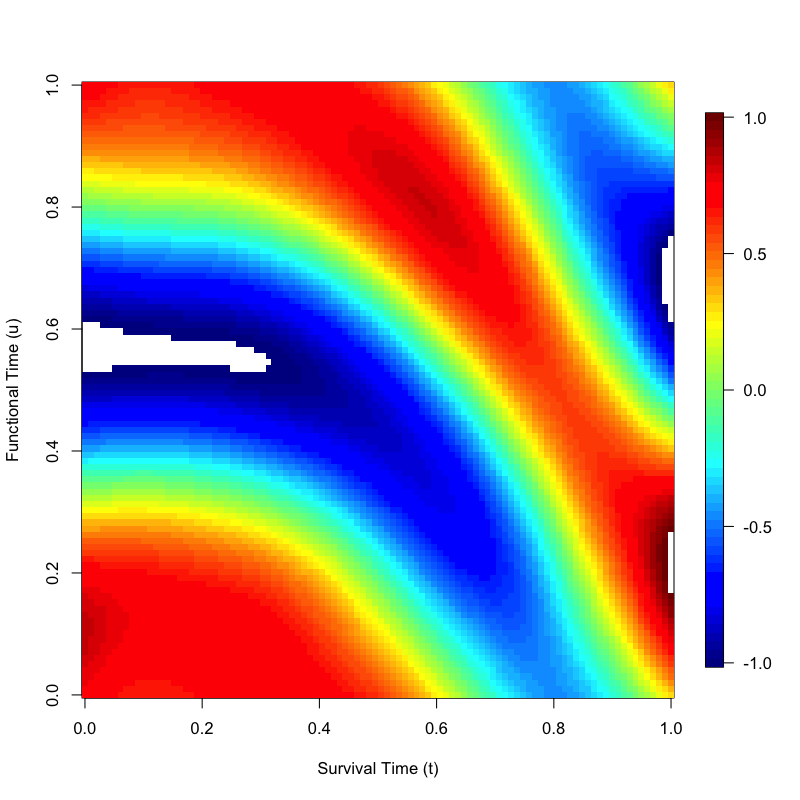} \\
 
 & w=$\infty$ & \includegraphics[width=0.2\textwidth]{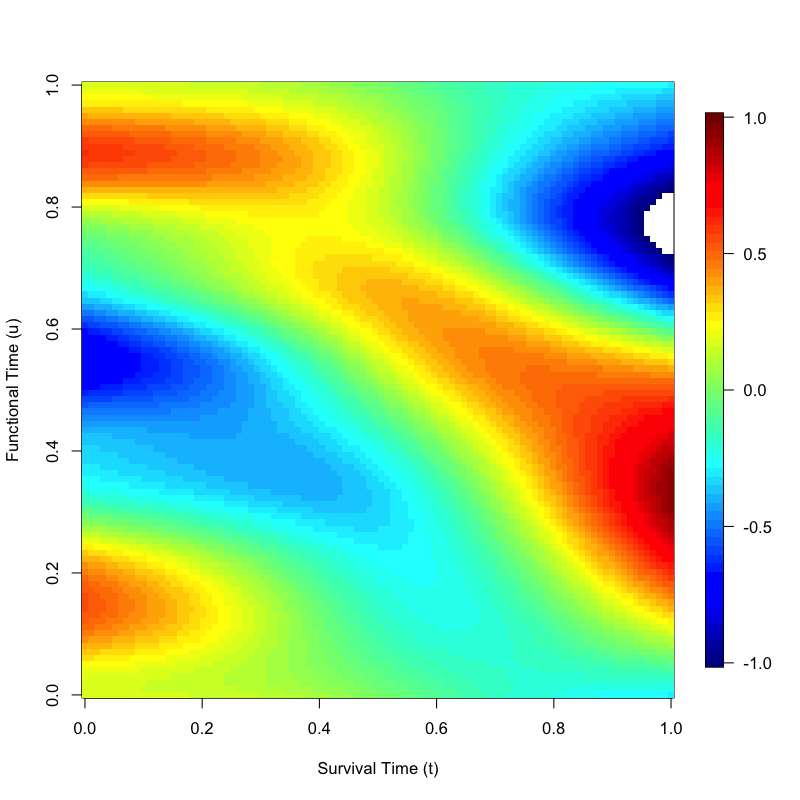} & \includegraphics[width=0.2\textwidth]{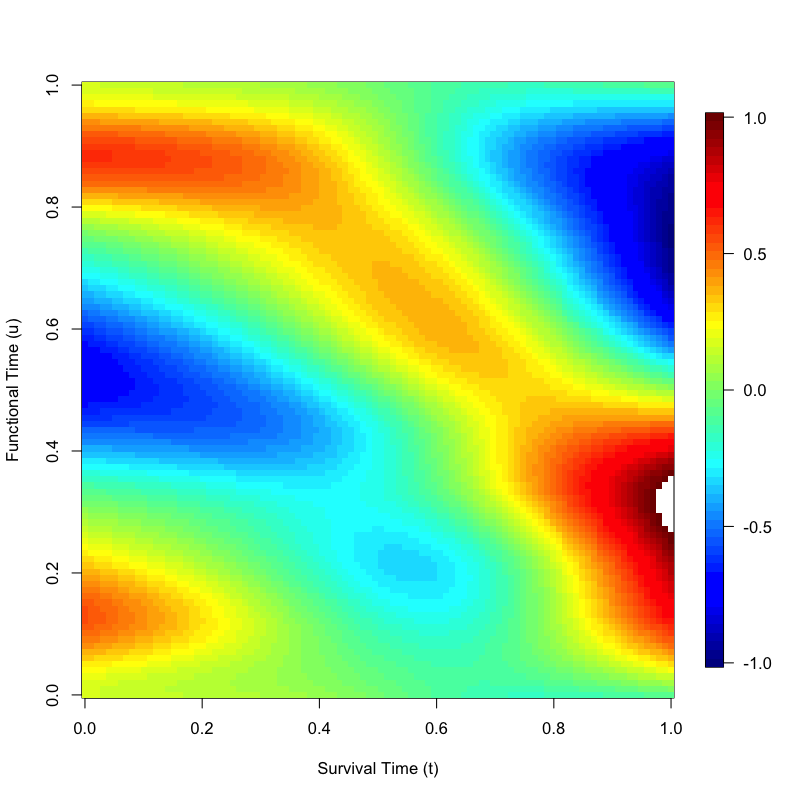} & \includegraphics[width=0.2\textwidth]{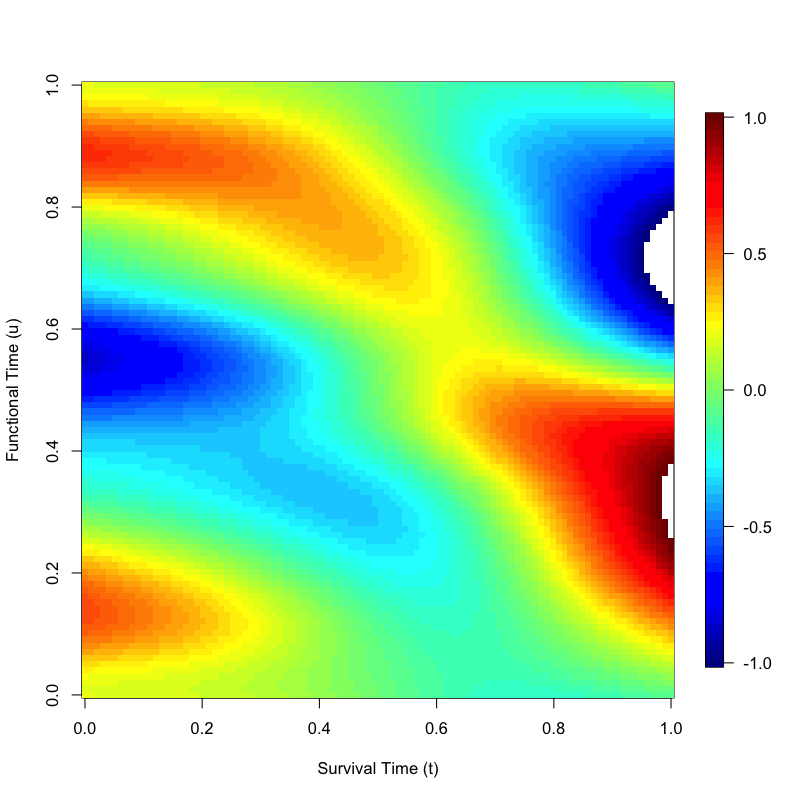} \\

 & Poisson & \includegraphics[width=0.2\textwidth]{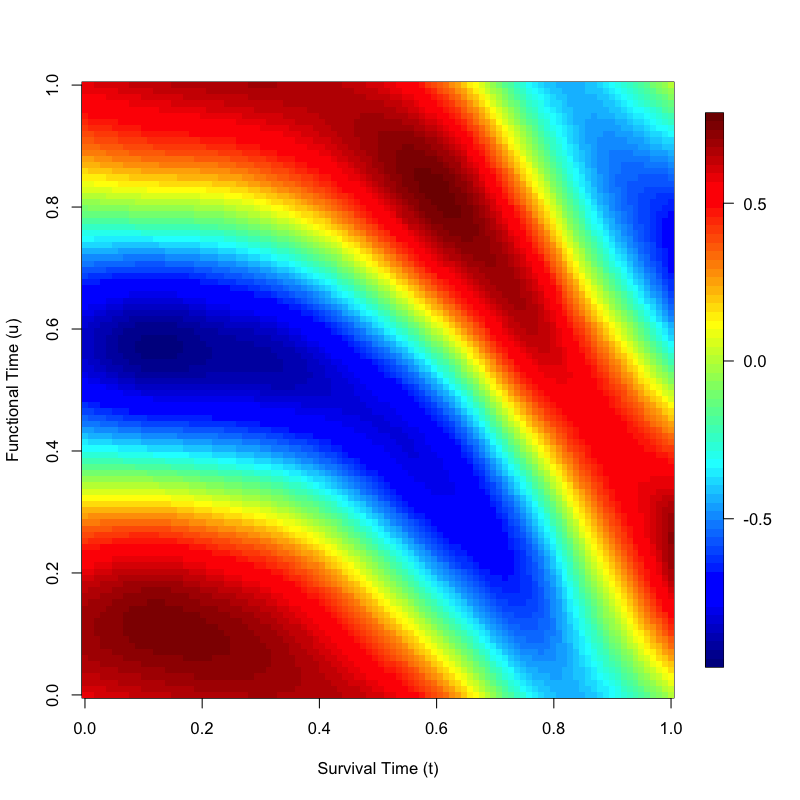} & \includegraphics[width=0.2\textwidth]{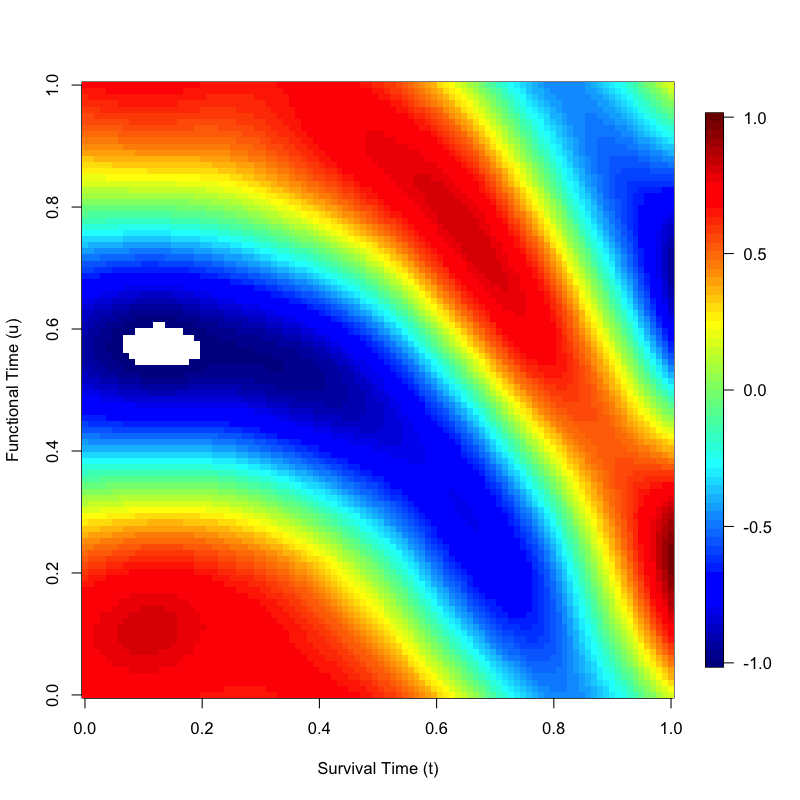} & \includegraphics[width=0.2\textwidth]{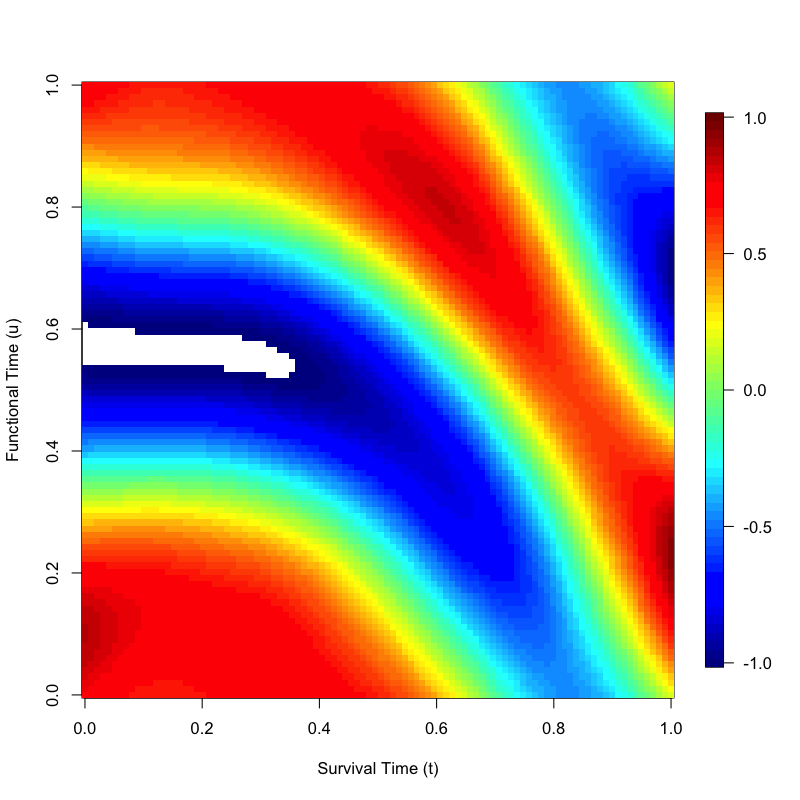} \\
 
 & CI & \includegraphics[width=0.2\textwidth]{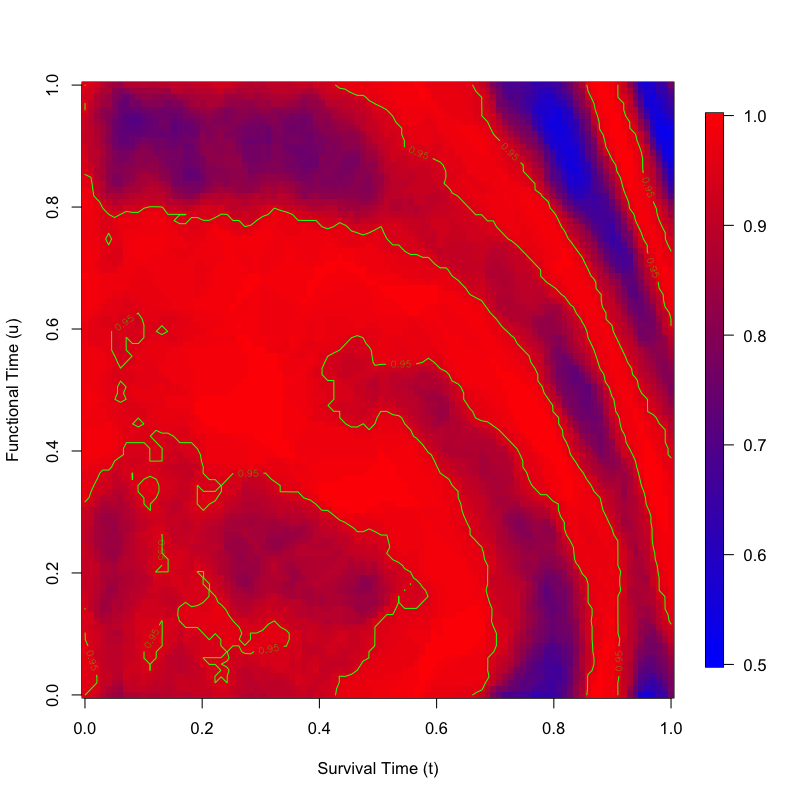} & \includegraphics[width=0.2\textwidth]{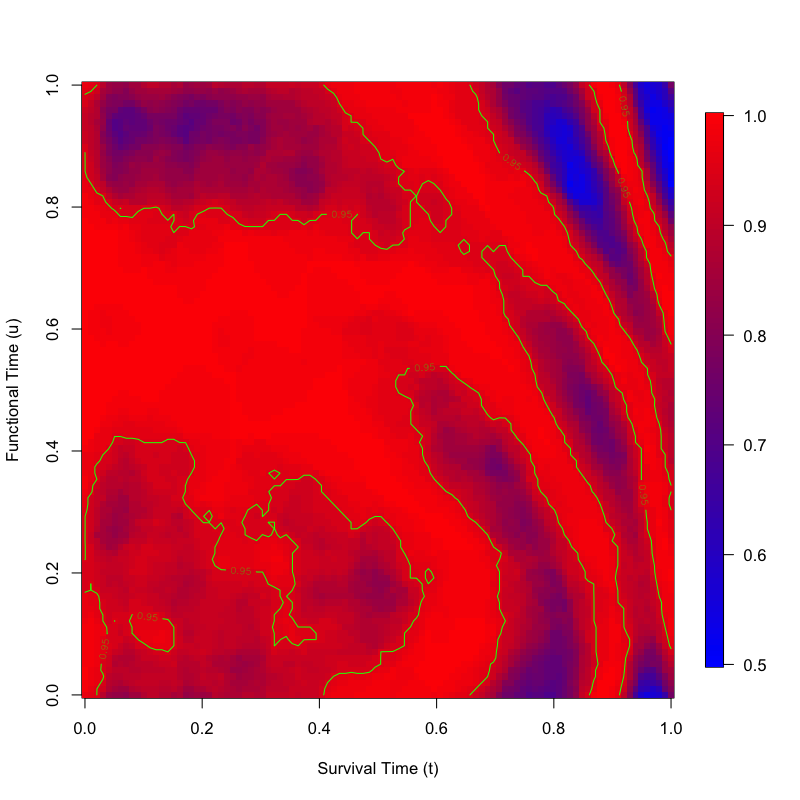} & \includegraphics[width=0.2\textwidth]{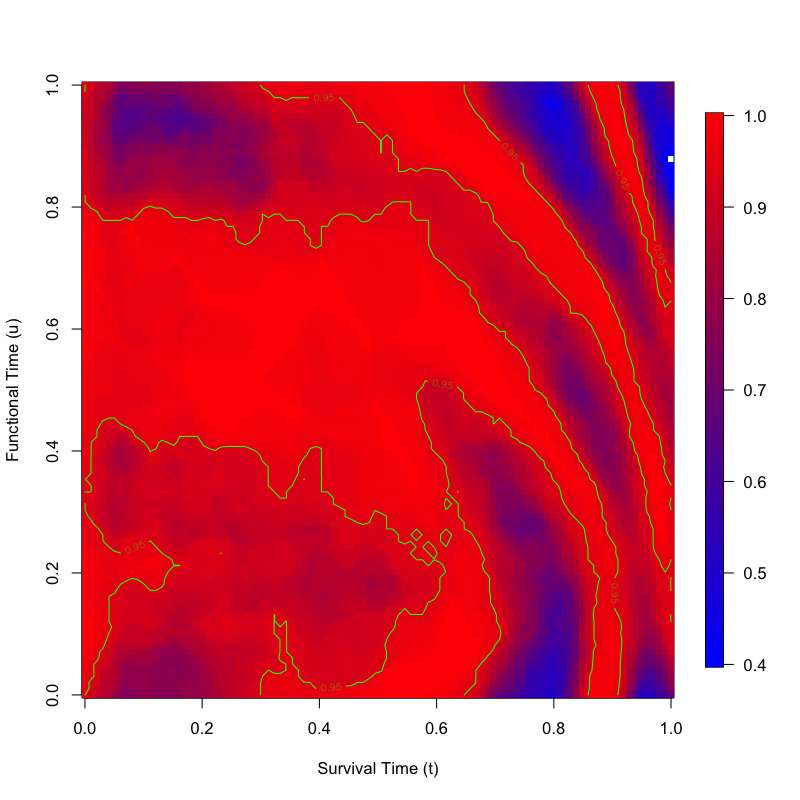} \\

 \midrule 

 Tests & Methods & N=2000 & N=3000 & N=4000 \\
 \cmidrule(lr){1-1}\cmidrule(lr){2-2}\cmidrule(lr){3-3}\cmidrule(lr){4-4}\cmidrule(lr){5-5}

 AMSE & w=0.04 & 0.169 & 0.137 & 0.133 \\

 & w=$\infty$ & 0.884 & 0.772 & 0.768 \\

 & Poisson & 0.159 & 0.126 & 0.119 \\
 
 Coverage Rate & CI & 92.0\% & 93.0\% & 94.0\% \\
\midrule
 Computation Time & w=0.04 & 40 seconds & 42 seconds & 47 seconds \\

& Poisson & 720 seconds & 924 seconds & 930 seconds \\
 \bottomrule
\end{tabular}
}
\end{table}

\newpage

\begin{table}[!ht]
\caption{The table of survival curves and histograms of event times for event indicators of 0 or 1 based on four different functional effects: $sin(2\pi u)/(t+0.5)$, $sin(2\pi u)/(t/2+1)$, $10cos\left\{4\pi(t-u)\right\}$, and $cos(2\pi(t^3-2/(u^2+1)))$}
\label{tab:table6}
\scalebox{0.95}{
\begin{tabular}{*{5}{c}}
\toprule
Plots & Function 1 & Function 2 & Function 3 & Function 4 \\

\midrule

Curves &
\includegraphics[width=0.2\textwidth, height=0.12\textheight]{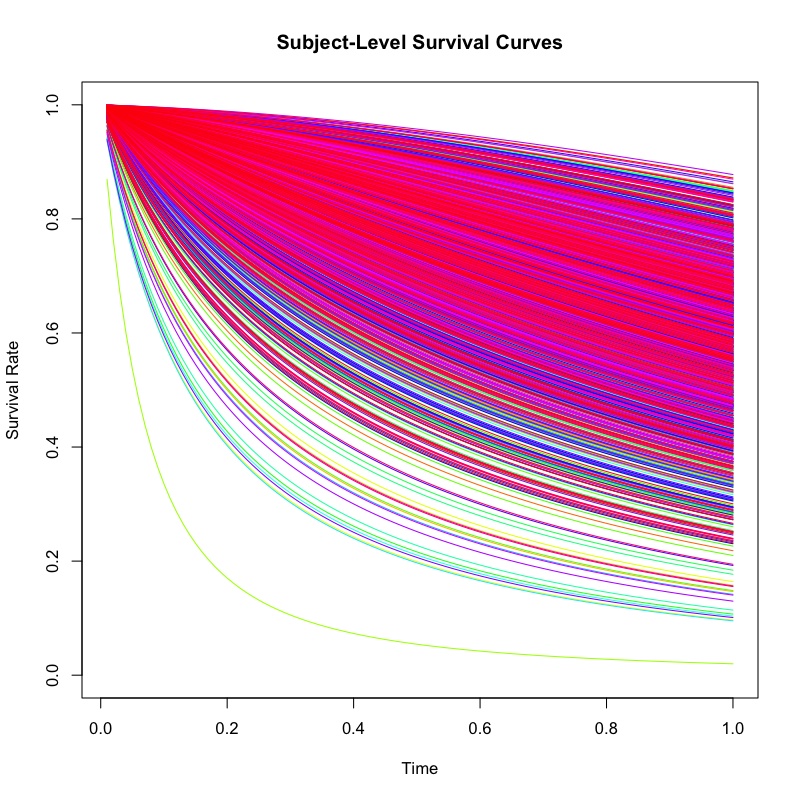} & \includegraphics[width=0.2\textwidth, height=0.12\textheight]{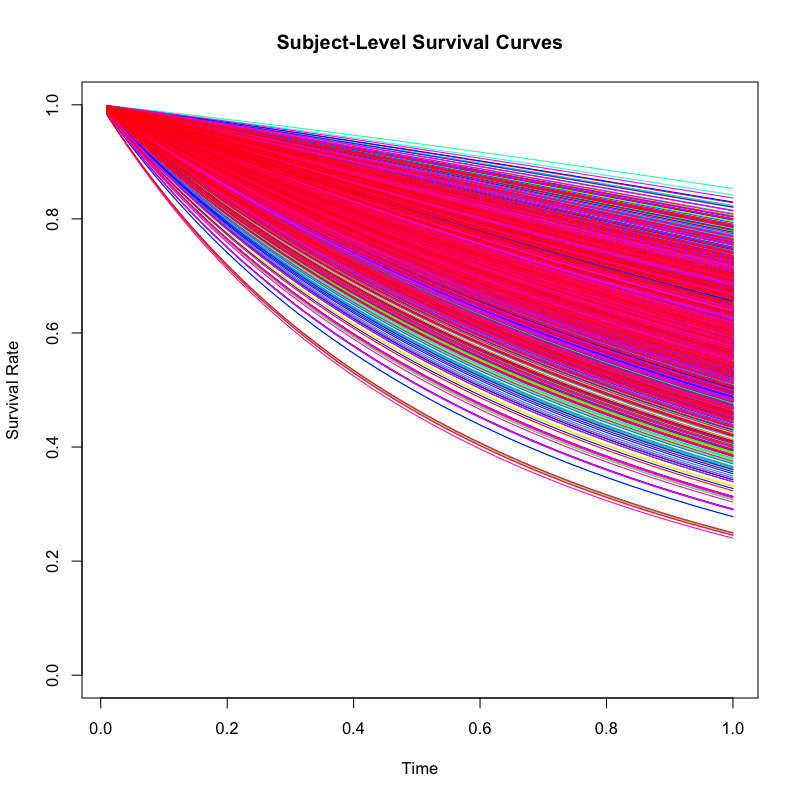} & \includegraphics[width=0.2\textwidth, height=0.12\textheight]{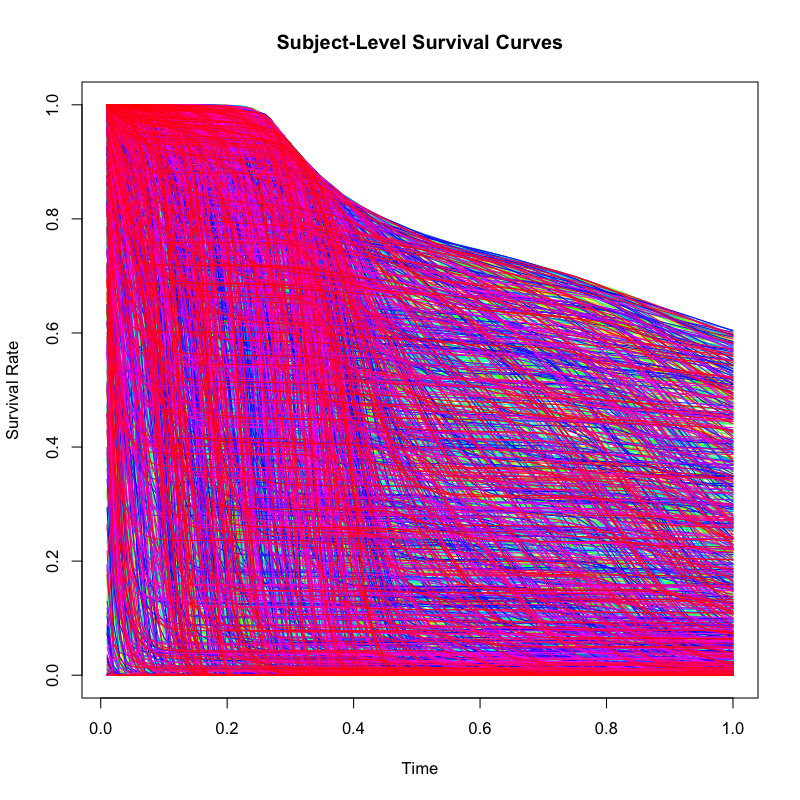} & \includegraphics[width=0.2\textwidth, height=0.12\textheight]{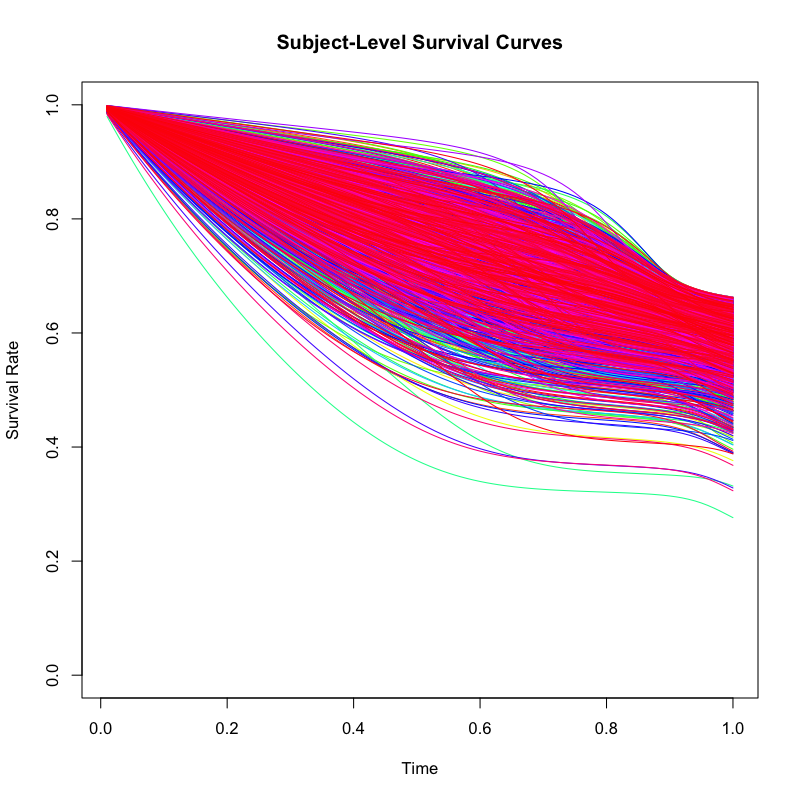} \\

 Censored &
 \includegraphics[width=0.2\textwidth, height=0.12\textheight]{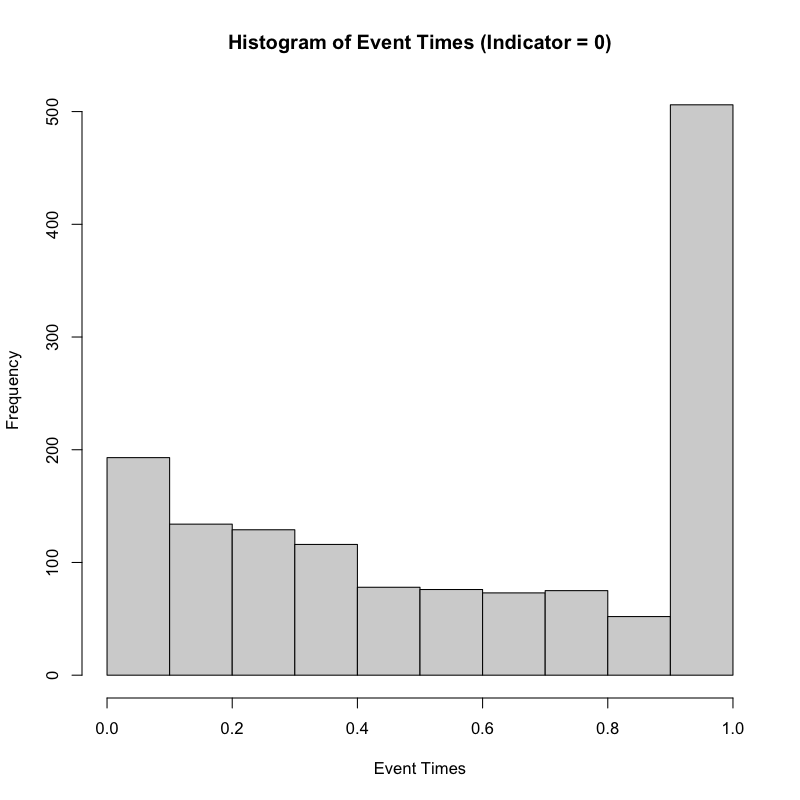} & \includegraphics[width=0.2\textwidth, height=0.12\textheight]{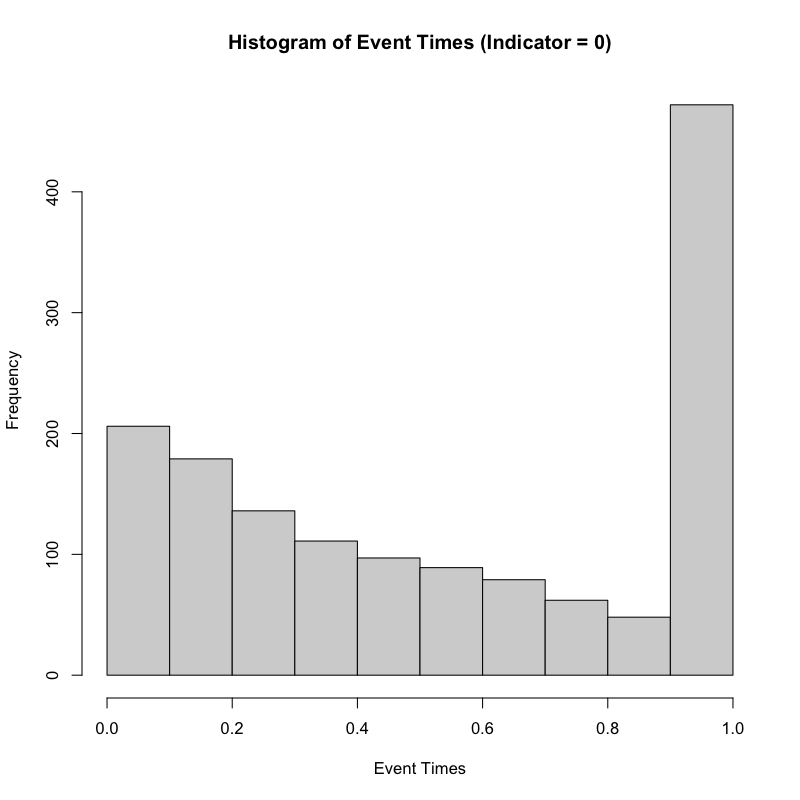} & \includegraphics[width=0.2\textwidth, height=0.12\textheight]{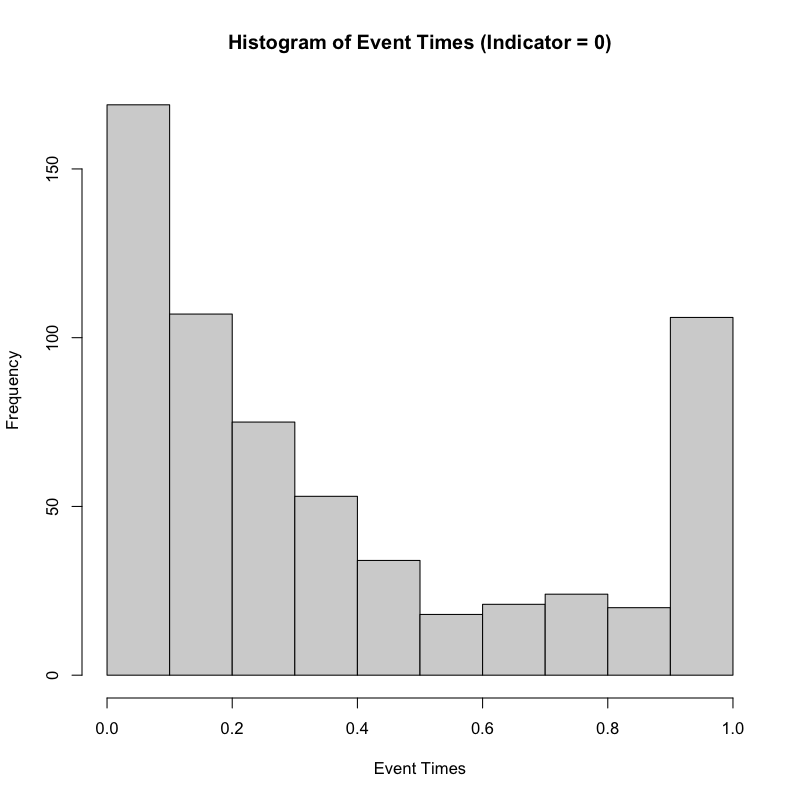} & \includegraphics[width=0.2\textwidth, height=0.12\textheight]{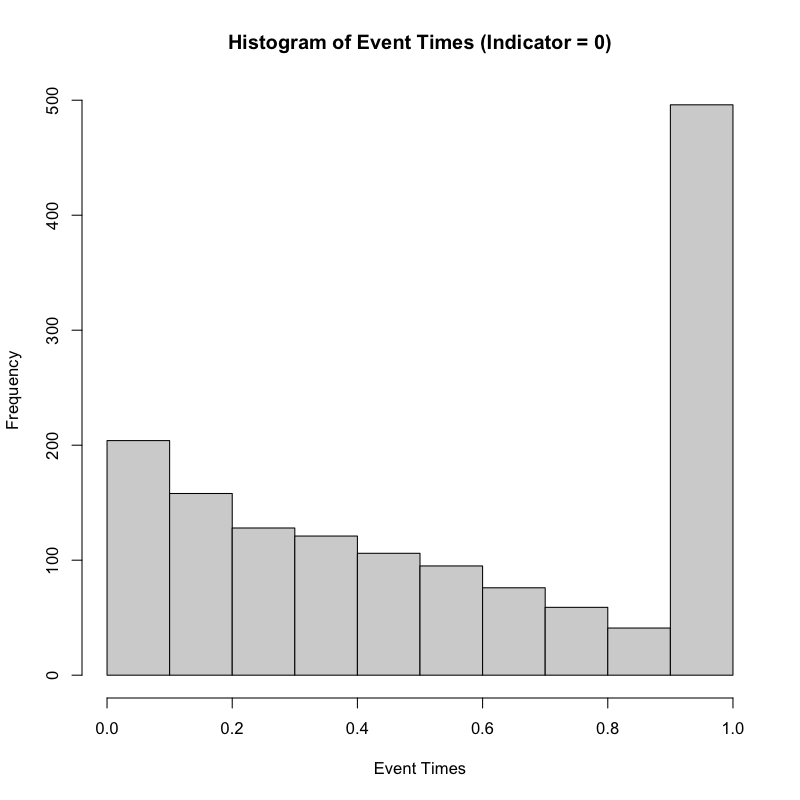} \\

 Non-Censored &
 \includegraphics[width=0.2\textwidth, height=0.12\textheight]{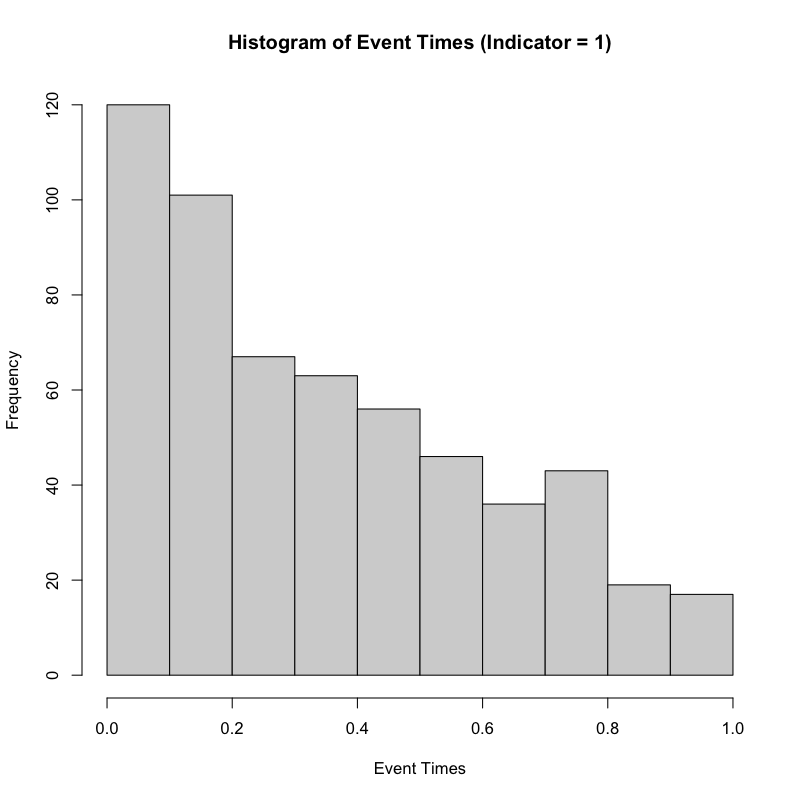} & \includegraphics[width=0.2\textwidth, height=0.12\textheight]{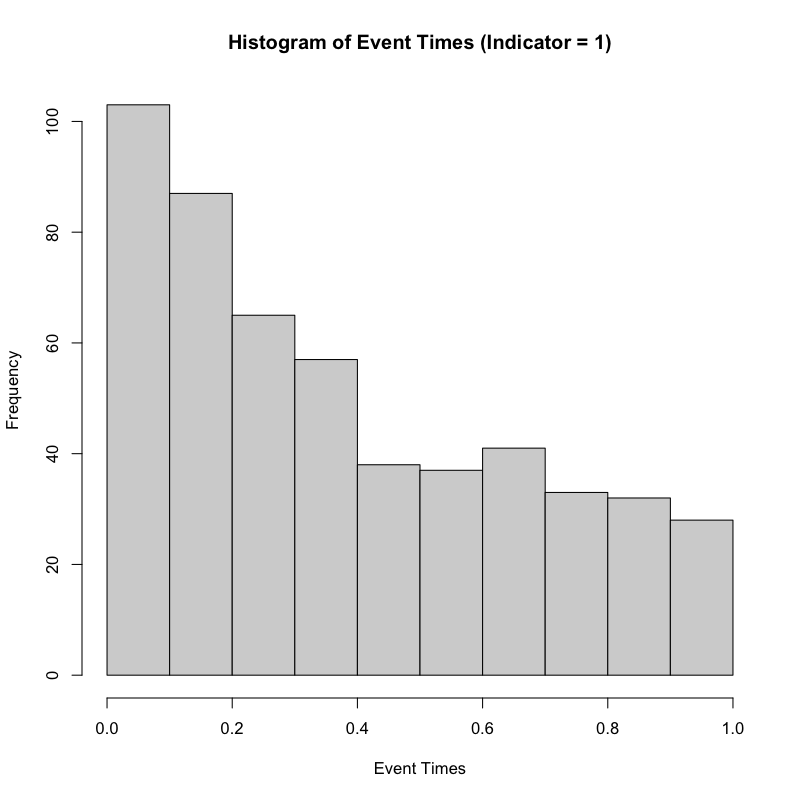} & \includegraphics[width=0.2\textwidth, height=0.12\textheight]{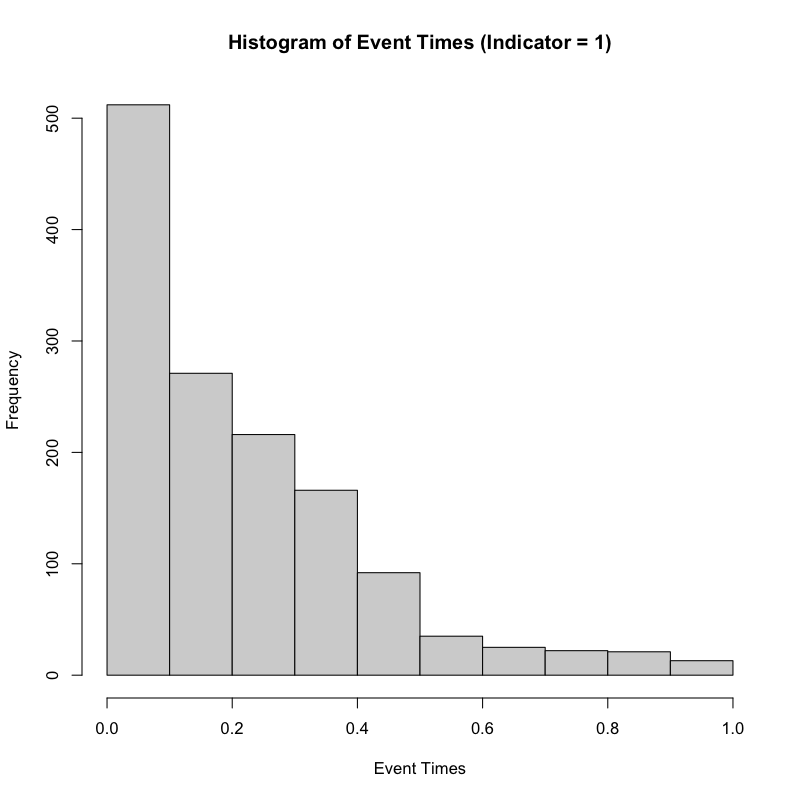} & \includegraphics[width=0.2\textwidth, height=0.12\textheight]{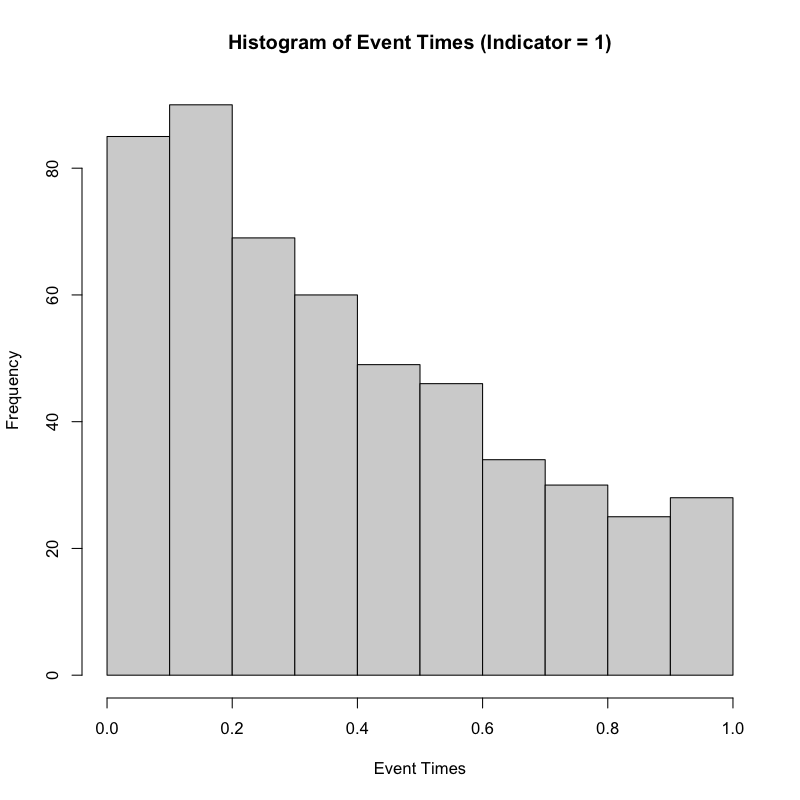} \\
 
 \midrule 

 Curves &
 \includegraphics[width=0.2\textwidth, height=0.12\textheight]{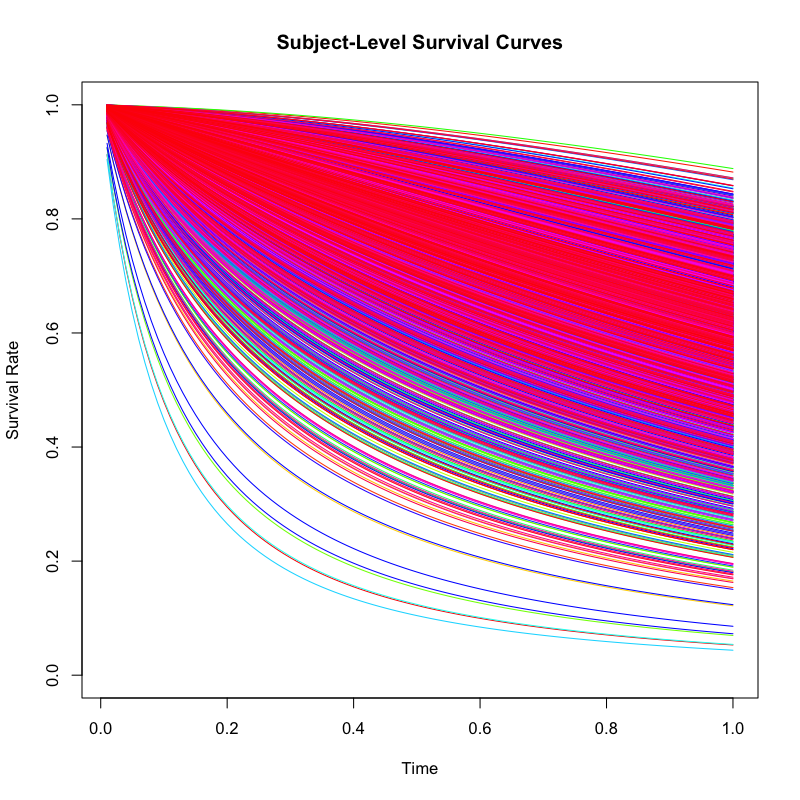} & \includegraphics[width=0.2\textwidth, height=0.12\textheight]{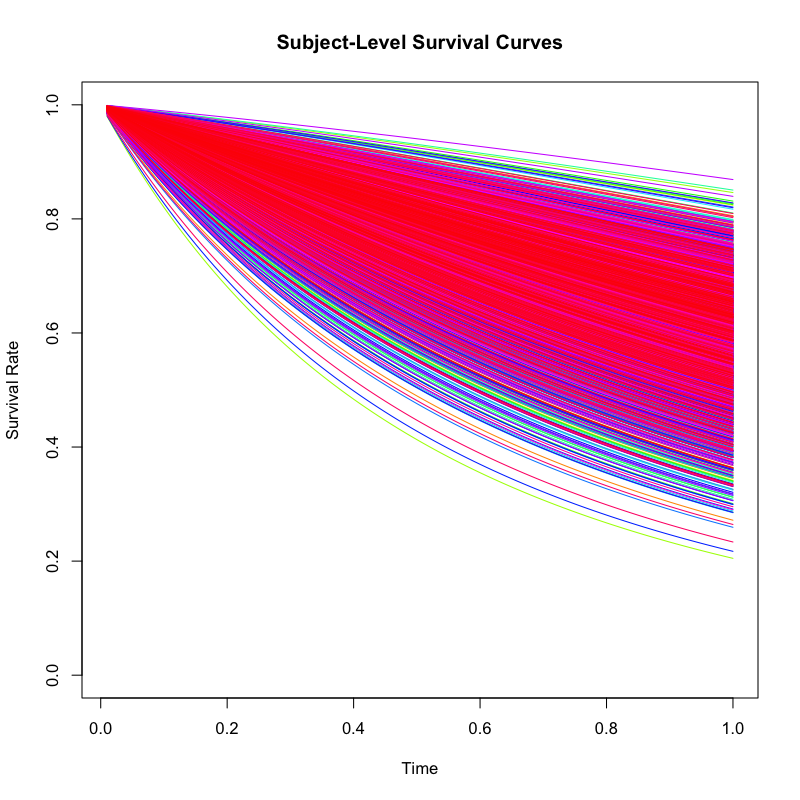} & \includegraphics[width=0.2\textwidth, height=0.12\textheight]{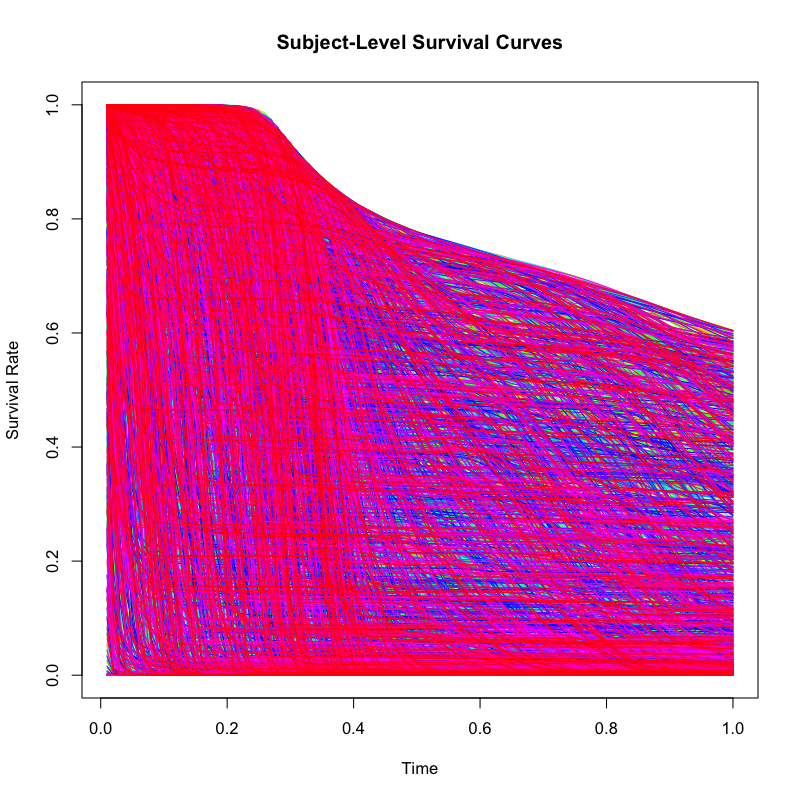} & \includegraphics[width=0.2\textwidth, height=0.12\textheight]{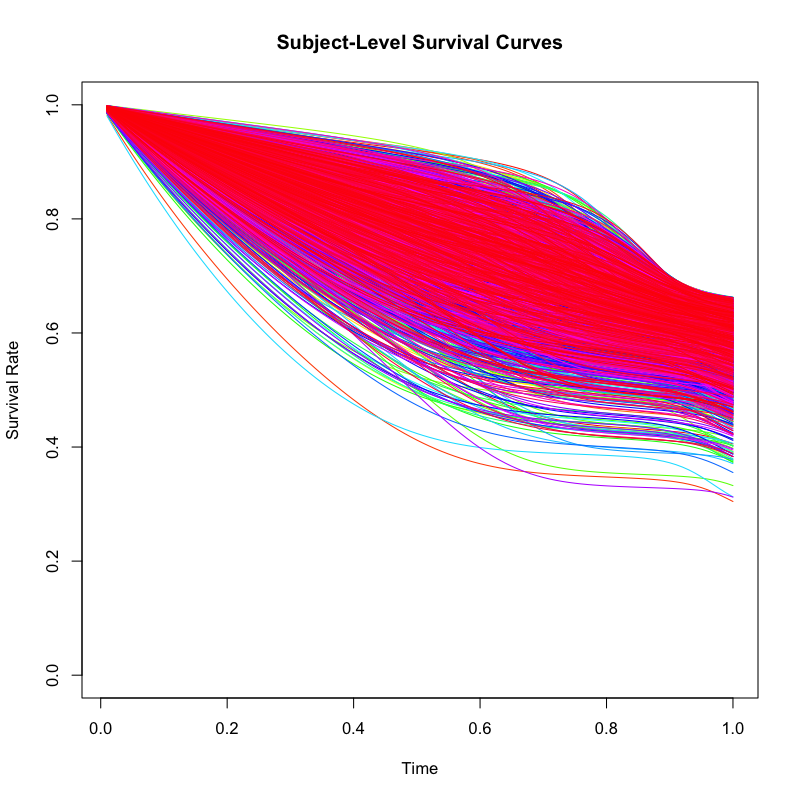} \\

 Censored &
 \includegraphics[width=0.2\textwidth, height=0.12\textheight]{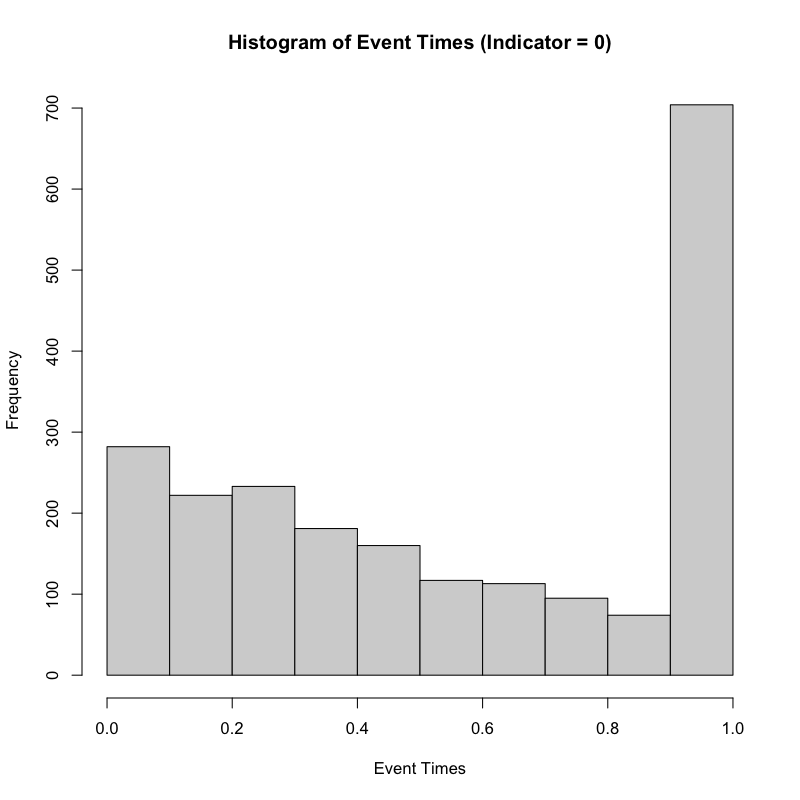} & \includegraphics[width=0.2\textwidth, height=0.12\textheight]{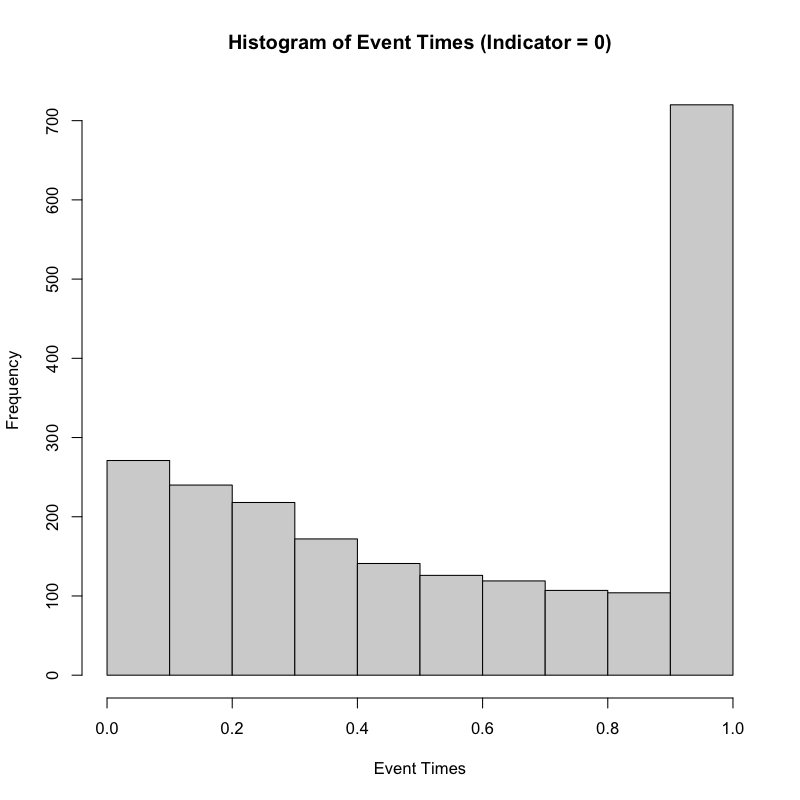} & \includegraphics[width=0.2\textwidth, height=0.12\textheight]{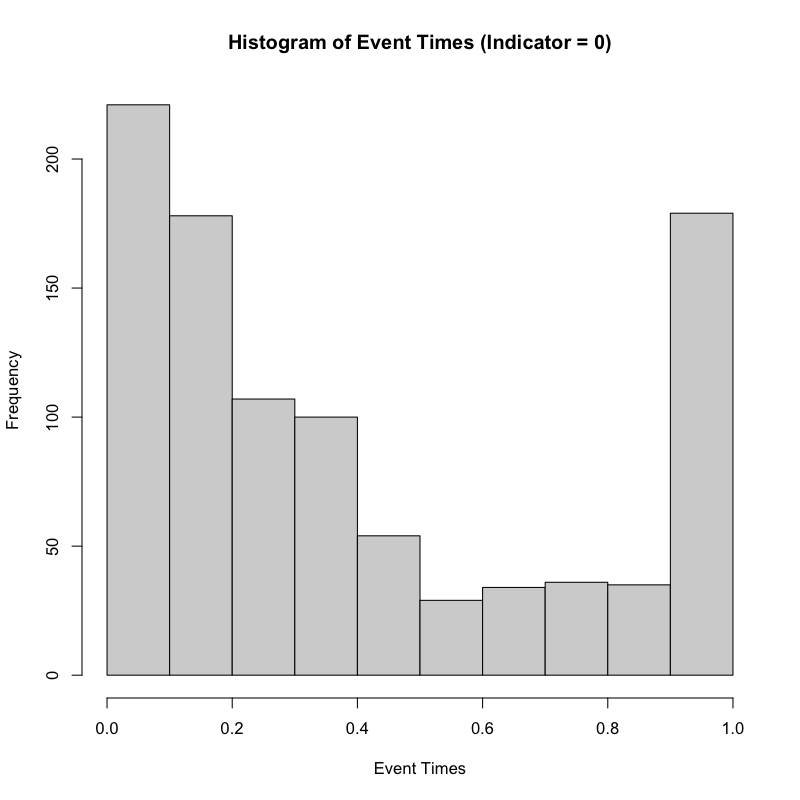} & \includegraphics[width=0.2\textwidth, height=0.12\textheight]{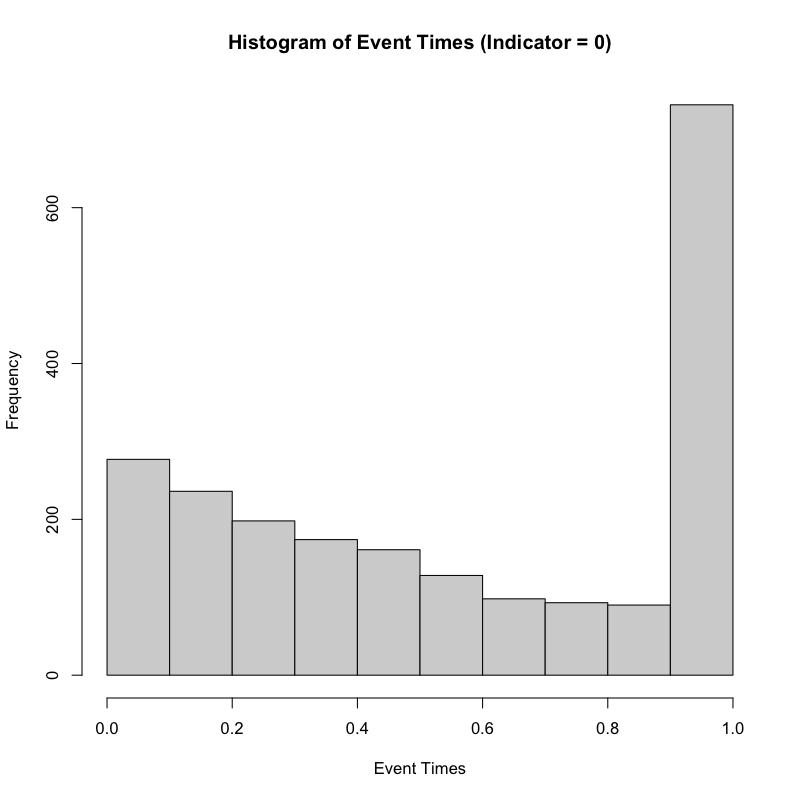} \\

 Non-Censored &
 \includegraphics[width=0.2\textwidth, height=0.12\textheight]{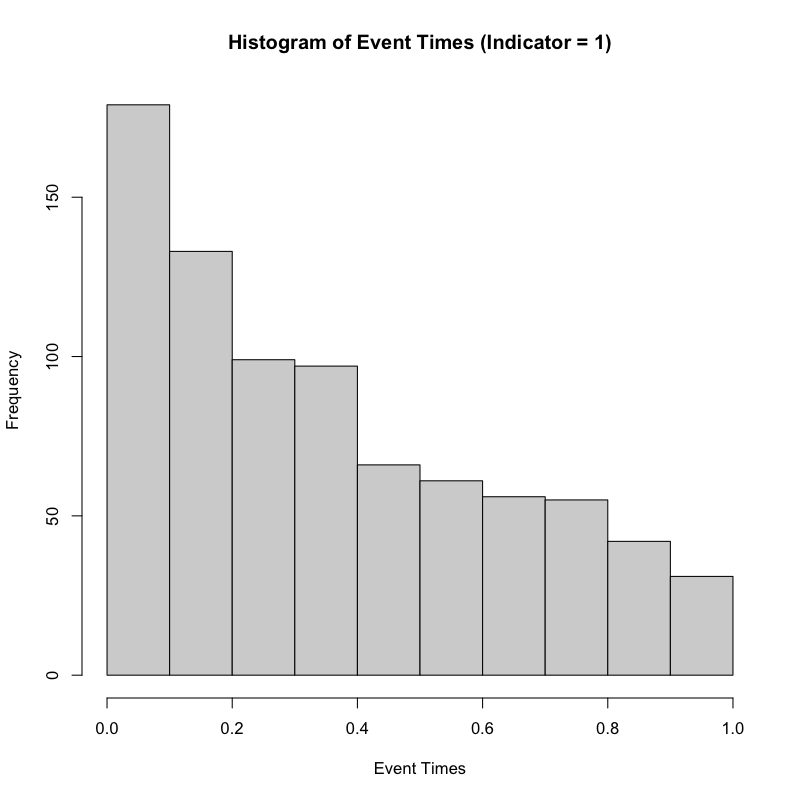} & \includegraphics[width=0.2\textwidth, height=0.12\textheight]{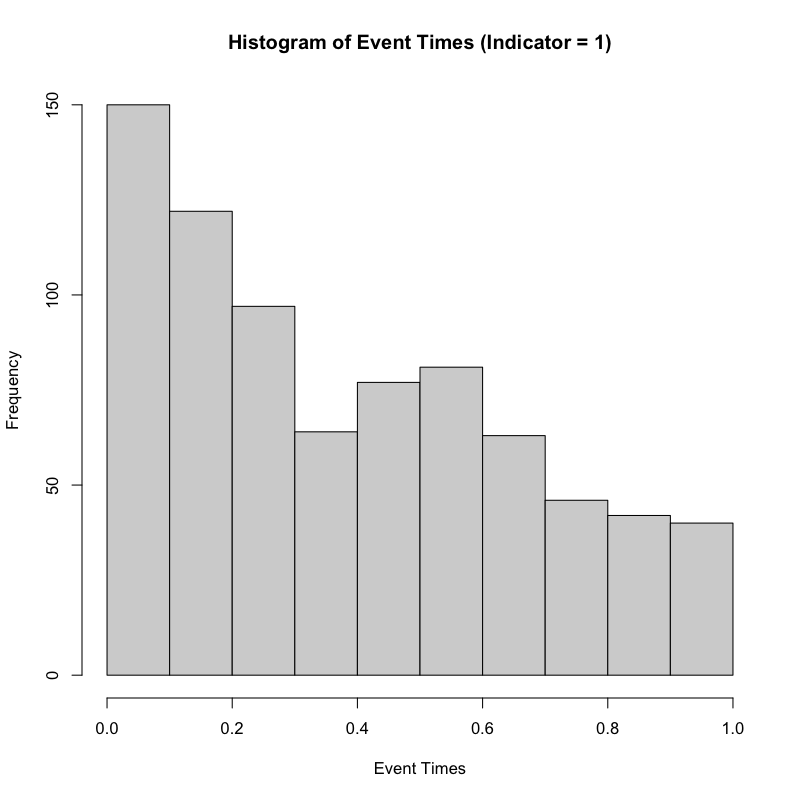} & \includegraphics[width=0.2\textwidth, height=0.12\textheight]{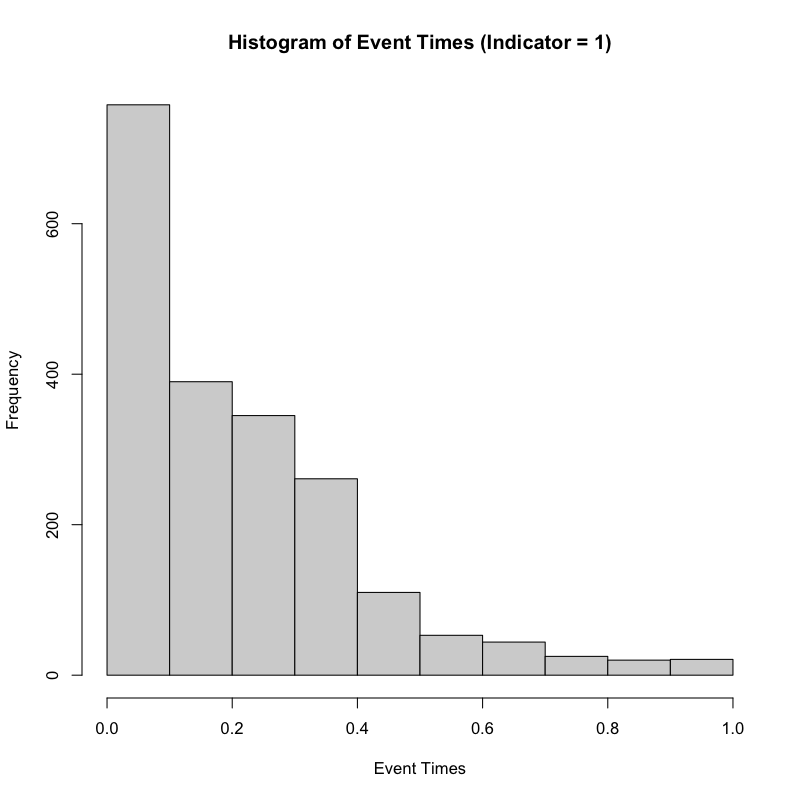} & \includegraphics[width=0.2\textwidth, height=0.12\textheight]{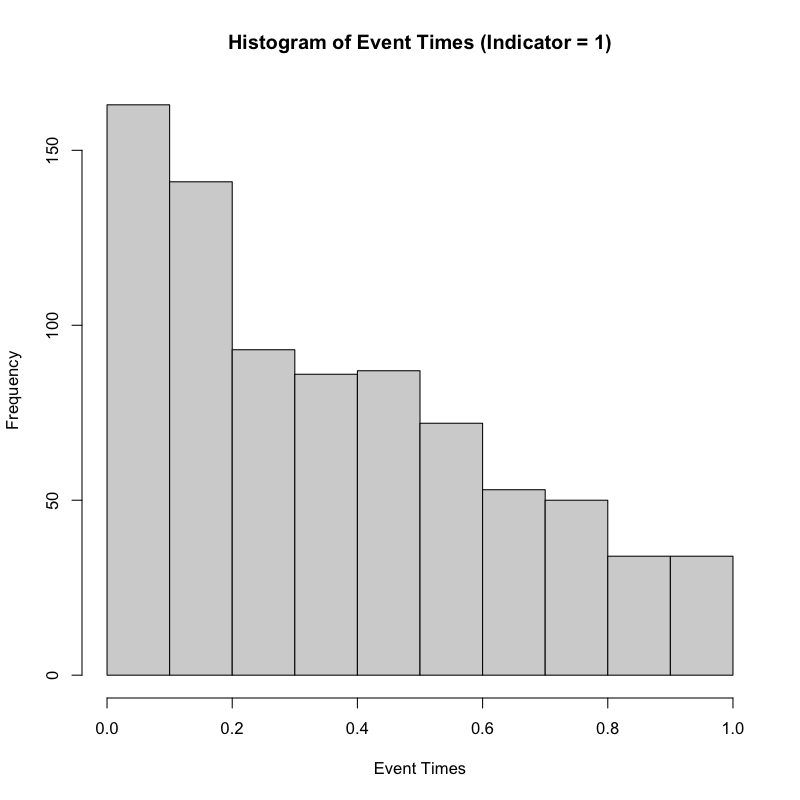} \\

 \bottomrule
\end{tabular}
}
\\
*The first three rows are based on sample size 2000. The second three rows are based on sample size 3000.
\end{table}

\end{document}